\documentclass[12pt]{iopart}
\usepackage{graphicx}
\begin{document}
\topical[Renormalization Group Methods for Reaction-Diffusion Problems]
{Applications of Field-Theoretic Renormalization Group Methods to 
Reaction-Diffusion Problems}


\author{Uwe C T\"auber \dag, 
        Martin Howard \ddag, and \\
        Benjamin P Vollmayr-Lee \P}

\address{\dag Department of Physics and
  Center for Stochastic Processes in Science and Engineering, 
  Virginia Polytechnic Institute and State University, 
  Blacksburg, Virginia 24061-0435, USA}
\address{\ddag Department of Mathematics, Imperial College London, 
  South Kensington Campus, London SW7 2AZ, UK}
\address{\P Department of Physics, Bucknell University, Lewisburg, 
  Pennsylvania 17837, USA}

\begin{abstract}
We review the application of field-theoretic renormalization group
(RG) methods to the study of fluctuations in reaction-diffusion
problems. We first investigate the physical origin of universality in
these systems, before comparing RG methods to other available analytic
techniques, including exact solutions and Smoluchowski-type
approximations. Starting from the microscopic reaction-diffusion 
master equation, we then pedagogically detail the mapping to a field
theory for the single-species reaction $k A \to \ell A$ ($\ell < k$). 
We employ this particularly simple but non-trivial system to introduce
the field-theoretic RG tools, including the diagrammatic perturbation
expansion, renormalization, and Callan--Symanzik RG flow equation. We
demonstrate how these techniques permit the calculation of universal
quantities such as density decay exponents and amplitudes via
perturbative $\epsilon = d_c - d$ expansions with respect to the upper
critical dimension $d_c$. With these basics established, we then 
provide an overview of more sophisticated applications to multiple
species reactions, disorder effects, L\'evy flights, persistence
problems, and the influence of spatial boundaries. We also analyze
field-theoretic approaches to nonequilibrium phase transitions
separating active from absorbing states. We focus particularly on the
generic directed percolation universality class, as well as on the 
most prominent exception to this class: even-offspring branching and
annihilating random walks. Finally, we summarize the state of the 
field and present our perspective on outstanding problems for the 
future.
\end{abstract}

submitted to \JPA \today

\maketitle

\section{Introduction}

Fluctuations and correlations in statistical systems are well known to
become large in the vicinity of a critical point. In this region,
fluctuations have a profound influence on the macroscopic properties 
of the system, leading to singular thermodynamic behavior 
characterized by universal critical exponents and scaling functions. 
These power-law singularities can be traced to an underlying emerging 
symmetry, namely scale invariance: at the critical point, the system 
possesses a diverging correlation length. Therefore renormalization 
group (RG) methods, which explicitly address the behavior of physical 
observables under scale transformations, have been employed with 
considerable success in describing critical fluctuations. The 
renormalization group provides a natural conceptual framework for 
explaining the occurrence of critical behavior, the emergence of 
universality, and the classification of different systems in terms of 
universality classes. Moreover, RG tools (especially in conjunction 
with series resummations or numerical implementations) also enable 
quantitative, controlled calculations of universal properties.

Most successful applications of the renormalization group address
systems in thermal equilibrium, where the Boltzmann--Gibbs probability
distribution provides a solid foundation for explicit calculations. 
However, many systems in nature cannot be cast into even an 
approximative equilibrium description, and a large variety of 
nonequilibrium systems, both relaxational and driven, also exhibit
critical behavior, which should presumably also be able to be analyzed
by RG techniques. Unfortunately, even for nonequilibrium steady 
states, we presently lack a general statistical framework to construct
the corresponding probability distributions and hence obtain the
relevant macroscopic quantities. Consequently, there are relatively
few cases where such explicit calculations can be developed, at least
to date. For systems that can be represented in terms of stochastic
partial differential equations of the Langevin type, there exists a
well-established mapping to a field-theoretic representation
\cite{Bausch76}. However, this inherently coarse-grained, mesoscopic
approach relies on an {\em a priori} identification of the relevant
slow degrees of freedom. Moreover, far from equilibrium there are no
Einstein relations that constrain the generalized stochastic forces or
noise terms in these Langevin equations. Thus, although the functional
form of the noise correlations may crucially affect the long-time,
long-wavelength properties, it often needs to be inferred from
phenomenological considerations, or simply guessed.

Fortunately, in certain cases an alternative approach exists which
allows these fundamental difficulties to be at least partially 
overcome. This method relies on the introduction of a `second
quantized' ladder operator formulation \cite{Doi76a,Grassberger80} for 
certain classical master equations, and on the coherent state 
representation to construct the statistical path integral 
\cite{Peliti85}. This in principle permits a mapping to a field theory
starting directly from a microscopically defined stochastic process
without invoking any further assumptions or approximations beyond
taking the appropriate continuum limit. In this way, the difficulties
of identifying the slow variables, and of guessing the noise
correlations, are circumvented, and a field theory can be
straightforwardly derived. Subsequently, the entire standard
field-theoretic machinery can then be brought to bear, and progress
made on understanding the role of fluctuations, and on identifiying 
and computing universal quantities. (As a cautionary note we add,
though, that the above-mentioned continuum limit may not always be
trivial and benign, and occasionally additional physical insight needs 
to be invoked to obtain an appropriate {\em effective} field theory 
description.) In this overview, we will concentrate on the application 
of field-theoretic RG techniques to the nonequilibrium dynamics of 
reaction-diffusion systems. Such models consist of classical particles 
on a lattice, which evolve by hopping between sites according to some 
specified transition probabilities. The particles can also interact, 
either being created or destroyed on a given site following prescribed 
reaction rules (for general reviews, see \cite{Kuzovkov88,
Ovchinnikov89}). However, in order to successfully employ RG methods 
to these systems, one necessary feature is a mean-field theory that is 
valid for some parameter range, typically when the spatial dimension 
$d$ exceeds an {\em upper critical dimension} $d_c$. This property 
renders the renormalization group flows accessible within perturbation 
theory, via a dimensional expansion in $\epsilon = d_c - d$. For 
reaction-diffusion models this requirement is usually easily 
satisfied, since the mean-field theory is straightforwardly given by 
the `classical' rate equations of chemical kinetics.

It is the purpose of this topical review to provide an introduction to
the methods of the field-theoretic RG in reaction-diffusion systems,
and to survey the body of work that has emerged in this field over the
last decade or so. Other theoretical and numerical simulation 
approaches will not be as systematically covered, though various
results will be mentioned as the context requires. A review by Mattis
and Glasser \cite{Mattis98} also concerns reaction-diffusion models
via Doi's representation, but does not address the RG methods
presented here. Recent overviews by Hinrichsen \cite{Hinrichsen00} and
\'Odor \cite{Odor04} are primarily concerned with classifying
universality classes in nonequilibrium reaction-diffusion phase
transitions via Monte Carlo simulations. We will also touch on this
topic in our review (for brief summaries of the RG approach to this
problem, see also Refs.~\cite{Taeuber02,Taeuber03}). The field theory 
approach to directed and dynamic isotropic percolation, as based on a 
mesoscopic description in terms of Langevin stochastic equations of
motion, is discussed in depth in Ref.~\cite{Janssen05}. As with the 
previous reviews, our presentation will concentrate on theoretical
developments. Unfortunately, experiments investigating fluctuations in
reaction-diffusion systems are deplorably rare. Three notable 
exceptions are: (i) the unambiguous observation of a $t^{-1/2}$ 
density decay (in an intermediate time window) for the 
diffusion-limited fusion process $A + A \to A$, as realized in the
kinetics of laser-induced excitons in quasi one-dimensional
N(CH$_3$)$_4$MnCl$_3$ (TMMC) polymer chains \cite{Kroon93}; (ii) the
demonstration of non-classical $A + B \to C$ kinetics, with an
asymptotic $t^{-3/4}$ density decay in three dimensions in a calcium /
fluorophore system \cite{Monson04}; and (iii) the identification of
directed percolation critical exponents in studies of spatio-temporal
intermittency in ferrofluidic spikes \cite{Rupp03}.

The review is organized as follows. The following section provides a
basic introduction to reaction-diffusion models and the various
approaches employed for their investigation. \Sref{sec:mapping}
describes the mapping of classical reaction-diffusion models onto a
field theory, while \sref{sec:renormalization} presents the RG
techniques in the context of the $k A \to \ell A$ $(\ell < k)$
annihilation reaction. A selection of other single-species reactions 
is treated in \sref{sec:furtherapp}, where variations such as L\'evy
flight propagation and the influence of disorder are also considered. 
In addition, this section covers the two-species annihilation 
reaction $A + B \to 0$ with homogeneous and segregated initial 
conditions, as well as with disorder and shear flow. Other 
multi-species reactions that exhibit similar asymptotic decay are also
discussed here. \Sref{sec:transitions} deals with directed 
percolation, branching-annihilating random walks, and other examples
of nonequilibrium phase transitions between active and inactive /
absorbing states. The influence of spatial boundaries is also
addressed here. Finally, in \sref{sec:future}, we give our perspective
on open problems for future studies.

\section{Basic Features of Reaction-Diffusion Systems}
\label{sec:basic}

\subsection{Models}

Our goal is to describe local reactions, of either a creation or
annihilation type, for which the particles rely on diffusion (or
nearest-neighbor hopping) to be brought within reaction range. Hence,
these processes are often referred to as {\em diffusion-limited}
reactions. Some single-species examples are the pair annihilation
reaction $A + A \to 0$, where `$0$' denotes a chemically inert
compound, and coagulation $A + A \to A$. The diffusive particle
propagation can be modeled as a continuous- or discrete-time random
walk, either on a lattice or in the continuum. Reactions occur when
particles are within some prescribed range; on a lattice, they can
also be required to occupy the same lattice site. In such systems a
single lattice site may be subject to an occupancy restriction (to,
say, at most $n_{\rm max}$ particles per site) or not, and, of course,
the lattice structure can be varied (e.g., square or triangular in two
dimensions). Computer simulations typically employ discrete time
random walks, whereas, for example, the analysis of the two-species
pair annihilation reaction $A + B \to 0$ by Bramson and Lebowitz
(discussed in \sref{subsec:relation}) uses a continuous time random
walk on a lattice with unlimited occupation number, but also with an
infinite reaction rate so that no site simultaneously contains $A$ and
$B$ particles \cite{Bramson88}. With such a variety of microscopic
models to represent a single reaction type, it is important to
determine which features are {\em universal} as opposed to those
properties that depend on the specific implementation of the processes
under consideration.

The most general single-species reaction-diffusion system can be
described by means of a set of reaction rules, the $i$th of which 
reads $k_i A \to \ell_i A$, with non-negative integers $k_i$, 
$\ell_i$, and where each process occurs with its own rate or
probability per time step. Notice that this includes the possibility
of reversible reactions, for example $2 A \leftrightarrow A$ as
represented by $(k_1,\ell_1)=(2,1)$ and
$(k_2,\ell_2)=(1,2)$. Similarly, directed percolation, which defines a
broad universality class of nonequilibrium phase transitions between
active and absorbing states, can be described by the reactions $A + A
\leftrightarrow A$ and $A \to 0$, where the critical point is reached 
through tuning the rates of the $A \to (0,2A)$ reactions (see
\sref{sec:transitions}).

More generic in chemical systems are two-species processes, for 
example $A + B \to 0$, which requires particles of {\em different}
types to meet in order for the reaction to occur. The different
particle species may or may not have the same diffusion constant. A
general multi-species reaction may be written as $\sum_j k_j A_j \to
\sum_j \ell_j A_j$, where $A_j$ labels the $j$th species, and the most
general reaction model is then a set of such multi-species processes.

With such generality available, it is possible to construct both
driven and relaxational systems. The former case, which includes
directed percolation, typically comprises both reactions that increase
and decrease the particle number. Depending on appropriate
combinations of the corresponding reaction rates, the ensuing
competition can, in the thermodynamic limit and at long times, either 
result in an `active' state, characterized by a finite steady-state 
density of particles, or a situation that evolves towards the empty 
lattice. For reactions that require at least the presence of a single 
particle, the latter case constitutes an `inactive' or `absorbing' 
state with vanishing fluctuations from which the system can never 
escape. The continuous transition from an active to an absorbing state 
is analogous to a second-order equilibrium phase transition, and 
similarly requires tuning of appropriate reaction rates as control 
parameters to reach the critical region. As in equilibrium, 
universality of the critical power laws emerges as a consequence of
the diverging correlation length $\xi$, which induces scale invariance
and independence in the critical regime of microsopic parameters. 
Alternatively, relaxational cases, such as the single-species pair 
annihilation reaction $A + A \to 0$, are ultimately decaying to an 
absorbing state: the empty lattice (or a single left-over particle). 
However, here it is the asymptotic decay law that is of interest, and 
the scaling behavior of the correlation functions in the universal
regime that is often reached at large values of the time variable $t$.

\subsection{The origin of universality in relaxational reactions}

Reaction-diffusion models provide a rather intuitively accessible
explanation for the origin of universality. The large-distance
properties of random walks are known to be universal, depending only
on the diffusion constant, a macroscopic quantity. Decay processes,
such as $A + A \to 0$, eventually result in the surviving particles
being separated by large distances, so at late times the probability
of a pair of particles diffusing to close proximity takes on a
universal form. For spatial dimension $d \leq 2$ random walks are
re-entrant, which enables a pair of particles to find each other with
probability one, even if they are represented by points in a
continuum. Therefore, at sufficiently long times the {\em effective}
reaction rate will be governed by the limiting, universal probability
of the pair diffusing from a large to essentially a zero relative
separation. As it turns out, this asymptotically universal reaction
rate is sufficient to fully determine the entire form of the leading
density decay power law, that is, both its {\em exponent} and the 
{\em amplitude} become universal quantities.

The situation is different in dimensions $d > 2$: the probability for
the pair of particles to come near each other still has a universal
form, but even in close proximity the reactants actually meet with
probability zero if they are described as point particles in a
continuum. For any reaction to occur, the particles must be given a
finite size (or equivalently, a finite reaction range), or be put on a
lattice. Since the ensuing finite effective reaction rate then clearly
depends on the existence and on the microscopic details of a
short-distance (ultraviolet) regulator, universality is (at least
partially) lost.

Consequently, for any two-particle reaction, such as $A + A \to (0,A)$
or $A + B \to 0$, we infer the upper critical dimension to be 
$d_c = 2$. There is some confusion on this point in the literature for
the two-species process $A + B \to 0$, for which sometimes $d = 4$ is
claimed to be the upper critical dimension. This is based on the
observation that for equal initial $A$ and $B$ particle densities the
asymptotic density power-law decay becomes $\sim t^{-d/4}$ for $d < 4$
and $t^{-1}$ for $d \geq 4$. However, this behavior is in fact fully 
exhibited within the framework of the {\em mean-field} rate equations, 
see \sref{subsec:rate_eq} below. Thus $d = 4$ does not constitute an 
upper critical dimension in the usual sense (namely, that for 
$d < d_c$ fluctuations are crucial and are not adequately captured 
through mean-field approximations). Yet surprisingly, in the 
two-species pair annihilation process there occurs no marked 
qualitative change at two dimensions for the case of equal initial 
densities. However, the critical dimension $d_c = 2$ strongly impacts 
the scenario with unequal initial densities, where the minority 
species decays exponentially with a presumably nonuniversal rate for 
all $d > 2$, exhibits logarithmic corrections in $d_c = 2$, and decays
according to a stretched exponential law \cite{Bramson88,Bramson91b} 
with universal exponent and probably also coefficient \cite{Bray02,
Blythe03} for $d < 2$ (see \sref{subsec:CSequation}). The critical 
dimension is similarly revealed in the scaling of the reaction zones, 
which develop when the $A$ and $B$ particles are initially segregated 
(\sref{subsec:twospecies}).

The upper critical dimension is not always two in diffusion-limited
processes. For three-particle reactions, such as $3A \to 0$, the
upper critical dimension becomes $d_c = 1$, via the same mechanism as
described above: for three point particles to meet in a continuous
space, they must be constrained to one dimension. For a $k$th order
decay reaction $k A \to \ell A$ (with $\ell < k$), the upper critical
dimension is generally found to be $d_c = 2 / (k-1)$ \cite{Lee94a}. 
Consequently, mean-field descriptions should suffice in any physical 
dimension $d \geq 1$ for $k > 3$. However, this simple argument is 
not necessarily valid once competing particle production processes 
are present as well. For example, for the universality class of 
directed percolation that describes generic phase transitions from 
active to absorbing states, the upper critical dimension is shifted 
to $d_c = 4$, as a result of combining particle annihilation and 
branching processes (see \sref{sec:transitions}). As mentioned before,
universal features near the transition emerge as a consequence of a 
diverging correlation length, just as for equilibrium critical points.

Lattice occupation restrictions typically do not affect universality
classes for relaxational reactions, since the asymptotically low
densities essentially satisfy any occupation restrictions. A few
exceptions are noteworthy:

\noindent (i) The asymptotic decay law in the $A + B \to 0$ reaction
with equal $A$ and $B$ densities and in $d<4$ dimensions depends on
the fluctuations in the initial conditions, which in turn are expected
to be sensitive to lattice occupation restrictions 
\cite{Lee95,ODonoghue01}.

\noindent (ii) In systems with purely second- or higher-order 
reactions, site occupation restrictions crucially affect the
properties of the active phase and the phase transition that separates
it from the absorbing state \cite{Howard97,Henkel04,Janssen04}, see
\sref{sec:transitions}.

\noindent (iii) One-dimensional multi-species systems can be
constructed in which the spatial ordering of the reaction products
specified by the model, which cannot subsequently be changed by the
dynamics, does affect the asymptotic properties \cite{Kwon00,Odor02}.

\subsection{Rate equations}
\label{subsec:rate_eq}

In general, kinetic rate equations are obtained by taking the rate of
change of a given species' density or concentration to be proportional
to the appropriate product of the reactant densities and the reaction
rate. This effectively constitutes a factorization of higher-order
correlation functions (the joint probability of finding a given number
of reactants at the same location at a given time), and hence
corresponds to a mean-field type approximation. For example, in the
process $k A\to \ell A$ the probability of a reaction is assumed to be
proportional to $a(t)^k$, where $a(t)$ denotes the overall (mean) $A$
particle density at time $t$. Such a description that entirely
neglects correlations and spatial variations is in general justified
only if the reactants remain uncorrelated and homogeneously
distributed (and well-mixed for different participating species)
throughout the system's temporal evolution. The corresponding rate
equation then reads
\begin{equation}
\label{eq:kArate_eq}
     \partial_t a(t) = - (k - \ell) \, \lambda \, a(t)^k ~ ,
\end{equation}
where $\lambda$ represents a reaction rate constant, and the loss rate 
is proportional to $k - \ell$, the number of particles removed by the
reaction. We assume that $\ell < k$, so $\partial_t a$ is negative. 
Notice that in contrast to $k$, the integer variable $\ell$ does not 
enter the functional form of the rate equation. With an initial 
density $a_0$, Eq.~\eref{eq:kArate_eq} is solved by
\begin{equation}
\label{eq:kArate_sol}
     a(t) = {a_0 \over [1 + a_0^{k-1} (k-1)(k-\ell) 
     \lambda t]^{1/(k-1)}} ~ ,
\end{equation} 
which for $t \gg 1 / \lambda a_0^{k-1}$ leads to the asymptotic decay 
$a \sim (\lambda t)^{-1/(k-1)}$, independent of the initial density 
$a_0$.

Next consider an inhomogeneous system with a {\em local} density
$a({\bf x},t)$ that is assumed to be slowly varying on the scale of 
the capture radius or lattice size. The rate equation approximation 
for uncorrelated reactants can still be applied; however, the local 
density may now evolve not just via the reactions but also through 
diffusive particle motion. Since the latter process is simply linear 
in the density, we directly add a diffusion term to the rate equation,
\begin{equation}
     \partial_t a({\bf x},t) = D \nabla^2 a({\bf x},t) 
     - (k - \ell) \lambda \, a({\bf x},t)^k ~ .
\end{equation}

For two-species reactions a pair of rate equations is required. For
example, the pair annihilation process $A + B \to 0$ is represented
through
\begin{equation}
\label{eq:ab_rate}
     \partial_t a = D_A \nabla^2 a - \lambda \, a b ~ , \qquad
     \partial_t b = D_B \nabla^2 b - \lambda \, a b ~ 
\end{equation}
for the local particle densities $a({\bf x},t)$ and $b({\bf x},t)$. If
both densities may be taken to be uniform throughout the temporal
evolution, then their decay is just described by $a(t) \sim b(t) \sim
(\lambda t)^{-1}$, as in Eq.~\eref{eq:kArate_sol} with $k=2$. However,
this assumption is not always justified \cite{Ovchinnikov78}. Notice
that the difference in $A$ and $B$ particle numbers is locally
conserved by the annihilation reaction.  Correspondingly, in the case
of equal diffusion constants ($D_A = D_B$), the difference field
$a({\bf x},t) - b({\bf x},t)$ satisfies the diffusion equation. Thus
spatial inhomogeneities relax rather slowly. For the case of 
{\em equal} initial $A$ and $B$ densities $n_0$, Toussaint and Wilczek 
(TW) observed that the fluctuations in the initial density of this
diffusive field in fact determine the long-time characteristics of the
two-species annihilation process \cite{Toussaint83}. With the
additional assumption of asymptotic particle segregation into
separated $A$ and $B$ rich domains, which makes the spatially averaged
$a$ and $b$ densities exactly half of the $|a-b|$ average, TW obtained
the long-time behavior
\begin{equation}
     a(t) \sim b(t) \sim {\sqrt{n_0}\over \sqrt\pi (8\pi Dt)^{d/4}} ~.
\end{equation}
This decay is considerably slower than the one predicted by the
uniform rate equation, and thus dominates for $t \to \infty$, provided
$d < 4$. Again we emphasize that the ensuing qualitative changes in 
four dimensions merely reflect the importance of spatial 
inhomogeneities within the mean-field rate equation.

\subsection{Relation between RG and other methods}
\label{subsec:relation}

The rate equations of the previous section still play an important 
role in the field-theoretic analysis that aims to systematically
include spatial flucuations and correlations. In particular, the rate
equation solution represents the zeroth-order term in a loop expansion
for the density, also called the tree diagram sum. Above the upper
critical dimension, the higher-order terms in the loop expansion only
serve to modify the rate constant in some nonuniversal way. Thus it
was shown in Ref.~\cite{Lee95} that the rate equations
\eref{eq:ab_rate} are asymptotically valid {\em without} approximation
when $d > 2$. For $d \leq 2$, the higher-order terms in a loop
expansion provide divergent corrections, which then must be regulated
(for example, by introducing a lattice) and RG methods brought to bear
to extract a systematic $\epsilon$ expansion. In this case, it is
found that the rate equation solution gives rise, under RG flow, to 
the leading-order term in the $\epsilon$ expansion. As we shall see in
subsequent sections, the structure of a loop expansion correcting the
rate equation solution holds even for more complicated situations, 
such as the reaction zones in the two-species process $A + B \to 0$
with initially segregated $A$ and $B$ particles, first analyzed in
Ref.~\cite{Galfi88}.

For the $A + B \to 0$ pair annihilation process, Bramson and Lebowitz
demonstrated rigorously that the TW decay exponent is exact for all 
$d < 4$ for a particular two-species model, finding bounds on the 
density amplitude \cite{Bramson88,Bramson91b,Bramson91a}. RG methods, 
as mentioned above, showed that the rate equations are asymptotically
correct for $d > 2$, and therefore established that the TW result for
the density decay, including its amplitude, was quantitatively correct
and universal for $2 < d < 4$. However, while the RG methods suggest
that the TW result might extend to $d \leq 2$, a demonstration that
this is indeed the case has not yet, in our opinion, been successfully
accomplished (see \sref{sec:furtherapp} for details). In this case, it
is rather the exact Bramson-Lebowitz result that lends credence to the
conjecture that the TW decay exponent applies for all dimensions 
$d < 4$.

Exact solutions with explicit amplitudes are available for some
reaction-diffusion models, usually in $d=1$. For example, by 
exploiting a duality with the voter model, Bramson and Griffeath
solved a particular version of the $A + A \to 0$ model in one and two
dimensions \cite{Bramson80}, with the result
\begin{equation}
\label{eq:bramson}
     a(t) \sim \cases{ 1 / (8 \pi Dt)^{1/2} ~ ,  & $d=1$ \\
     \ln (Dt) / (8 \pi D t)  ~ ,  & $d=2$ . }
\end{equation}
The RG calculation for $A + A \to 0$ generalizes these results for the
decay exponent to arbitrary dimensions (albeit of limited use for 
$d \leq 2$) \cite{Peliti86} and provides a quantitative expression for
the decay amplitude as an expansion in $\epsilon = d_c - d$
\cite{Lee94a}. This expansion is poorly convergent in one dimension
($\epsilon = 1$), but, importantly, the RG analysis does demonstrate
that the density amplitude is universal. Thus, the exact solution and
the RG analysis contribute in a complementary manner to a full
understanding of the problem, and tell us that any variation of this
model, such as allowing reactions to occur whenever reactants are
within some fixed number of lattice spacings, will result
asymptotically in precisely the density decay law (\ref{eq:bramson}). 
Furthermore, the RG expression for the amplitude in $d = 2$ (so 
$\epsilon \to 0$) is explicit, and matches the exact solution 
(\ref{eq:bramson}). This serves both to demonstrate universality of 
the result and to provide a check on the RG method.  

There are many other exact results available in one dimension, and it
is beyond the scope of this review to provide a complete survey of
this field. We restrict ourselves to mentioning a few important
classes. Many exact techniques exploit a mapping from the microscopic
master equation onto a quantum spin chain \cite{Alcaraz94,Henkel97,
Schutz01,Stinchcombe01}. In several important cases, the ensuing spin 
system turns out to be integrable and a variety of powerful techniques 
can be brought to bear, such as mapping to free-fermion systems 
\cite{Lushnikov86,Grynberg96,Schuetz95,Bares99a}. These quantum 
systems may also be studied by real space RG methods 
\cite{Hooyberghs00}. The connection between the $A+A\to 0$ reaction 
and the one-dimensional Glauber dynamic Ising model has also been
usefully exploited \cite{Racz85,Family91,Mobilia01b}. For steady-state
situations, the asymmetric exclusion process, which has various
reaction-diffusion generalizations, can be solved via a Bethe ansatz
or suitable matrix ansatz (see Ref.~\cite{Derrida98} and references
therein). Another useful technique is the empty interval method
\cite{benAvraham90}, which has recently been generalized to a wide
range of problems \cite{Masser01,Mobilia01d,Aghamo03}. Techniques for 
dimensions $d > 1$ have also been developed (see 
Ref.~\cite{Mobilia01a} and references therein). We remark that these 
mappings generally require that each lattice site has finite occupancy 
(usually zero or one). Although such restrictions may originate from
physical considerations (e.g., modeling fast on-site reactions) they 
do limit the investigation of universality. Furthermore, very little 
is currently known using spin chain mappings about the dynamics of 
multi-species reactions.

Besides exact solutions, another important method is Smoluchowski
theory \cite{Smolu16,Chand43}, which constitutes a type of improved 
rate equation approximation. Whereas rate equations represent a 
one-point mean-field theory, i.e., a closed equation for the particle 
density, Smoluchowski theory may be viewed as a two-point mean-field
approximation, namely a closed set of equations for the density as
well as the pair correlation function. One test particle is taken to 
be fixed at the origin, and the remaining reactants are effectively
treated as a non-interacting diffusion field, with the boundary
condition that their density vanishes at the capture radius of our
original particle, and tends to some fixed value at infinity. The
resulting diffusion flux toward the fixed particle is subsequently
used to define an effective reaction rate, dependent on the density at
infinity. The reactive processes are now approximately incorporated by
assuming that the densities evolve via the usual rate equation but
with the new (time-dependent for $d\leq 2$) effective rate constants.

This approximation works surprisingly well, in that it actually 
predicts the correct decay exponents for the $A + A \to 0$ reaction,
and even captures the logarithmic correction at $d_c = 2$. The
amplitudes, however, are incorrect for $d < d_c$, but yield reasonable
numerical values \cite{Torney83}, and are accurate for $d = d_c$
\cite{Gildner04}. Remarkably therefore, this improved mean-field 
theory yields correct scaling exponents even below the upper critical 
dimension. As we shall see in \sref{sec:renormalization}, this can be 
traced to the fact that in the corresponding field theory there 
appears no propagator renormalization, and hence no anomalous 
dimension for the diffusion constant or the fields (see also 
Ref.~\cite{Krishnamurthy02}). Consequently the density decay exponent 
turns out to be sufficiently constrained (essentially through 
dimensional analysis) that it is determined exactly, i.e., to all 
orders of the $\epsilon$ expansion, within the RG. Yet the density 
amplitude requires an explicit perturbative calculation via the loop 
(and thus $\epsilon$) expansion. A reasonable approximation such as 
the Smoluchowski theory incorporates the correct dimensional analysis
that fully determines the decay exponent, but fails quantitatively for
the amplitude calculation (except at the marginal dimension). 
Furthermore, mixed reactions, such as those considered in 
\sref{sec:furtherapp}, may display decay exponents that are not simply
fixed by dimensional analysis but rather rely on the details of the 
particle correlations. In these cases, Smoluchowski theory is also 
insufficient to obtain the correct exponents, although here the 
Smoluchowski exponents have been shown to be the same as those from 
the RG improved tree level \cite{Howard96a}. However, unlike with 
field-theoretic methods, there is no obvious systematic way to improve
on Smoluchowski's self-consistent approach, with the goal to include 
higher-order correlations.

\section{Mapping to Field Theory}
\label{sec:mapping}

\subsection{The model}

We illustrate the mapping to a field theory representation first for
the $A + A \to 0$ single-species pair annihilation reaction, and then
generalize to other cases. We will consider particles on a lattice 
(say a hypercubic lattice with lattice constant $h$) performing a
continuous time random walk, where they hop to a neighboring site at
some uniform rate $D / h^2$ (such that $D$ becomes the usual diffusion
constant in the continuum limit). The particles do not interact,
except whenever two or more particles occupy the same site, in which
case they annihilate with fixed reaction rate $\lambda$. The state of
the system is then characterized by the probability $P(\{n\},t)$ at
time $t$ of a particular configuration uniquely specified by the
string of site occupation numbers $\{n\}=(n_1,n_2,\dots)$. The 
system's stochastic dynamics is captured through a master equation for
the time-dependent configuration probability $P$.  For pure diffusion,
it assumes the form
\begin{eqnarray}
     \fl \partial_t P(\{n\},t) = {D \over h^2} \sum_{<ij>} \biggl[
     &(n_i+1) P(\dots,n_i+1,n_j-1\dots,t) - n_i P(\{n\},t) \nonumber\\
     &+(n_j+1) P(\dots,n_i-1,n_j+1\dots,t) - n_j P(\{n\},t) \biggr] ~,
\label{eq:meq_diffusion}
\end{eqnarray}
where the summation extends over pairs of nearest-neighbor sites. The
first term in the square bracket represents a particle hopping from
site $i$ to $j$, and includes both probability flowing into and out of
the configuration with site occupation numbers $\{n\}$ as a 
consequence of the particle move. The second term corresponds to a hop
from site $j$ to $i$. The multiplicative factors of $n$ and $n+1$ are
a result of the particles acting independently.

Combining diffusion with the annihilation reaction gives
\begin{eqnarray}
     &\partial_t P(\{n\},t) = ({\rm diffusion~term}) \nonumber \\
     &\ + \lambda \sum_i \Bigl[ (n_i+2) (n_i+1) P(\dots,n_i+2,\dots,t) 
     - n_i (n_i-1) P(\{n\},t) \Bigr] ~ , 
\label{eq:mastereq}
\end{eqnarray}
where the $({\rm diffusion~term})$ denotes the right-hand side of
Eq.~\eref{eq:meq_diffusion}. All that remains is to specify the
initial probability $P(\{n\},t=0)$. For uniform, random initial
conditions the particle distribution will be a Poissonian on each site
$i$, i.e.,
\begin{equation}
\label{eq:poisson}
     P(\{n\},0) = \prod_i \left( {\bar n_0^{n_i} \over n_i!} 
     e^{-\bar n_0} \right) ,
\end{equation}
where $\bar n_0$ denotes the average number of particles per site.

\subsection{Doi's second-quantized representation}
\label{subsec:Doi}

Stochastic classical particle models with local reactions can be
rewritten in terms of ladder operators familiar from quantum 
mechanics, as shown by Doi \cite{Doi76a}, thus taking advantage of the
algebraic structure of second quantization. This representation
exploits the fact that all processes just change the site occupation
numbers by an integer. Since we have not implemented any site
occupation restrictions, we introduce for each lattice site 
$i,j,\ldots$ creation and annihilation operators subject to `bosonic'
commutation relations
\begin{equation}
     [\hat a_i,\hat a^\dagger_j] = \delta_{ij} ~ , \qquad
     [\hat a_i,\hat a_j] = [\hat a^\dagger_i,\hat a^\dagger_j] = 0 ~ .
\end{equation}
The `vacuum' (empty lattice) $|0\rangle$ is characterized by
$a_i | 0 \rangle = 0$ for all $i$, and on each site $i$ we define the 
state vector $| n_i \rangle = (\hat a_i^\dagger)^{n_i} | 0 \rangle$ 
(note that the normalization differs from the standard 
quantum-mechanical convention). It is then straightforward to show 
that
\begin{equation}
     \hat a_i|n_i\rangle=n_i |n_i-1\rangle ~ , \qquad 
     \hat a^\dagger_i|n_i\rangle=|n_i+1\rangle ~ .
\end{equation}

Next we employ these on-site vectors to incorporate the state of the
entire stochastic particle system at time $t$ in the quantity
\begin{equation}
\label{eq:statedef}
     | \phi(t) \rangle = \sum_{\{n\}} P(\{n\},t) \, \prod_i 
     (\hat a^\dagger_i)^{n_i} | 0 \rangle ~ .
\end{equation}
As a result, the first-order temporal evolution of the master equation
is cast into an `imaginary-time Schr\"odinger equation'
\begin{equation}
\label{eq:schroedinger}   
     - \partial_t | \phi(t) \rangle = \hat H | \phi(t) \rangle ~ ,
\end{equation}
where the non-Hermitian time evolution operator (`quasi-Hamiltonian') 
for the processes in Eq.~\eref{eq:mastereq} becomes 
\begin{equation}
\label{eq:hamiltonian}
     \hat H = {D \over h^2} \sum_{<ij>} (\hat a_i^\dagger 
     - \hat a_j^\dagger) (\hat a_i - \hat a_j) - \lambda \sum_i 
     \Bigl[ 1 - (\hat a^\dagger_i)^2 \Bigr] \hat a_i^2 ~ .
\end{equation}
\Eref{eq:schroedinger} is formally solved by $| \phi(t) \rangle =
\exp(-\hat H t) | \phi(0) \rangle$, with the initial state determined 
by Eqs.~\eref{eq:poisson} and \eref{eq:statedef}.

The equations of motion for $P(\{n\},t)$ and its moments in this
representation are of course identical to those following directly
from the master equation. Yet at this point we may see the advantage
of using Doi's formalism: the original master equation was complicated
by factors of $n$ and $n^2$ which are now absent. The
`second-quantized' formalism provides a natural framework for
describing independent particles that may be changing in number. 

A different approach is to write down an appropriate Fokker--Planck
equation \cite{vanKampen81,Risken84,Gardiner85} for the processes
under consideration, although this is somewhat awkward due to the
presence of reaction processes where particles are created and / or
destroyed. However, a more fundamental problem in this approach is
encountered in situations with low densities. In such cases the 
Kramers--Moyal expansion used in the derivation of the Fokker--Planck 
equation break down, and it is certainly not valid to terminate the 
expansion in the usual way after the second term \cite{Gardiner85}. 
Alternatively, if the Fokker--Planck equation is derived from a 
coarse-grained Langevin equation, its validity may again be 
questionable, since both the relevant slow variables and their 
stochastic noise correlations must often be inferred 
phenomenologically.

We also note that an alternative formalism exists which again starts
from a classical master equation and leads to a path integral
representation. This method uses the Poisson representation, and
assumes that the state of the system at time $t$ can be expanded into
a superposition of multivariate uncorrelated Poissonians
\cite{Gardiner85,Gardiner77,Gardiner78}. However, as shown in
Ref.~\cite{Droz94}, this approach is actually equivalent to Doi's
formalism, as presented here. An analogous representation in terms of
Pauli spin matrices may also be used, which then replace the bosonic
ladder operators considered here. This corresponds to a master 
equation in which only single occupancy is allowed per particle
species at each lattice site. These techniques can be especially
useful in one dimension, where the resulting second-quantized
formulation represents certain quantum spin chains, which are often
integrable \cite{Alcaraz94,Henkel97,Schutz01}. However, our primary
motive in introducing the second-quantized representation here is to
map the problem to a field theory, and for this purpose the bosonic
formalism developed above is more suitable.

To find ensemble averages $\overline A$ of observables at time $t$ we
cannot just use the standard quantum-mechanical matrix element, since
this would involve two factors of the probabilities $P$. Rather, we
need a projection state $\langle {\cal P}|$, defined by the conditions
$\langle {\cal P} | \hat a_i^\dagger = \langle {\cal P} |$ and 
$\langle {\cal P} | 0 \rangle = 1$, leading to
\begin{equation}
     \langle{\cal P}| = \langle 0 | e^{\sum_i \hat a_i} ~ .
\end{equation}
From the above properties it follows that
\begin{equation}
\label{eq:quantumavg}
     \fl \qquad \overline{A}(t) = \sum_{\{n\}} P(\{n\},t) \, A(\{n\}) 
     = \sum_{\{n\}} P(\{n\},t) \, \langle {\cal P} | \hat A \prod_i 
     (\hat a_i^\dagger)^{n_i} | 0 \rangle = 
     \langle {\cal P} | \hat A | \phi(t) \rangle ~ ,
\end{equation}
where the operator $\hat A$ is given by the function $A(\{n\})$ with
the substitution $n_i \to \hat a_i^\dagger \hat a_i$. It is naturally
not surprising that the expression for averages here differs from
usual quantum mechanics, since we consider, after all, an entirely
classical model. Notice too that there is no requirement for $\hat H$
to be Hermitian. In fact, particle annihilation and creation reactions
obviously lead to non-Hermitian ladder operator combinations. 
Furthermore 
\begin{equation}
     1 = \langle {\cal P} | \phi(t) \rangle
     = \langle {\cal P} | \exp(-\hat H t) | \phi(0) \rangle
\end{equation}
shows that probability conservation is enforced through the condition
$\langle {\cal P} | \hat H = 0$. Any quasi-Hamiltonian $\hat H$
derived from a probability conserving master equation will necessarily
satisfy this property.

The factor $\exp(\sum_i\hat a_i)$ in the projection state can be
commuted through to the right in Eq.~\eref{eq:quantumavg}, with the
net effect of taking $\hat a_i^\dagger \to 1 + \hat a_i^\dagger$
\cite{Grassberger80}. While this has the advantage of simplifying the
expression for $\overline{A}$, we choose not to follow this procedure
because there are cases, such as branching and annihilating random
walks with an even number of offspring particles, where such a shift
is undesirable as it masks an important symmetry. Furthermore, as
shown below, the same effect may be obtained in the field theory, when
desired, by a corresponding field shift. 

We note that probability conservation is reflected in the fact that 
the time evolution operator $\hat H$ must vanish on replacing all 
$\hat a_i^\dagger \to 1$. In addition, because of the projection 
state, for any operator $\hat A$ there exists a corresponding operator
$\hat A'$, with the same average $\overline{A}$, that is obtained by 
normal-ordering $\hat A$ and then replacing the creation operators 
$\hat a_i^\dagger$ by unity (the projection state eigenvalue). For 
example, the density operator $\hat a_i^\dagger \hat a_i$ may be 
replaced with $\hat a_i$, and the two-point operator 
$\hat a_i^\dagger \hat a_i \, \hat a_j^\dagger \hat a_j$ becomes 
$\hat a_i \delta_{ij} + \hat a_i \hat a_j$.

\subsection{Coherent state representation and path integrals}
\label{subsec:coherent}

From the second-quantized representation a field theory can be
obtained via the very same path integral techniques as developed for
true quantum many-particle systems. A general discussion of these
methods can be found in standard textbooks \cite{Schulman81,Negele88},
and a presentation specific to reaction-diffusion models is given by
Peliti \cite{Peliti85}. For completeness, we present the basic method
here, with a few supplementary observations.

First, the (stochastic) temporal evolution is divided into $N$ slices 
of ultimately infinitesimal size $\Delta t = t/N$ via
\begin{equation}
\label{eq:trotter}
     | \phi(t) \rangle = \exp(-\hat H t) | \phi(0) \rangle = 
     \lim_{\Delta t \to 0} \exp(- \hat H \Delta t)^{t/\Delta t} 
     | \phi(0) \rangle ~ .
\end{equation}
The second-quantized operators are then mapped onto ordinary complex
numbers (in the bosonic case) or Grassmann variables (for fermions) by
inserting a complete set of coherent states at each time slice before
the limit $\Delta t\to 0$ is taken.

Coherent states are right eigenstates of the annihilation operator,
$\hat a |\phi\rangle = \phi |\phi\rangle$, with complex eigenvalue
$\phi$. Explicitly, $|\phi\rangle = \exp(-\frac 1 2 |\phi|^2 + \phi
\hat a^\dagger) | 0 \rangle$. Their duals are left eigenstates of the 
creation operator: $\langle \phi | \hat a^\dagger = \langle \phi |
\phi^*$. A useful overlap relation is
\begin{equation}
\label{eq:overlapid}
     \langle \phi_1 | \phi_2\rangle = \exp \left( - \frac 1 2 
     |\phi_1|^2 - \frac 1 2 |\phi_2|^2 + \phi_1^* \phi_2 \right) .
\end{equation}
The coherent states are over-complete, but nonetheless may be used to
form a resolution of the identity operator. With our convention for 
the states $|n\rangle$, we have for a single lattice site
\begin{equation}
\label{eq:single-site-identity}
     {\bf 1} = \sum_n {1\over n!} \, |n\rangle \langle n| 
     = \sum_{m,n}  {1\over n!} \, |n\rangle \langle m| \, \delta_{mn} 
     = \int {d^2 \phi \over \pi} \, |\phi\rangle \langle \phi| ~ ,
\end{equation}
where we have used 
\begin{equation}
\label{eq:Kdelta}
     \delta_{mn} = {1\over\pi m!} \int d^2\phi \, e^{-|\phi|^2} 
     \phi^{*m} \phi^n ~ ,
\end{equation}
with the integration measure $d^2\phi = d(\Re \phi)d(\Im \phi)$. The
above expression generalizes straightforwardly to multiple lattice
sites according to
\begin{equation} 
\label{eq:identity}
     {\bf 1} = \int \prod_i \left( d^2 \phi_i \over \pi \right) 
     |\{\phi\}\rangle \langle\{\phi\}| ~ ,
\end{equation}
where $\{\phi\} = (\phi_1,\phi_2,\dots)$ denotes a set of coherent
state eigenvalues, one for each annihilation operator $\hat a_i$, and
$|\{\phi\}\rangle = |\phi_1\rangle \otimes |\phi_2\rangle \otimes 
\dots$.

A mathematical subtlety can arise with this representation of the
identity. Expressions such as $\exp(-\hat H \Delta t)$ are defined in
terms of their power series, with an implied sum. The identity 
operator \eref{eq:single-site-identity} contains an integral, and in
many cases these sums and integrals do not commute. Consider, for
example, $\langle 0 | \exp(\lambda \hat a^k) | 0 \rangle = 1$ with the
identity \eref{eq:single-site-identity} inserted:
\begin{equation}
\label{eq:math_issue}
     1 = \langle 0| \sum_{n=0}^\infty {\lambda^n (\hat a^k)^n\over n!}
     \int {d^2 \phi \over \pi} |\phi\rangle \langle \phi|0 \rangle = 
     \sum_{n=0}^\infty {\lambda^n \over n!} \int {d^2 \phi \over \pi} 
     \exp(-|\phi|^2) (\phi^k)^n ~ ,
\end{equation}
where Eq.~\eref{eq:overlapid} was used for $\langle 0|\phi\rangle$. By
Eq.~\eref{eq:Kdelta} the sum over $n$ correctly yields $1 + 0 + 0 
\dots$. However, naively exchanging integration and summation gives
\begin{equation}
\label{eq:math_issue2}
     1 = \int {d^2\phi\over\pi} \exp(-|\phi|^2 + \lambda \phi^k) ~ ,
\end{equation}
which does not exist for $k>2$ or for $k=2$ and $|\lambda| > 1$. 
Nevertheless, it is convenient to represent expressions such as
\eref{eq:math_issue} formally by \eref{eq:math_issue2}, with the
understanding that an explicit evaluation implies a perturbative
expansion in powers of $\lambda$. As a consequence, our formal
expressions for the field theory actions will appear to be unstable,
but are actually well-defined within the framework of perturbation
theory. We remark that it will be necessary to treat the diffusion
terms in $\hat H$ non-perturbatively. This is possible because the
expansion in powers of the diffusion constant, when acting on the
coherent state $| \phi \rangle$, is uniformly convergent in the $\phi$
plane (a sufficient condition for commuting sums and integrals), as
will be shown below.

The projection state $\langle \cal P |$ is proportional to the dual
coherent state with all eigenvalues $\phi_i = 1$, which for brevity we
will call $\langle 1 |$. Also, for Poisson initial conditions,
$| \phi(0) \rangle$ is proportional to the coherent state 
$| \bar n_0 \rangle$ with all $\phi_i = \bar n_0$. Now label each time
slice in \eref{eq:trotter} by a time index $\tau$ that runs in steps 
of $\Delta t$ from time zero to $t$ and insert the identity 
\eref{eq:identity} between each slice into Eq.~\eref{eq:quantumavg}:
\begin{eqnarray}
\label{eq:insert}
     \overline{A}(t) = {\cal N}^{-1} \lim_{\Delta t\to 0} \int 
     &\biggl( \prod_{i,\tau} d^2\phi_{i,\tau} \biggr) \, 
     \langle 1 | \hat A | \{\phi\}_t \rangle \nonumber \\ 
     &\left[ \prod_{\tau=\Delta t}^t \langle\{\phi\}_\tau | 
     \exp(-\hat H \Delta t) | \{\phi\}_{\tau - \Delta t} \rangle 
     \right] \langle \{\phi\}_0 | \bar n_0 \rangle ~ .
\end{eqnarray}
The normalization factor ${\cal N}$ is to be determined later by
requiring the identity operator to average to unity. Note that a set
of states has been inserted at the time slices at $0$ and $t$ as well,
for reasons that will become clear below. We now proceed to analyze
the various contributions.

First, with our caveat above about implied perturbative calculations,
we may take
\begin{equation}
\label{eq:expH}
     \fl \quad \langle\{\phi\}_\tau| \exp(-\hat H \Delta t) 
     | \{\phi\}_{\tau - \Delta t} \rangle = \langle \{\phi\}_\tau | 
     \{\phi\}_{\tau - \Delta t} \rangle \exp\Bigl( - 
     H(\{\phi^*\}_\tau, \{\phi\}_{\tau-\Delta t}) \Delta t \Bigr) ~ ,
\end{equation}
where $H(\{\phi^*\}_\tau, \{\phi\}_{\tau-\Delta t}) = \langle 
\{\phi\}_\tau | \hat H | \{\phi\}_{\tau-\Delta t}\rangle$. This 
function is straightforwardly obtained by normal ordering $\hat H$ and
acting the $\hat a_i$ to the right and the $\hat a_i^\dagger$ to the
left, whereupon the creation / annihilation operators become
respectively replaced with the coherent state eigenvalues $\phi_i^*$ /
$\phi_i$. The remaining overlap in Eq.~\eref{eq:expH} factors:
\begin{equation}
     \langle\{\phi\}_\tau | \{\phi\}_{\tau - \Delta t}\rangle =
     \prod_i \, \langle \phi_{i,\tau} | \phi_{i,\tau - \Delta t} 
     \rangle ~ ,
\end{equation}
where, according to Eq.~\eref{eq:overlapid}, for each lattice site
\begin{equation}
\label{eq:siteoverlap}
     \fl \qquad \langle \phi_{i,\tau} | \phi_{i,\tau-\Delta t} \rangle
     = \exp(-\phi^*_{i,\tau} [\phi_{i,\tau} - \phi_{i,\tau-\Delta t}])
     \exp\left( \frac 1 2 |\phi_{i,\tau}|^2 - \frac 1 2 
     |\phi_{i,\tau-\Delta t}|^2 \right) .
\end{equation}
Stringing together a product of these states for increasing $\tau$
will cause the second exponential term to cancel except at the initial
and final times. The first exponential yields a factor $\exp[ -
\phi_{i,\tau^*} (d\phi_{i,\tau}/dt) \Delta t + O(\Delta t^2)]$ for 
each time slice $\tau$ and lattice site $i$.

The operator $\hat A$ is assumed to be a function of the annihilation
operators $\hat a_i$ only, by the procedure described at the end of
\sref{subsec:Doi}. Therefore $\langle 1 | \hat A | \{\phi\}_t\rangle =
\langle 1 | \{\phi\}_t \rangle A(\{\phi\}_t)$, where the latter 
function is obtained from $\hat A$ through the replacement 
$\hat a_i \to \phi_{i,t}$. The matrix element is multiplied with the 
remaining exponential factors at time $t$ from 
Eq.~\eref{eq:siteoverlap}, giving
\begin{equation}
     \langle 1 | \{\phi\}_t \rangle \prod_i \exp\left( \frac 1 2
     |\phi_{i,t}|^2 \right) \propto \exp\left( \sum_i \phi_{i,t} 
     \right) .
\end{equation}
The initial term also picks up a factor from 
Eq.~\eref{eq:siteoverlap},
\begin{equation}
     \langle \{\phi\}_0|\bar n_0 \rangle \prod_i \exp\left( 
     -\frac 1 2|\phi_{i,0}|^2 \right) \propto \exp\left( \sum_i 
     \left[ \bar n_0\phi^*_{i,0} - |\phi_{i,0}|^2 \right] \right) .
\end{equation} 

We may now take the limit $\Delta t \to 0$. The $O(\Delta t)$ time
difference in the $\phi^*_\tau$ and $\phi_{\tau-\Delta t}$ arguments
of $H$ is dropped with the provision that, in cases where it matters,
the $\phi^*$ field should be understood to just follow the $\phi$
field in time. Indeed, this was found to be an essential distinction
in a numerical calculation of the path integral \cite{Beccaria97}, and
it will also play a role in the treatment of the initial conditions
below. The $O(\Delta t)$ terms in the product over $\tau$ in
Eq.~\eref{eq:insert} will become the argument of an exponential:
\begin{equation}
\label{eq:insert2}
     \overline{A}(t) = {\cal N}^{-1} \int \biggl( \prod_i {\cal D}
     \phi_i {\cal D}\phi^*_i \biggr) \, A(\{\phi\}_t) \, 
     \exp[ - S(\{\phi^*\},\{\phi\})_0^t ] ~ .
\end{equation}
Here $S$ denotes the action in the statistical weight, which reads
explicitly
\begin{eqnarray}
\label{eq:discreteS}
     S(\{\phi^*\},\{\phi\})_0^{t_f} = \sum_i \Biggl( &- \phi_i(t_f) - 
     \bar n_0 \phi^*_i(0) + |\phi_i(0)|^2 \nonumber \\
     &+ \int_0^{t_f} dt \Bigl[ \phi^*_i \, \partial_t \phi_i + 
     H(\{\phi^*\},\{\phi\}) \Bigr] \Biggr) ,
\end{eqnarray}
where we have renamed the final time $t \to t_f$ for clarity. The
${\cal D}\phi_i{\cal D}\phi^*_i$ represent functional differentials
obtained from $\prod_\tau d\phi_{i,\tau}d\phi^*_{i,\tau}$ in the limit
$\Delta t\to 0$. At last, the normalization factor is now fixed via
${\cal N} = \int \prod_i {\cal D}\phi_i {\cal D}\phi^*_i \,
\exp[ - S(\{\phi^*\},\{\phi\})] $.

Before specifying $H$, let us discuss the initial and final terms in
the action \eref{eq:discreteS} in more detail. The initial terms are
of the form $\exp( \phi^*(0) [\phi(0) - \bar n_0] )$, which implies
that the functional integral over the variables $\phi^*(0)$ will 
create $\delta$ functions that impose the constraints 
$\phi(0) = \bar n_0$ at each lattic site. Thus the initial terms may 
be dropped from the action \eref{eq:discreteS} in lieu of a constraint
on the initial values of the fields $\phi_i$ \cite{Peliti85}. However,
a path integral with such an implied constraint is not directly 
amenable to a perturbation expansion, so an alternative approach was 
developed \cite{Lee94c}. All calculations will be performed 
perturbatively with respect to a reference action $S_0$ composed of 
the bilinear terms $\propto \phi^*\phi$ in $S$. As will be 
demonstrated below, any such average will give zero unless every 
factor $\phi$ in the quantity to be averaged can be paired up with an 
earlier $\phi^*$ (that is, the propagator only connects earlier 
$\phi^*$ to later $\phi$). The initial terms in the action 
\eref{eq:discreteS} can be treated perturbatively by expanding the 
exponential. Recalling that the time ordering of the product 
$\phi^*(0) \phi(0)$ has $\phi$ slightly earlier than $\phi^*$, we see 
that all terms in the perturbative expansion will give zero, which is 
equivalent to simply dropping the $\phi^*\phi$ initial term from the 
action \eref{eq:discreteS}. The remaining initial state contribution 
$\exp[- \bar n_0 \phi^*(0)]$ then replaces an implied constraint as 
the means for satisfying the assumed random (Poissonian) initial 
conditions.

Prior to commenting on the final term $-\phi_i(t_f)$ in the action
\eref{eq:discreteS}, we now proceed to take the continuum limit via
$\sum_i \to \int h^{-d} d^dx$, $\phi_i(t) \to \phi({\bf x},t) h^d$,
and $\phi^*_i(t) \to \tilde\phi({\bf x},t)$. The latter notation
indicates that we shall treat the complex conjugate field
$\tilde\phi({\bf x},t)$ and $\phi({\bf x},t)$ as independent 
variables. This is especially appropriate once we apply a field shift
$\tilde\phi \to 1 + \bar\phi$, which, in addition to modifying the
form of $H$, has the effect of replacing
\begin{equation}
     \int_0^{t_f} dt \, \tilde\phi \, \partial_t \phi \> \to \> 
     \phi(t_f) - \phi(0) 
     + \int_0^{t_f} dt \, \bar\phi \, \partial_t \phi ~ .
\end{equation}
Thus the final term $- \phi({\bf x},t_f)$ in the action is cancelled,
which simplifies considerably the perturbative calculations, but
introduces a new initial term. However, the latter will again vanish
when perturbatively averaged against the bilinear action $S_0$, as
described above. For many of the problems discussed here such a field
shift will be employed. Lastly, the remaining initial time
contribution reads $\bar n_0 \to n_0 h^d$, where $n_0$ denotes the
number density per unit volume. Notice that we have (arbitrarily)
chosen $\phi({\bf x},t)$ to have the same scaling dimension as a
density. While the continuum limit could have been defined differently
for the fields $\phi$ and $\tilde\phi$, our prescription ensures that
the `bulk' contributions to the action must vanish as 
$\tilde\phi \to 1$ owing to probability conservation.

At this point, let us explicitly evaluate $H$ for diffusion-limited
pair annihilation, $A + A \to 0$. Since the time evolution operator
\eref{eq:hamiltonian} is already normal ordered, we obtain directly
\begin{equation}
\label{eq:hamiltonian2}
     H(\{\phi^*\},\{\phi\}) = {D \over h^2} \sum_{<ij>} 
     (\phi_i^*-\phi_j^*) (\phi_i-\phi_j) 
     - \lambda \sum_i (1 - \phi_i^{*2}) \phi_i^2 ~ .
\end{equation}
We now proceed from a lattice to the continuum limit as outlined
above, replacing the finite lattice differences in 
Eq.~\eref{eq:hamiltonian2} with spatial gradients. The resulting field
theory action, prior to any field shift, reads
\begin{equation}
\label{eq:2Aunshact}
     \fl \quad S[\tilde\phi,\phi] = \int d^dx \left\{ -\phi(t_f) + 
     \int_0^{t_f} dt \left[ \tilde\phi \left( \partial_t - D \nabla^2 
     \right) \phi - \lambda_0 \left( 1 - \tilde\phi^2 \right) \phi^2 
     \right] - n_0 \tilde\phi(0) \right\} ,
\end{equation} 
where $\lambda_0 = \lambda h^d$. After applying the field shift 
$\tilde\phi \to 1 + \bar\phi$, we obtain
\begin{equation}
\label{eq:2Aaction}
     \fl \qquad S[\bar\phi,\phi] = \int d^dx \left\{ \int_0^{t_f} dt 
     \left[ \bar\phi \left( \partial_t - D \nabla^2 \right) \phi + 
     \lambda_1 \bar\phi \phi^2 + \lambda_2 \bar\phi^2 \phi^2 \right] 
     - n_0 \bar\phi(0) \right\} ~ ,
\end{equation} 
with $\lambda_1 = 2 \lambda_0$ and $\lambda_2 = \lambda_0$. Finally, 
we remark again that these actions are defined through the 
perturbation expansion with respect to the nonlinearities, as
discussed above. However, the diffusion terms are uniformly 
convergent, resumming to give $\exp(-D|\nabla\phi|^2)$, which is
bounded for all $\phi$. Hence, the diffusion part of the action may be
treated non-perturbatively.

\subsection{Generalization to other reactions}

This procedure to represent a classical stochastic master equation in
terms of a field theory can be straightforwardly generalized to other
locally interacting particle systems, e.g., the $k$th order decay
reaction $k A \to \ell A$ with $\ell < k$. The appropriate master
equation for identical particles will result in the time evolution
operator
\begin{equation}
     \hat H = \hat H_D-\sum_i \lambda_0 \left[ (\hat a^\dagger_i)^\ell
     - (\hat a^\dagger_i)^k \right] \hat a^k_i ~ , 
\end{equation} 
where $\hat H_D$ denotes the unaltered diffusion part as in
Eq.~\eref{eq:hamiltonian}. Following the method described above, and
performing the field shift $\tilde\phi \to 1 + \bar\phi$ eventually
results in the field theory action
\begin{equation}
\label{eq:kAaction}
     S[\bar\phi,\phi] = \int d^dx \left\{ \int_0^{t_f} dt \left[ 
     \bar\phi \left( \partial_t - D \nabla^2 \right) \phi 
     + \sum_{i=1}^k \lambda_i \, \bar\phi^i \phi^k \right] 
     - n_0 \bar\phi(0) \right\} ~ ,
\end{equation}
with $\lambda_i = \lambda_0 \left(k\atop i\right) - \lambda_0 \left( 
\ell \atop i \right)$ for $i \leq \ell$, and $\lambda_i = \lambda_0 
\left( k \atop i \right)$ for $i > \ell$ (note that always 
$\lambda_k = \lambda_0$). Also, the integer $k$ determines which 
vertices are present, while $\ell$ only modifies coefficients. In the 
simplest case, $k = 2$, we recover $\lambda_1 = 2 \lambda_0$ for pair 
annihilation $A + A \to 0$, whereas $\lambda_1 = \lambda_0$ for pair 
coagulation $A + A \to A$. One variant on the $A + A \to 0$ reaction 
would be to allow for mixed pair annihilation and coagulation. That 
is, whenever two $A$ particles meet, with some probability they 
annihilate according to $A + A \to 0$, or otherwise coagulate, 
$A + A \to A$. In the master equation these competing processes are 
represented by having both reaction terms present, with reaction rates 
$\lambda^{(\ell)}$ (where $\ell = 0,1$ indicates the number of 
reaction products) in the correct proportions. The end result is an 
action of the form \eref{eq:kAaction} above, but with a coupling ratio 
$\lambda_1 / \lambda_2$ that interpolates between $1$ and $2$. 

The description of multi-species systems requires, at the level of the
master equation, additional sets of occupation numbers. For example,
the master equation for the two-species pair annihilation reaction 
$A + B \to 0$ employs a probability $P(\{m\},\{n\},t)$ where
$\{m\},\{n\}$ respectively denote the set of $A/B$ particle occupation
numbers. Various forms of occupation restrictions could be included in
the master equation, e.g., Bramson and Lebowitz \cite{Bramson88}
consider a model in which a given site can have only $A$ or only $B$
particles. Here we will consider unrestricted site occupation. The
$A$ and $B$ particles diffuse according to 
Eq.~\eref{eq:meq_diffusion}, though possibly with distinct diffusion 
constants. Combining the reactions then gives
\begin{eqnarray}
     \fl \quad \partial_t P = ({\rm diffusion~terms}) + \lambda \sum_i 
     \Bigl[ & (m_i+1)(n_i+1) P(\dots,m_1+1,\dots,n_1+1,\dots,t) 
     \nonumber \\ &- m_i n_i P(\{m\},\{n\},t) \Bigr] ~ .
\end{eqnarray}
The second-quantized formulation then requires distinct creation and
annihilation operators for each particle species. The state vector is
therefore constructed as
\begin{equation}
  | \phi(t) \rangle = \sum_{\{m\},\{n\}} P(\{m\},\{n\},t) \prod_i 
  (\hat a_i^\dagger)^{m_i} (\hat b_i^\dagger)^{n_i} | 0 \rangle ~ , 
\end{equation}
and the master equation again assumes the form \eref{eq:schroedinger},
with the time evolution operator
\begin{equation}
\label{eq:ab_hamiltonian}
     \fl \ \hat H = {D_A \over h^2} \sum_{<ij>} (\hat a_i^\dagger - 
     \hat a_j^\dagger) (\hat a_i - \hat a_j) + {D_B \over h^2}
     \sum_{<ij>} (\hat b_i^\dagger - \hat b_j^\dagger) 
     (\hat b_i-\hat b_j) - \lambda \sum_i \Bigl( 1 - \hat a^\dagger_i 
     \hat b^\dagger_i \Bigr) \hat a_i \hat b_i ~ .
\end{equation}
In the mapping to the field theory we must then involve two sets of
coherent states, resulting in two independent fields $a(x,t)$ and
$b(x,t)$. Hence, after shifting both $\tilde a \to 1 +\bar a$ and
$\tilde b \to 1 + \bar b$, the action reads:
\begin{eqnarray}
\label{eq:ab_action}
     \fl \qquad S[\bar a,a,\bar b,b] = \int d^dx \Biggl\{ \int_0^{t_f} 
     dt &\Biggl[ \bar a \left( \partial_t - D_A \nabla^2 \right) a 
     + \bar b \left( \partial_t - D_B \nabla^2 \right) b \nonumber \\
     &+ \lambda_0 (\bar a + \bar b) a b + \lambda_0 \, \bar a \bar b a 
     b \Biggr] - a_0 \bar a(0) - b_0 \bar b(0) \Biggr\} ~ .
\end{eqnarray}

Further generalizations are straightforward: for each new particle 
species additional occupation numbers, second-quantized operators, and
fields are to be introduced. The details of the reaction are coded 
into the master equation, though after some practice, it is actually 
easier to directly start with the Doi time evolution operator, as it 
is a more efficient representation. The general result is as follows: 
For a given reaction, two terms appear in the quasi-Hamiltonian (as in 
the original master equation). The first contribution, which is 
positive, contains both an annihilation and creation operator for each 
reactant, normal-ordered. For example, for the $A + A \to 0$ and 
$A + A \to A$ reactions this term reads $\hat a^{\dagger 2} \hat a^2$, 
whereas one obtains for the $A + B \to 0$ reaction $\hat a^\dagger 
\hat b^\dagger \hat a \hat b$. These contributions indicate that the 
respective second-order processes contain the particle density 
products $a^2$ and $a b$ in the corresponding classical rate 
equations. The second term in the quasi-Hamiltonian, which is 
negative, entails an annihilation operator for every reactant and a 
creation operator for every product, normal-ordered. For example, in 
$A + A \to 0$ this term would be $\hat a^2$, whereas for $A + A \to A$ 
it becomes $\hat a^\dagger \hat a^2$, and for $A + B + C \to A + B$ it 
would read $\hat a^\dagger \hat b^\dagger \hat a \hat b \hat c$. These 
terms thus directly reflect the occurring annihilation and creation 
processes in second-quantized language.

\subsection{Relation to stochastic partial differential equations}
\label{subsec:spde}

In some cases, the field theory developed above can be cast into a 
form reminiscent of stochastic partial differential equations (SPDE) 
with multiplicative noise. Consider the action \eref{eq:2Aaction} for 
single-species pair annihilation $A + A \to 0$: apart from the quartic 
term $\lambda_2\bar\phi^2\phi^2$ every term in $S$ is linear in the 
$\bar\phi$ field. The quartic term can in fact also be `linearized' by 
means of introducing an auxiliary field, where
\begin{equation}
\label{eq:lintrafo}
     \exp(- \lambda_2 \bar\phi^2 \phi^2) \propto \int d\eta \exp(- 
     \eta^2 / 2) \exp(i \eta \, \sqrt{2\lambda_2} \, \bar\phi \phi) ~.
\end{equation}
Substituting this relation into the action results in three 
fluctuating fields, namely $\bar\phi$, $\phi$, and $\eta$, but with 
the benefit that $\bar\phi$ appears just linearly. Therefore, 
performing the functional integral $\int {\cal D} \bar\phi \, 
\exp(\bar\phi [\dots])$ simply yields a functional Dirac $\delta$ 
function, ensuring that all configurations $\phi({\bf x},t)$ satisfy 
the corresponding constraint given by its argument. As a result, the 
field $\phi$ is determined by a stochastic partial differential 
equation:
\begin{equation}
\label{eq:pde}
     \partial_t \phi = D \nabla^2 \phi - 2 \lambda_0 \phi^2 + 
     i \sqrt{2\lambda_0} \, \phi \, \eta ~ ,
\end{equation}
where $\eta$ represents a stochastic Gaussian variable with unit 
variance, i.e., $\langle \eta \rangle = 0$, $\langle \eta({\bf x},t)\, 
\eta({\bf x}',t') \rangle = \delta({\bf x}-{\bf x}') \, \delta(t-t')$.
Notice that the above procedure is just the reverse of the standard 
field theory representation of a Langevin-type SPDE \cite{Bausch76}.

It is not surprising that the noise term vanishes as the field
$\phi \to 0$, since the shot noise should diminish as the number of 
particles decreases. Once the absorbing state with zero particles is
reached, all deterministic as well as stochastic kinetics ceases.
However, the appearance of the imaginary noise appears rather strange,
and even when the stochastic force is redefined as 
$\zeta = i \sqrt{2\lambda_0} \, \phi \, \eta$, its variance becomes 
formally negative: $\langle \zeta({\bf x},t) \, \zeta({\bf x}',t') 
\rangle = - 2 \lambda_0 \, \phi({\bf x},t)^2 \delta({\bf x}-{\bf x}') 
\, \delta(t-t')$. Recall, however, that the field $\phi$ is complex, 
and therefore cannot be simply interpreted as the particle density. 
Furthermore, there is evidence suggesting that the SPDE \eref{eq:pde} 
is numerically unstable \cite{Beccaria97,Deloubriere01}.

Interestingly, though, one may obtain an SPDE for a real density field 
as follows \cite{Janssen01}. Starting with the unshifted action 
\eref{eq:2Aunshact}, we apply the nonlinear Cole--Hopf transformation 
$\tilde\phi = e^{\tilde\rho}$, $\phi = \rho \, e^{- \tilde\rho}$, such 
that $\tilde\phi \phi = \rho$, and where the Jacobian is unity. This 
yields $\tilde\phi \, \partial_t \phi = \partial_t [\rho (1 - 
\tilde\rho)] + \tilde\rho \, \partial_t \rho$, and, omitting boundary 
contributions, $-D \tilde\phi \, \nabla^2 \phi = -D \phi \nabla^2 
\tilde\phi = -D \rho [\nabla^2 \tilde\rho + (\nabla \tilde\rho)^2]$ 
for the diffusion term. Finally, the annihilation reaction is 
represented through $- \lambda_0 (1 - \tilde\phi^2) \phi^2 = \lambda_0 
\rho^2 (1 - e^{-2 \tilde\rho}) = 2 \lambda_0 \, \tilde\rho \rho^2 - 
2 \lambda_0 \, \tilde\rho^2 \rho^2 \ldots$, if we expand the 
exponential. Hence we see that the quadratic term in the field 
$\tilde\rho$ is now of the opposite sign to before, and therefore 
corresponds to real, rather than imaginary, noise. However, the 
truncation of this expansion at second order is not justifiable, and, 
furthermore, a consistent description of the annihilation kinetics in 
terms of the fields $\rho$ and $\tilde\rho$ comes with the price of 
having to incorporate `diffusion noise', i.e. the nonlinear coupling 
$-D \rho (\nabla \tilde\rho)^2$.

At any rate, this analysis and the previous discussions show that 
simply writing down a mean-field rate equation for the annihilation 
reaction and then adding {\em real} Gaussian noise does not in general 
yield an appropriate SPDE. An even more significant observation is 
that only two-particle reactions can straightforwardly be cast in the 
form of an SPDE, since the linearization \eref{eq:lintrafo} requires 
that the field $\bar \phi$ appears quadratically. For example, the 
triplet annihilation reaction $3 A \to 0$ reaction cannot simply be 
represented as the corresponding rate equation plus real, 
multiplicative noise.

\section{Renormalization Group Method}
\label{sec:renormalization}

In this section, we describe the basic methodology for performing
perturbative RG calculations in the context of reaction-diffusion
field theories. While our aim is to present these techniques in a
pedagogical manner, our discussion cannot be entirely self-contained
here. For additional details, specifically with respect to 
perturbation theory and its representation in terms of Feynman
diagrams, reference should also be made to the standard field theory
literature \cite{Amit84,Itzykson89,ZinnJustin93}.

\subsection{Diagrammatic expansion}

The diagrammatic expansion for performing field theory calculations is
constructed in the standard way: the part of the action that is 
bilinear in the fields is identified as a Gaussian reference action
$S_0$. All other terms are evaluated perturbatively by expanding the
exponential $\exp( - S + S_0)$ and averaging with statistical weight
$\exp(- S_0)$. These Gaussian averages decompose into products of pair 
correlation functions, and can be represented symbolically through
Feynman diagrams. Propagators, represented by lines in the Feynman
graphs, correspond to the pair correlators of fields that are averaged
together. The nonlinear couplings are graphically depicted as vertices
that connect propagators together.

We illustrate this procedure explicitly for the $k$th order 
single-species annihilation reactions $k A \to \ell A$ with 
$\ell < k$. In this case, we may take the diffusive part of the action
as $S_0$. Thus, the propagator becomes just the diffusion Green
function. For a single-species reaction, the diffusion action reads
\begin{equation}
\label{eq:S0}
     \fl S_0 = \int d^dx \int_{-\infty}^\infty dt \, \bar\phi \left( 
     \partial_t - D \nabla^2 \right) \phi = \int {d^dp \over (2\pi)^d}
     {d\omega \over 2\pi} \, \bar\phi({-\bf p},-\omega)
     \left( -i\omega + D p^2 \right) \phi({\bf p},\omega) ~ ,
\end{equation}
where we have used the time and space Fourier transformed fields
\begin{equation}
     \phi({\bf p},\omega) = \int d^dx \int dt \, 
     \exp(-i{\bf p}\cdot{\bf x} + i\omega t) \, \phi({\bf x},t) ~ ,
\end{equation}
and extended the time integration range to the entire real axis (as
will be justified below). Next we define the propagator as 
$G({\bf x},t) = \langle \bar\phi({\bf x},t) \phi(0,0) \rangle_0$. Its 
Fourier transform has the form $\langle \bar\phi({\bf p},\omega) 
\phi({\bf p}',\omega') \rangle_0 = G_0({\bf p},\omega) \, (2\pi)^d 
\delta({\bf p}+{\bf p}') \, 2\pi \delta(\omega+\omega')$, with the 
$\delta$ functions originating from spatial and temporal translation
invariance. Explicitly, we infer from the action \eref{eq:S0}
\begin{equation}
\label{eq:fprop}
     G_0({\bf p},\omega) = {1 \over -i\omega + Dp^2} ~ ,
\end{equation}
which follows from straightforward Gaussian integration. In the 
complex frequency plane the function $G_0({\bf p},\omega)$ has a 
single pole at $\omega = -i D p^2$. Upon performing the inverse
temporal Fourier transform from $\omega$ to $t$ we find
\begin{equation}
\label{eq:tprop}
     G_0({\bf p},t) = \exp \left( -D p^2 t \right) \, \Theta(t) ~ ,
\end{equation}
where $\Theta(t)$ denotes Heaviside's step function. Mathematically,
its origin is that the sign of $t$ determines whether the integration
contour is to be closed in the upper or lower frequency half plane. 
Physically, it expresses causality: the unidirectional propagator only
connects earlier $\bar\phi$ fields to later $\phi$ fields (as 
advertised before in \sref{subsec:coherent}). Since there exists no 
earlier source of $\bar\phi$ fields, the time integral in $S_0$ may as
well be extended from $[0,t_f]$ to all times, as claimed above.  
Obviously, for multi-species systems there is a distinct propagator of 
the form \eref{eq:fprop}, \eref{eq:tprop} for each particle type.

\begin{figure}[htb]
\begin{center}
\includegraphics[width=4.0in]{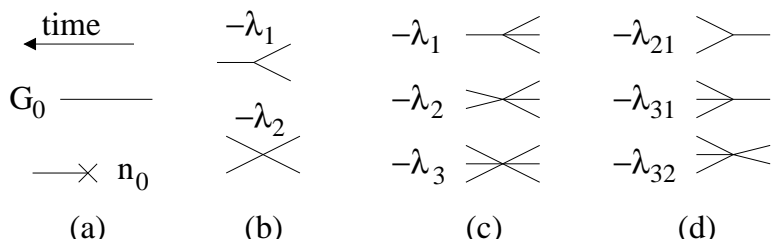}
\caption{Various diagrammatic components: (a) the propagator $G_0$ and
the initial density $n_0$, (b) vertices for $2 A \to \ell A$, (c)
vertices for $3 A \to \ell A$, and (d) vertices for branching 
reactions, such as $A \to 2 A$. Our convention throughout is that time
increases to the left.}
\label{fig:diagramparts}
\end{center}
\end{figure}

The vertices in the Feynman diagrams originate from the perturbative
expansion of $\exp(- S + S_0)$. For example, the term $\lambda_i 
\bar\phi^i \phi^k$ in the action \eref{eq:kAaction} requires $k$ 
{\em in}coming propagators, that is, averages connecting $k$ earlier
$\bar\phi$ fields to the $k$ later $\phi$ `legs' attached to the
vertex, and $i$ {\em out}going propagators, each of which links a
$\bar\phi$ to a later $\phi$ field. The diagrammatic representation of
the propagator and some vertices is depicted in 
\fref{fig:diagramparts}. Propagators attach to vertices as 
distinguishable objects, which implies that a given diagram will come
with a multiplicative combinatorial factor counting the number of ways
to make the attachments. (Note that we do not follow the convention of
defining appropriate factorials with the nonlinear couplings in the
action to partially account for this attachment combinatorics.) Any
contribution of order $m$ in the coupling $\lambda_i$ is represented
by a Feynman graph with $m$ corresponding vertices. Moreover, if we
are interested in the perturbation expansion for cumulants only, we
merely need to consider fully connected Feynman diagrams, whose
vertices are all linked through propagator lines. Lastly, the
so-called vertex functions are given in terms of one-particle
irreducible diagrams that do not separate into disjoint subgraphs if
one propagator line is `cut'.

The perturbative expansion of the initial state contribution 
$\exp[- n_0 \bar\phi(t=0)]$ creates $\bar\phi$ fields at $t = 0$. A 
term of order $n_0^m$ comes with a factor of $1/m!$, but will have 
$m!$ different ways to connect the initial $\bar\phi$ fields to the
corresponding Feynman graph, so the end result is that these 
factorials always cancel for the initial density. A similar
cancellation of factorials happens for the vertices; for example, a
term of order $\lambda_1^m$ does have the $1/m!$ cancelled by the
number of permutations of the $m$ $\lambda_1$ vertices in the
diagram. Notice, however, these combinatorial factors are distinct
from those that arise from the different possibilities to attach
propagators.

\begin{figure}[htb]
\begin{center}
\includegraphics[width=3.25in]{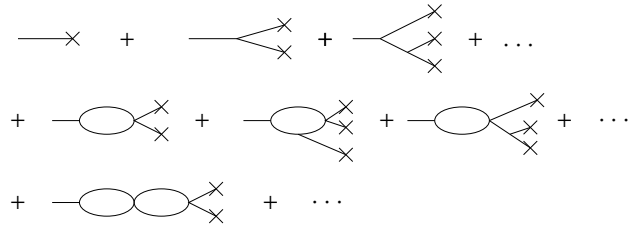}
\caption{Feynman graphs that contribute to the mean particle density
for the pair annihilation and coagulation reactions $A + A \to 0$ and
$A + A \to A$. The first row depicts tree diagrams, the second row
one-loop diagrams, and the third row two-loop diagrams.}
\label{fig:density}
\end{center}
\end{figure}

Since the systems of interest are frequently translationally invariant
(in space and time), the mathematical expressions represented by the
Feynman graphs are often most conveniently evaluated in Fourier
space. To calculate, say, the mean particle density $\langle \phi(t)
\rangle$ according to Eq.~\eref{eq:insert2}, one needs diagrams with a
single $\phi$ field at time $t$ which terminates the graph on the 
left. All diagrams that end in a single propagator line will thus
contribute to the density, see \fref{fig:density}. Since 
$\langle \phi(t) \rangle$ is spatially uniform, the final propagator 
must have ${\bf p} = 0$. Similarly, in momentum space the initial 
density terms are of the form $n_0 \bar\phi({\bf p} = 0, t = 0)$, so 
the propagators connected to these also come with zero wavevector. The
$\lambda_i$ vertices are to be integrated over position space, which 
creates a wavevector-conserving $\delta$ function. Diagrams that 
contain loops may have `internal' propagators with ${\bf p} \neq 0$, 
but momentum conservation must be satisfied at each vertex. These 
internal wavevectors are then to be integrated over, as are the 
internal time or frequency arguments.

To illustrate this procedure, consider the second graph in the first
row, and the first diagram in the second row of \fref{fig:density}, to
whose loop we assign the internal momentum label $\bf p$:
\begin{eqnarray}
\label{eq:I02}
     I_{02} &= \int_0^t dt_1 \, G_0(0,t-t_1) \, (-\lambda_1) \, 
     G_0(0,t_1)^2 \, n_0^2 ~ , \\ 
\label{eq:I12}
     I_{12} &= \int_0^t dt_2 \int_0^{t_2} dt_1 \, G_0(0,t-t_2) \, 
     (-\lambda_1) \nonumber \\
     &\times \int {d^dp\over(2\pi)^d} \ 2 \, G_0({\bf p},t_2-t_1) \, 
     G_0(-{\bf p},t_2-t_1) \, (-\lambda_2) \, G_0(0,t_1)^2 \, n_0^2
\end{eqnarray}
(the indices here refer to the number of loops and the factors of
initial densities involved, respectively). The factor $2$ in the
second contribution originates from the number of distinguishable ways
to attach the propagators within the loop. Noting that 
$G_0(p = 0, t > 0) = 1$, and that the required ${\bf p}$ integrations 
are over Gaussians, these expressions are clearly straightforward to
evaluate,
\begin{equation}
\label{eq:Ires}
     I_{02} = -\lambda_1 n_0^2 \, t ~ , \qquad 
     I_{12} = {8 \lambda_1 \lambda_2 \, n_0^2 \over (8 \pi D)^{d/2}} 
     \, {t^{2-d/2} \over (2-d) (4-d)}
\end{equation}
(for $d \not= 2,4)$. Consequently, the effective dimensionless 
coupling associated with the loop in the second diagram is 
proportional to $(\lambda_2 / D^{d/2}) \, t^{1-d/2}$.

Hence, in low dimensions $d < 2$, the perturbation expansion is benign
at small times, but becomes ill-defined as $t \to \infty$, whereas the
converse is true for $d > 2$. In two dimensions, the effective 
coupling diverges as $(\lambda_2 / D) \ln (D t)$ for both $t \to 0$
and $t \to \infty$. The `ultraviolet' divergences for $d \geq 2$ in
the short-time regime are easily cured by introducing a short-distance
cutoff in the wavevector integrals. This is physically reasonable
since such a cutoff was anyhow originally present in the form of the
lattice spacing (or particle capture radius). The fluctuation
contributions will then explicitly depend on this cutoff scale. Thus,
in dimensions $d > d_c = 2$, perturbation theory is applicable in the
asymptotic limit; this implies that the overall scaling behavior of 
the parameters of the theory cannot be affected by the analytic loop
corrections, which can merely modify amplitudes. In contrast, the 
physically relevant `infrared' divergences (in the long-time, 
long-distance limit) in low dimensions $d \leq 2$ are more serious and 
render a `naive' perturbation series meaningless. However, as will be 
explained in the following subsections, via exploiting scale 
invariance and the exact structure of the renormalization group, one 
may nevertheless extract fluctuation-corrected power laws by means of 
the perturbation expansion.

\begin{figure}[htb]
\begin{center}
\includegraphics[width=2.25in]{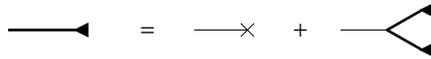}
\caption{Graphical representation of the Dyson equation for the 
particle density in the $A + A \to 0$ and $A + A \to A$ reactions.}
\label{fig:densityDyson}
\end{center}
\end{figure}

Feynman graphs that contain no loops are called {\em tree} diagrams. 
For the particle density calculation illustrated in 
\fref{fig:density}, these tree diagrams are formed with only 
$\lambda_1$ and $n_0$ vertices, and we denote the sum of all those
tree contributions by $a_{\rm tr}(t)$. For example, for the 
$A + A \to (0,A)$ pair reactions, we may construct this entire series 
iteratively to all orders as shown graphically in 
\fref{fig:densityDyson}. Thus we arrive at a self-consistent Dyson 
equation for the particle density. More generally, for the 
single-species reactions $k A \to \ell A$ with $\ell < k$ the vertex 
on the right-hand side is connected to $k$ full tree density lines. 
Since all propagators in tree diagrams come with ${\bf p} = 0$, the 
corresponding analytical expression for the Dyson equation reads
\begin{equation}
     a_{\rm tr}(t) = n_0-\lambda_1 \int_0^t dt'\, a_{\rm tr}(t')^k ~.
\end{equation}
Upon taking a time derivative, this reduces to the mean-field rate
equation \eref{eq:kArate_eq}, with the correct initial condition. 
Furthermore, $\lambda_1 = (k-\ell) \lambda_0$, i.e., the rate constant
is properly proportional to the number of particles removed by the 
reaction. Evidently, therefore, the tree-level approximation is 
equivalent to simple mean-field theory, and any fluctuation 
corrections to the rate equation must emerge from Feynman graphs that 
incorporate higher-order vertices $\lambda_i$ with $i>1$, i.e., 
diagrams with loops. We note that the mean-field rate equations also 
follow from the stationarity conditions, i.e., the `classical field 
equations', for the action $S$ (regardless of performing any field 
shifts). For example, taking $\delta S / \delta \phi = 0 = \delta S / 
\delta \bar\phi$ for the action \eref{eq:kAaction} results in
$\bar\phi = 0$ and $\phi = a_{\rm tr}(t)$.

Following up on our earlier discussion, we realize that the loop
fluctuation contributions cannot alter the asymptotic power laws that
follow from the tree diagrams in sufficiently large dimensions 
$d > d_c$, where mean-field theory should therefore yield accurate 
scaling exponents. However, note that, for $d > d_c$, there will 
generally be non-negligible and non-universal (depending on the 
ultraviolet cutoff) fluctuation corrections to the amplitudes. Recall 
that the (upper) critical dimension $d_c$ can be readily determined as
the dimension where the effective coupling associated with loop 
integrals becomes dimensionless.

\subsection{Renormalization}
\label{subsec:renor}

As we have seen in the above example \eref{eq:I12}, when one naively
tries to extend the diagrammatic expansion beyond the tree 
contributions to include the corrections due to loop diagrams, one
encounters divergent integrals. We are specifically interested in the
situation at low dimensions $d \leq d_c$: here the {\em infrared} (IR)
singularities (apparent as divergences as external wavevectors 
${\bf p} \to 0$ and either $t \to \infty$ or $\omega \to 0$) emerging 
in the loop expansion indicate substantial deviations from the 
mean-field predictions. Our goal is to extract the correct asymptotic 
power laws associated with these `physical' infrared singularities in 
the particle density and other correlation functions. To this end, we
shall turn to our advantage the fact that power laws reflect an
underlying {\em scale invariance} in the system. Once we have found a
reliable method to determine the behavior of any correlation function
under either length, momentum, or time scale transformations, we can
readily exploit this to construct appropriate scaling laws.

There exist well-developed tools for the investigation and subsequent
renormalization of {\em ultraviolet} (UV) singularities, which stem
from the large wavenumber contribution to the loop integrals. In our
models, these divergences are superficial, since we can always
reinstate short-distance cutoffs corresponding to microscopic lattice
spacings. However, any such {\em regularization} procedure introduces
an explicit dependence on the associated regularization scale. Since
it does not employ any UV cutoff, dimensional regularization is 
especially useful in higher-loop calculations. Yet even then, in order
to avoid the IR singularities, one must evaluate the integrals at some
finite momentum, frequency, or time scale. In the following, we shall
denote this normalization momentum scale as $\kappa$, associated with 
a length scale $\kappa^{-1}$, or, assuming purely diffusive 
propagation, time scale $t_0 = 1 / (D \kappa^2)$. Once the theory has 
been rendered finite with respect to the UV singularities via the 
renormalization procedure, we can subsequently extract the dependence 
of the relevant renormalized model parameters on $\kappa$. This is 
formally achieved by means of the Callan-Symanzik RG flow equations. 
Precisely in a regime where scale invariance holds, i.e., in the 
vicinity of a RG fixed point, the ensuing ultraviolet scaling 
properties also yield the desired algebraic behavior in the infrared. 
(For a more elaborate discussion of the connections between UV and IR 
singularities, see Ref.~\cite{Janssen05}.)

The renormalization procedure itself is, in essence, a resummation of
the naive, strongly cutoff-dependent loop expansion that is 
subsequently well-behaved as the ultraviolet regulator is removed. 
Technically, one defines renormalized effective parameters in the 
theory that formally absorb the ultraviolet poles. When such a 
procedure is possible --- i.e., when the field theory is
`renormalizable', which means only a {\em finite} number of 
renormalized parameters need to be introduced --- one obtains in this
way a unique continuum limit. Examining the RG flow of the 
scale-dependent parameters of the renormalized theory, one encounters
universality in the vicinity of an IR-stable fixed point: there the 
theory on large length and time scales becomes independent of 
microscopic details. The preceding procedure is usually only 
quantitatively tractable at the lowest dimension that gives UV 
singularities, i.e., the upper critical dimension $d_c$, which is also 
the highest dimension where IR divergences appear. In order to obtain 
the infrared scaling behavior in lower dimensions $d < d_c$, we must 
at least initially resort to a perturbational treatment with respect 
to the marginal couplings in the theory, which are dimensionless at 
$d_c$ (and perhaps subsequently resum the perturbation series). The 
scaling exponents can then be obtained in a controlled manner in a 
dimensional expansion with respect to the small parameter $\epsilon = 
d_c - d$.
 
We can follow standard procedures (see, e.g., Refs.~\cite{Amit84,
Itzykson89, ZinnJustin93}) for implementing the renormalization 
program. First, we must identify the primitive UV divergences, the 
sub-components of the diagrams that are responsible for these 
singularities. This is most conveniently done using the vertex 
functions $\Gamma^{(m,n)}$, which represent a sum of all possible 
one-particle irreducible Feynman graphs that are attached to $n$ 
incoming and $m$ outgoing propagator lines (of course, for a 
multispecies vertex function, separate indices are required for the
incoming and outgoing lines of each species). In frequency and
wavevector space these subdiagrams enter multiplicatively, which means
that once the vertex function divergences are resolved, the general
diagrammatic expansion will be well-behaved. Which vertex functions
are primitively divergent can be ascertained by direct power counting:
$[\Gamma^{(m,n)}] = \kappa^\alpha$, where $\kappa$ denotes some
reference wavevector, such as the normalization scale mentioned above. 
The scaling dimension of the vertex function $\Gamma^{(m,n)}$ is just 
that of the coupling $\lambda_{mn}$ from a term $\lambda_{mn} 
\bar\phi^m \phi^n$ in the action, since at the tree level 
$\Gamma^{(m,n)} \sim \lambda_{mn}$. Loop diagram fluctuation 
corrections, however, require additional nonlinear couplings, which in
turn determine the primitive degree of divergence for the associated
momentum space integrals. We refer to the standard field theory texts
for a general discussion of the ensuing vertex function dimensional
analysis, but provide a brief outline of the procedure for our
situation.

The action $S$ itself must be dimensionless, so any term in the
integrand of $S$ must have scaling dimension $\kappa^{d+2}$. Consider
the $k A \to \ell A$ annihilation reaction with $\ell<k$. The 
interaction vertices $\lambda_i \bar\phi^i \phi^k$ in the action
\eref{eq:kAaction} correspond to $k$ incoming and $i=1 \dots k$
outgoing lines. With our choice of taking the continuum limit, the
scaling dimension of the fields are $[\phi] = \kappa^d$ and
$[\bar\phi] = \kappa^0$. Hence we obtain $[\Gamma^{(i,k)}] =
[\lambda_i] = \kappa^{2-(k-1)d}$ for all $i$ (recall that $\lambda_i
\propto \lambda_0$ of the unshifted theory). Let us now investigate 
the lowest-order loop correction to the vertex function 
$\Gamma^{(i,k)}$, which contains $k$ internal propagator lines, and 
thus is proportional to $\lambda_i \, \lambda_k$. The involved 
momentum integral therefore must scale as $[\lambda_k]^{-1} = 
\kappa^{(k-1)d-2}$. By choosing as $\kappa$ the inverse short-distance 
cutoff, we see that if the exponent here is non-negative, the vertex 
function contains a primitive UV divergence (as $\kappa \to \infty$) 
and must be renormalized. (The converse is true for the IR 
singularities, which emerge in the limit $\kappa \to 0$.) The $l$-th 
order corrections to the bare vertex function must scale as 
$\kappa^{l[(k-1)d-2]}$. Consequently, for a given $k$, the vertex 
functions $\Gamma^{(i,k)}$ become primitively UV-divergent, and 
increasingly so in higher loop orders, for $d > d_c = 2/(k-1)$ for all 
$i$. The IR singularities, on the other hand, become successively 
worse in higher orders of the perturbation expansion for $d < d_c$. At 
the critical dimension, the loop diagrams carry logarithmic UV and IR 
divergences, independent of the loop order.

Hence this scaling discussion already reveals the upper critical
dimension $d_c = 2$ for pair annihilation and coagulation, $A+A \to 
(0,A)$, in agreement with our analysis following Eq.~\eref{eq:Ires}, 
whereas $d_c = 1$ for the triplet reactions $3A \to \ell A$. All
higher-order reactions should be adequately described by mean-field
theory, as represented by the tree Feynman graphs. But in general,
since the assignment of scaling dimensions to the fields $\phi$ and
$\bar\phi$ is somewhat arbitrary, one needs to be more careful and
first determine the effective couplings in the perturbational 
expansion and from there infer the upper critical dimension. For 
example, the field theory for directed percolation, see 
\sref{sec:transitions} below, incorporates the vertices $\lambda_{12} 
\bar\phi \phi^2$ and $\lambda_{21} \bar\phi^2 \phi$. The effective 
coupling then turns out to be the product $u \sim \lambda_{12} 
\lambda_{21}$, whose scaling dimension is $[u] = \kappa^{4-d}$, which 
indicates that actually $d_c = 4$ in this case.

Once the primitive divergences are identified, they are used to define
renormalized coupling constants into which the UV singularities are
absorbed. For the $k A \to \ell A$ decay reaction, this procedure is
unusually simple, and in fact the entire perturbation series can be
summed to all orders. First, we note that since all vertices in the
action \eref{eq:kAaction} have $k \geq 2$ incoming lines, one cannot
construct any loop diagram for the vertex function $\Gamma^{(1,1)}$
that corresponds to the inverse propagator. Hence the bare diffusion
propagator \eref{eq:fprop} or \eref{eq:tprop} is not affected by
fluctuations and the tree-level scaling $x \sim p^{-1} \sim 
(D t)^{1/2}$ remains intact \cite{Peliti86}. The absence of field and
diffusion constant renormalization in this case constitutes the
fundamental reason that improved mean-field theories of the 
Smoluchowski type, and also simple scaling approaches, are capable of
obtaining correct density decay exponents. Hence, for pair 
annihilation or coagulation, one would predict that the density of 
surviving particles at time $t$ is given in terms of the 
diffusion length by $a(t) \sim x^{-d} \sim (D t)^{-d/2}$.

\begin{figure}[htb]
\begin{center}
\includegraphics[width=3.25in]{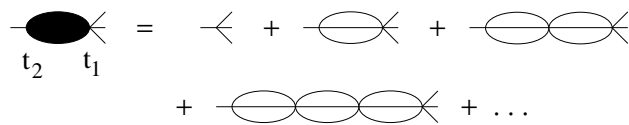}
\caption{The set of primitively divergent diagrams contributing to 
$\Gamma^{(1,3)}(t_2-t_1)$ for the $3 A \to \ell A$ reaction, 
$\ell \leq 2$.}
\label{fig:renormalization}
\end{center}
\end{figure}

This leaves us with the renormalization of the vertex couplings 
$\lambda_m$, encoded in the vertex functions $\Gamma^{({m,k})}$. Since
all $\lambda_m$ are proportional to the reaction rate $\lambda_0$, 
there is essentially only a single renormalized coupling $g_R$ (as is 
obvious if we work with the unshifted action). As a representative 
example, we depict the ensuing diagrammatic expansion for the case 
$m=1$ and $k=3$ in \fref{fig:renormalization}. It will be sufficient 
to work with ${\bf p}=0$. In $({\bf p},t)$ space we obtain explicitly
\begin{eqnarray}
\label{eq:coupling}
     \Gamma^{({m,k})}(t_2-t_1) = &\lambda_m \delta(t_2-t_1) 
     - \lambda_m \lambda_0 I(t_2-t_1) \nonumber \\
     &+ \lambda_m \lambda_0^2 \int_{t_1}^{t_2} \! dt' \, I(t_2-t') 
     I(t'-t_1) - \dots \ ,
\end{eqnarray}
where we have used $\lambda_k = \lambda_0$, and $I(t)$ is given by the
following integral with $k-1$ loops over $k$ propagators:
\begin{equation}
\label{eq:loopres}
     \fl \qquad I(t) = k! \int \prod_{i=1}^k \left( 
     {d^dp_i\over(2\pi)^d} \right) (2\pi)^d \delta\left( \sum_{i=1}^k 
     {\bf p}_i \right) \exp\biggl( -\sum_{i=1}^k D p_i^2 t \biggr) = 
     B_k (D t)^{-d/d_c} ~ ,
\end{equation}
with $B_k = k! \, k^{-d/2} (4\pi)^{-d/d_c}$ and $d_c = 2/(k-1)$. 
Taking the Laplace transform
\begin{equation}
\label{eq:laplace}
     \widetilde\Gamma^{({m,k})}(s) = 
     \int_0^\infty \Gamma^{({m,k})}(t) \, e^{-s t}\, dt ~ ,
\end{equation}
the convolution theorem renders \eref{eq:coupling} into a geometric 
sum:
\begin{equation}
\label{eq:verseries}
     \widetilde\Gamma^{(m,k)}(s) = 
     {\lambda_m \over 1 + \lambda_0 \widetilde{I}(s)} ~ , \qquad 
     \widetilde{I}(s) = B_k \, \Gamma(\epsilon/d_c) D^{-d/d_c} 
     s^{-\epsilon/d_c} ~ ,
\end{equation}
where $\epsilon = d_c - d$, and $\Gamma(x)$ is Euler's gamma function. 
For $\epsilon > 0$, the loop integral $\widetilde{I}(s)$ displays the 
expected IR divergence as $s \to 0$. On the other hand, the UV 
singularity (for finite $s$) at $d_c$ appears as an $\epsilon$ pole in 
the gamma function $\Gamma(\epsilon/d_c) \propto d_c / \epsilon$.

Recall that the bare effective coupling $\lambda_0/D$ has scaling
dimension $\kappa^{2-(k-1)d} = \kappa^{2\epsilon/d_c}$. Hence we may
introduce the {\em dimensionless} parameter $g_0 = (\lambda_0/D)\,
\kappa^{-2\epsilon/d_c}$ with some arbitrary momentum scale $\kappa$. 
Next we define its {\em renormalized} counterpart by the value of the 
vertex function $\widetilde\Gamma^{({k,k})}$ at vanishing external 
wavevectors and Laplace argument $s = 1/t_0 = D \kappa^2$, which sets 
our {\em normalization point}. Thus,
\begin{equation}
\label{eq:gR}
     \fl \qquad g_R = \widetilde\Gamma^{({k,k})}(s) 
     \Big\vert_{s=D\kappa^2} \, \kappa^{-2\epsilon/d_c} / D = Z_g \, 
     g_0 ~ , \qquad Z_g^{-1} = 1 + g_0 \, B_k \, \Gamma(\epsilon/d_c)
\end{equation}
according to Eq.~\eref{eq:verseries}. Formally, a perturbative 
expansion in terms of $g_0$ can be readily exchanged for an expansion
in $g_R$ by straight substitution $g_0 = g_R / [1 - g_R \, B_k \,
\Gamma(\epsilon/d_c)]$. As $\epsilon \to 0$, however, this 
substitution becomes singular, since the multiplicative 
renormalization constant $Z_g^{-1}$ that was introduced to absorb the
UV pole diverges in this limit. It is crucial to note that the
renormalized coupling explicitly depends on the normalization scale
$\kappa$.  Indeed, this is borne out by calculating the associated
{\em RG $\beta$ function}
\begin{equation}
\label{eq:beta}
     \fl \quad \beta_g(g_R) = \kappa \frac{\partial}{\partial \kappa} 
     \, g_R = g_R \left[ - \frac{2 \epsilon}{d_c} - \kappa 
     \frac{\partial}{\partial \kappa} \ln Z_g^{-1} \right] = 2 g_R 
     \Biggl[ - \frac{\epsilon}{d_c} + B_k \, \Gamma\left( 
     1+\frac{\epsilon}{d_c}\right) g_R \Biggr] ~ .
\end{equation}
Notice that $\beta_g$, here computed to all orders in perturbation
theory, is regular as $\epsilon \to 0$ when expressed in terms of
renormalized quantities.  For the simple annihilation models, the
$\beta$ function is exactly quadratic in $g_R$, which is not typical
and is due to the geometric sum in the primitively divergent vertex
function, which reduces the fluctuation contributions effectively to
the one-loop graph. Eq.\eref{eq:beta} has the standard structure that
the linear coefficient is of order $\epsilon=d_c-d$ while the 
quadratic coefficient is order unity. The theory is manifestly
scale-invariant (independent of $\kappa$) at either $g_R = 0$ (which
here corresponds to pure diffusion, no reactions) or at the special
{\em fixed-point} value for the coupling given by the nontrivial zero
of the $\beta$ function
\begin{equation}
\label{eq:fpoint}
     g_R^* = [B_k \, \Gamma(\epsilon/d_c)]^{-1} ~ , 
\end{equation}
which is of order $\epsilon$.

We remark that the above renormalization structure directly applies to
multispecies annihilation reactions as well. Consider, for example, 
the action \eref{eq:ab_action} for $A+B \to 0$. Once again, the
vertices permit no propagator renormalization, and the perturbation
expansion for the vertex functions that define the renormalized
reaction rate takes the same form as in \fref{fig:renormalization},
with incoming $A$ and $B$ lines, and the series of internal loops
formed with precisely these two distinct propagators. Consequently,
we just recover the previous results \eref{eq:gR}, \eref{eq:beta}, and
\eref{eq:fpoint} with $k=2$.

\subsection{Callan-Symanzik equation and loop expansion for the 
            density}
\label{subsec:CSequation}

In order to employ the renormalization group machinery to obtain
asymptotic (long-time, large-distance) expressions for the particle
density and its correlations, we next develop the Callan-Symanzik
equation for the density. By means of perturbation theory in the
IR-finite regime and the subsequent substitutions of $g_R$ for
$\lambda_0$, the density can be calculated as an explicit function of
time $t$, initial density $n_0$, diffusion constant $D$, renormalized
coupling $g_R$, and arbitrary normalization point $\kappa$ (or the
equivalent time scale $t_0 = 1/D \kappa^2$). Since $\kappa$ does not
appear at all in the unrenormalized theory, we must have $\kappa \,
da(t,n_0,D,\lambda_0)/d\kappa=0$ for the particle density at fixed
{\em bare} parameters, or equivalently, after rewriting in terms of
the renormalized density,
\begin{equation}
\label{eq:CSeq1}
     \left[ \kappa \, \frac{\partial}{\partial \kappa} + \beta_g(g_R) 
     \,\frac{\partial}{\partial g_R} \right] a(t,n_0,D,\kappa,g_R) = 0
\end{equation}
with the $\beta$ function \eref{eq:beta}. Yet dimensional analysis
tells us that $a(t,n_0,D,\kappa,g_R) = \kappa^d \hat
a(t/t_0,n_0/\kappa^d,g_R)$, reducing the number of independent
variables to three. Consequently, Eq.~\eref{eq:CSeq1} yields the {\em
Callan-Symanzik} (CS) equation
\begin{equation}
\label{eq:CSeq}
     \left[ 2Dt \, {\partial\over\partial (Dt)} - d n_0 \, {\partial
     \over \partial n_0} + \beta_g(g_R) \, {\partial \over \partial 
     g_R} + d \, \right] a(t,n_0,D,\kappa,g_R) = 0 \ .
\end{equation}

This partial differential equation is solved by the standard method of
characteristics, where one introduces a flow parameter via $\kappa \to
\kappa \ell$. The IR asymptotic region then corresponds to $\ell \to 
0$. For our purposes, it is most convenient to directly
employ $(\kappa \ell)^2 = 1/D t$ or $\ell^2 = t_0/t$. Thereby we find
\begin{equation}
\label{eq:CSsolution}
     a(t,n_0,t_0,g_R) = (t_0/t)^{d/2} \ 
     a\Bigl( \tilde n_0(t) , \tilde g_R(t) \Bigr) ~ ,
\end{equation}
with the running initial density
\begin{equation}
\label{eq:n0tilde}
     \tilde n_0(t) = (t/t_0)^{d/2} \, n_0 ~ ,
\end{equation}
and the running coupling $\tilde g_R$ defined by the solution of the 
characteristic equation
\begin{equation}
\label{eq:floweq}
     \fl \qquad\qquad \ell \, {d\tilde g_R(\ell) \over d\ell} = 
     - 2 t \, {d\tilde g_R(t) \over dt} = \beta_g(\tilde g_R) ~ , 
     \qquad \tilde g_R(\ell=1) = \tilde g_R(t=t_0) = g_R ~ .
\end{equation}
The method of characteristics requires a known value of the function,
in this case the density, for some value of the running parameters. 
Since we chose the normalization point $s = 1/t_0 > 0$ outside the 
IR-singular region, we may use perturbation theory to calculate the 
right-hand side of Eq.~\eref{eq:CSsolution}. The Callan-Symanzik 
equation then allows us to transport this result into the 
perturbatively inaccessible asymptotic region.

Because of the simple form of the $\beta$ function \eref{eq:beta}, the
running coupling can be found exactly by integrating the flow equation
\eref{eq:floweq}:
\begin{equation}
\label{eq:gRtilde}
     \tilde g_R(t) = g_R^* \left[ 1 + {g_R^*-g_R \over g_R} \left( 
     \frac{t_0}{t} \right)^{\epsilon/d_c} \right]^{-1} \qquad 
     (\epsilon \not= 0) ~ . 
\end{equation}
Thus we see for $\epsilon > 0$ that the running renormalized coupling
approaches the RG fixed point \eref{eq:fpoint} as $t \to \infty$,
independent of its initial value $g_R$. Thus, {\em universal} behavior
emerges in the asymptotic regime, whose scaling properties are 
governed by the IR-stable fixed point $g_R^*$. Therefore an expansion
in powers of $g_0$ is converted, via the CS equation, to an expansion
in powers of $\epsilon$. For this purpose, we merely need to invert
Eq.~\eref{eq:gR} to find $g_0$ in terms of $g_R$: $g_0 = g_R / [1 -
g_R/g_R^*] = g_R + g_R^2 / g_R^* + \dots$. Notice that $g_R = g_R^*$
formally corresponds to $g_0 \sim \lambda_0 / D = \infty$, i.e., the
annihilation reactions are indeed diffusion-limited. Above the 
critical dimension ($\epsilon < 0$), $\tilde g_R(t) \to 0$
algebraically $\sim t^{-|\epsilon|/d_c}$, whereas precisely at $d_c$
the running coupling tends to zero only logarithmically,
\begin{equation}
\label{eq:dcgRt}
     \tilde g_R(t) = {g_R \over 1 + B_k \, g_R \ln (t/t_0)} \qquad 
     (\epsilon = 0) ~ .
\end{equation}

We may use these findings to already make contact with both the rate
equation and Smoluchowski approximations. For $d > d_c$, the effective
reaction rate $\lambda(t) \sim D (\kappa \ell)^{2 \epsilon/d_c} \tilde
g_R(\ell) = D (D t)^{-\epsilon/d_c} \tilde g_R(t) \to \rm const.$
asymptotically, as implicitly taken for granted in mean-field theory. 
Below the critical dimension, however, $\lambda(t \gg t_0) \sim D 
(D t)^{-\epsilon/d_c} g_R^*$ or its density-dependent counterpart 
$\lambda(a) \sim D a^{2 \epsilon / (d \, d_c)}$ decrease precisely as 
in the Smoluchowski approach. At $d_c$, we have instead $\lambda(t) 
\sim D / \ln (t/t_0)$ or $\lambda(a) \sim D / \ln (1/a)$. Replacing 
$\lambda \to \lambda(t)$ or $\lambda(a)$ in the mean-field rate 
equations \eref{eq:kArate_eq} then immediately yields the results 
\eref{eq:bramson} for $k=2$, whereas $a(t) \sim [\ln (Dt) / Dt]^{1/2}$ 
for $k=3$ at $d_c = 1$.

\begin{figure}[htb]
\begin{center}
\includegraphics[width=3.25in]{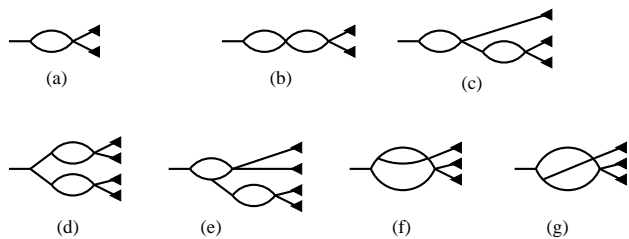}
\caption{One-loop and two-loop Feynman diagrams for the particle 
density, shown for the pair annihilation reaction $k=2$.}
\label{fig:loops}
\end{center}
\end{figure}

While we have now established a systematic expansion in terms of 
$g_R$, a perturbative calculation in powers of $n_0$ is not useful,
since $\tilde n_0(t)$ diverges for $t \gg t_0$, Eq.~\eref{eq:n0tilde}. 
It is thus imperative to calculate to all orders in the initial 
density $n_0$. To this end, we proceed to group the Feynman graphs for 
the particle density according to the number of closed loops involved. 
First, we obtain the tree diagrams represented by the Dyson equation 
in \fref{fig:densityDyson}. When substituted into the right-hand side 
of Eq.~\eref{eq:CSsolution} the limit $\tilde n_0 \to \infty$ will 
give a finite result, with leading corrections $\sim 1/\tilde n_0 \sim 
t^{-d/2}$ that vanish asymptotically. Explicitly, replacing the bare 
with flowing renormalized quantities in Eqs.~\eref{eq:CSsolution} and 
\eref{eq:kArate_sol} at the RG fixed point gives
\begin{equation}
\label{eq:rendensity}
     \fl \qquad a(t) = {n_0 \over [1 + n_0^{2/d_c} (k-1)(k-\ell) g_R^* 
     \, (D t)^{d/d_c}]^{d_c/2}} ~ 
     \to ~ \widetilde A_{k\ell} \, (D t)^{-d/2} ~ ,
\end{equation}
with {\em universal} amplitude $\widetilde A_{k\ell} = [(k-1) (k-\ell)
g_R^*]^{-1/(k-1)}$. 

\begin{figure}[htb]
\begin{center}
\includegraphics[width=3.25in]{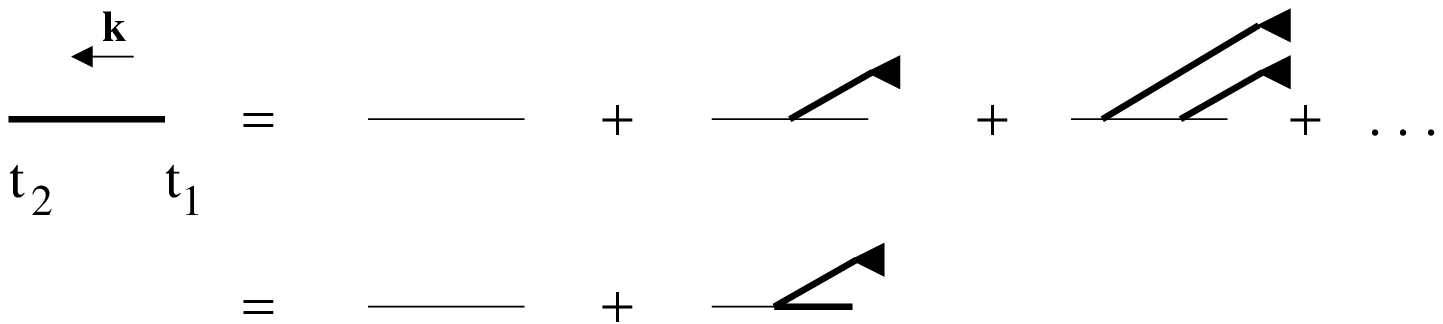}
\caption{Response function for k=2}
\label{fig:responsefunction}
\end{center}
\end{figure}

Next, we consider one- (a) and two-loop (b)--(g) diagrams for the
density, depicted in \fref{fig:loops} (all for the case $k=2$, but the
generalization to arbitrary $k$ is obvious). In order to sum over all
powers of $n_0$, the propagators in these diagrams are replaced with
{\em response functions}, which include sums over all tree-level
dressings. These are depicted in \fref{fig:responsefunction}, along
with the Dyson equation they satisfy. For $d \leq d_c$ each vertex
coupling asymptotically flows to the $O(\epsilon)$ fixed point
\eref{eq:fpoint}, so the loop expansion corresponds to an ordering in
successive powers of $\epsilon = d_c-d$. However, each order of the
expansion, under RG flow, comes in with the same $t^{-d/2}$ time
dependence. Thus, the loop expansion confirms that the exponent is
given explicitly by the tree-level result, and provides an epsilon
expansion for the {\em amplitude} of the density decay.

As mentioned above, the very same renormalizations hold for 
multi-species annihilation reactions, for example the pair process
$A+B \to 0$. Consequently, in the case of unequal initial $A$ and $B$
densities (with $a_0 < b_0$, say) in dimensions $d > d_c = 2$, the
mean-field result that the minority species vanishes exponentially
$a(t) \sim \exp (- \lambda t)$ is recovered, whereas for $d < 2$ the
direct replacement $\lambda \to \lambda(t) \sim D (Dt)^{-1+d/2}$
correctly yields a stretched exponential decay $a(t) \sim \exp[-{\rm
const.} \, (Dt)^{d/2}]$, while at $d_c=2$ the process is slowed down
only logarithmically, $a(t) \sim \exp[-{\rm const.} \, Dt/\ln (Dt)]$. 
The asymptotic $B$ particle saturation density is approached with the 
same time dependences. The amplitudes in the exponentials were 
computed exactly by other means in dimensions $d \leq 2$ by Blythe
and Bray \cite{Blythe03}.

\section{Further Applications}
\label{sec:furtherapp}

Now that we have established the basic field-theoretic RG machinery
necessary to systematically compute exponents and amplitudes, we can
summarize some more sophisticated applications. We deal first with
systems without phase transitions, before moving on in
\sref{sec:transitions} to describe reaction-diffusion systems that
display nonequilibrium phase transitions between active and absorbing
states. Our aim in this section will be to give a brief outline of the
results available using RG methods, rather than to delve too deeply
into calculational details.

\subsection{Single-species reactions}
\label{subsec:onespecies}

$\bullet$ The $k A \to \ell A$ reaction with $\ell < k$:

\noindent
The RG treatment for the general single-species annihilation reactions
$kA \to \ell A$ ($\ell < k$) was explicitly covered in the previous
sections. The upper critical dimension of these reactions is
$d_c=2/(k-1)$, and we note in particular that the reactions $A+A\to 0$
and $A+A\to A$ are in the same universality class \cite{Peliti86}. We
also emphasize again that, for $d\leq d_c$, the amplitudes and
exponents are universal, independent of the initial conditions (apart
from highly specialized initial conditions, such as those in
Ref.~\cite{Droz95}, where the particles were initially positioned in
pairs).

\

\noindent
$\bullet$ $A+A\rightarrow (0,A)~$ with particle input $0 \to A$:

\noindent
Droz and Sasv\'ari \cite{Droz93} studied the steady state of the
combined $A+A \to (0,A)$ and $0 \to A$ reactions, focusing 
particularly on how the density scales with $J$, the particle input 
rate. This process appears as an interaction $J\bar\phi$ in the 
action. Power counting gives $[J]=\kappa^{2+d}$ and straightforward 
arguments show that for $d < d_c = 2$, and for sufficiently small 
values of $J$, the density scales as $J^{d/(d+2)}$, and that the 
characteristic relaxation time behaves as $\tau \sim J^{-2/(d+2)}$. 
Finally, these findings were combined to reproduce the standard 
density scaling as $t^{-d/2}$. Rey and Droz extended this approach to 
provide explicit perturbative calculations of the density scaling 
function \cite{Rey97}.

\

\noindent
$\bullet$ Disordered systems:

\noindent
Another important variation on these simple reaction-diffusion models
is to include quenched disorder in the transport. Various models of
quenched random velocity fields in the $A+A\to 0$ reaction have been
investigated using RG techniques, including uncorrelated (Sinai) 
disorder \cite{Richardson99a} and also long-ranged correlated 
potential disorder \cite{Park98}. We consider first the case of (weak)
Sinai disorder, with velocity correlator $\langle v^{\alpha}(x) 
v^{\beta}(y) \rangle = \Delta\delta_{\alpha,\beta} \delta(x-y)$ 
analyzed by Richardson and Cardy \cite{Richardson99a}. An effective 
action is found by averaging over this disorder, after which one must 
renormalize the disorder strength and diffusion constant in addition 
to the reaction rate. Unlike the case of pure diffusive transport, it 
turns out that the amplitude for the asymptotic density decay rate as 
a function of time is nonuniversal for $d < 2$: $n \sim C_d t^{-d/z}$, 
with $z = 2+2\epsilon^2+O(\epsilon^3)$ and $\epsilon = 2-d$, but where
$C_d$ must be nonuniversal on dimensional grounds. It is only when 
rewriting the density as a function of the disorder-averaged diffusion 
length that a universal scaling relation emerges: $n \sim B_d \langle 
r^2 \rangle^{-d/2}$, where $B_d$ {\em is} universal. Results were also 
obtained for $d=2$, where the effects of the uncorrelated disorder are 
not strong (only the amplitude, but not the exponent, of the 
asymptotic density decay is altered). Interestingly, for weak 
disorder, it was found that the amplitude of the density decay is 
reduced, implying that the effective reaction rate is faster than for 
the case of purely diffusing reactants. Physically, this results from 
the disorder `pushing' particles into the same region of space, thus 
speeding up the kinetics. Theoretically, this originates in a 
disorder-induced renormalization of the reaction rate. However, as the
disorder is increased it was also shown that the reaction rate would 
then begin to decrease. This stems from a disorder-induced 
renormalization of the diffusion, which works to slow the kinetics, 
i.e. operates in the opposite direction to the disorder-induced 
reaction rate renormalization.

The related case of long-ranged potential disorder, where the random
velocity field can be considered as the gradient of a random 
potential, was analyzed using RG methods in two dimensions by Park and
Deem \cite{Park98}. In this case, rather more drastic effects were
found, with an altered decay exponent from the case of purely 
diffusing reactants. Physically, this results from the different 
nature of the disordered landscape \cite{Kravtsov85}, where for 
long-ranged potential disorder, but not of the Sinai type, deep 
trapping wells exist where, in order to escape the trap, a particle 
must move in an unfavorable direction. Park and Deem employed replicas 
to analyze the effect of long-ranged disorder, where the correlation 
function of the quenched random potential behaves as $\gamma/k^2$. 
They obtained that the asymptotic density decay was modified to 
$t^{\delta-1}$, where $\delta$ is defined in the absence of reaction 
by the anomalous diffusion relation $\langle r(t)^2\rangle \sim 
t^{1-\delta}$. Here, $\delta$ was found to be a nonuniversal exponent 
depending on the strength of the disorder. The amplitude of the decay 
also turned out to be a nonuniversal quantity.  

We also mention that `superfast' reactivity has been found in $d=2$
for the $A+A\to 0$ reaction in a model of turbulent flow with 
potential disorder \cite{Deem98b}. This case was also investigated 
numerically \cite{Tran99}. RG methods indicated that this regime 
persists in a more realistic time-dependent model for the random 
velocity field \cite{Tran99}. The case of $A+A\to 0$ also in a
time-dependent random velocity field, but generated now by a
stochastically forced Navier-Stokes equation, was considered in
Ref.~\cite{Hnatich00}.

\

\noindent
$\bullet$ L\'evy flights in reactive systems:

\noindent
Replacing diffusive propagation with long-ranged L\'evy flights
constitutes another important modification to the dynamics of reactive
systems. Such L\'evy flights are characterized by a probability for a 
particle's jump length $\ell$ decaying for large $\ell$ as $P \sim 
\ell^{-d-\sigma}$. For $\sigma < 2$ this results in a mean-square 
displacement in one dimension growing as $t^{1/\sigma}$, faster than 
the $t^{1/2}$ law of diffusion. Naturally, one expects that altering 
the dynamics of the system in this way will modify the kinetics, as 
one is effectively making the system better mixed with decreasing 
$\sigma$. The propagator for L\'evy processes becomes 
$G_0({\bf p},\omega) = (-i\omega + D_Lp^\sigma)^{-1}$, meaning that 
time scales acquire scaling dimension $\kappa^{-\sigma}$ rather than  
$\kappa^{-2}$ (for $\sigma < 2$). Consequently, power counting for the
$A+A\to (0,A)$ reaction gives $[\lambda] = \kappa^{\sigma - d}$, 
implying that $d_c = \sigma$. Once again only the reaction rate is 
renormalized, which then flows to an $O(\epsilon=\sigma-d$) fixed 
point under the RG. Dimensional analysis subseqently fixes the 
asymptotic density decay rate as $t^{-d/\sigma}$, for $d < \sigma$ 
\cite{Hinrichsen99}.

Note that the upper critical dimension is now a function of the L\'evy
index $\sigma$. This feature has been exploited by Vernon 
\cite{Vernon03} to compute the density amplitude for the $A+A \to 0$ 
reaction with L\'evy flights to first order in $\epsilon = \sigma-d$. 
$\sigma$ was then set to be slightly larger than unity and the 
behavior of the system was studied numerically in $d = 1$. This 
ensures that $\epsilon = \sigma - d$ is a genuinely small expansion 
parameter (i.e., $\epsilon \ll 1$) in the physical dimension $d = 1$. 
This contrasts with the case of $A+A\to 0$ with standard diffusion 
where, in order to access $d = 1$, $\epsilon = 2-d$ must be set to 
unity. As we have seen, in that situation the expansion for the 
density amplitude agrees only rather poorly with numerics 
\cite{Lee94a}. However, for the L\'evy flight case, Vernon 
demonstrated that the accuracy of the expansion indeed improves with 
decreasing $\epsilon$ (i.e. decreasing $\sigma$ towards unity). This 
ability to vary the value of $d_c$ has also been used to probe the 
behavior of directed percolation and branching-annihilating random 
walks \cite{Hinrichsen99,Janssen99,Vernon01}, see 
\sref{sec:transitions} below. We also mention that the reaction 
$A+A \to 0$ with L\'evy flights and quenched disorder was studied
using RG methods in Ref.~\cite{Chen02}. Finally, the authors of 
Ref.~\cite{Park99} used RG techniques to investigate the case of 
short-ranged {\em diffusion}, but with long-ranged {\em reactive} 
interactions.

\subsection{Two-species reactions}
\label{subsec:twospecies}

$\bullet$ The homogeneous $A + B \to 0$ reaction:

\noindent
The two-species decay reaction is perhaps the most relevant to 
chemical systems. It is also considerably more complicated to analyze, 
since the $A+B \to 0$ pair annihilation process leaves the local 
density difference field $a-b$ unchanged. This conservation law 
provides a slow mode in the dynamics that is crucial in determining 
the long-time behavior of the system. We consider first the case where
the $A$ and $B$ particles are initially mixed together throughout the 
system. If their initial densities are unequal, say $b_0 > a_0$, the 
asymptotic dynamics will approach a steady concentration of $b_0 - 
a_0$ of $B$ particles, with very few isolated $A$ particles suriving. 
In this situation, exact results indicate an exponentially decaying 
$A$ particle density for $d > 2$, logarithmic corrections to an 
exponential in $d = 2$, and a stretched exponential $\exp(-c\sqrt{t})$
form for $d = 1$ \cite{Bramson88,Bray02,Blythe03}, where $c$ is a 
constant. As briefly discussed in \sref{subsec:CSequation} above, 
these results correspond in the RG framework to the standard 
renormalization of the reaction rate.

In contrast, when starting from equal initial densities, the
fluctuations in the initial conditions for the difference field $a-b$
decay to zero slowly, by diffusion. This case was studied by Toussaint
and Wilczek \cite{Toussaint83} based on the idea that after a time
$t$, on length scales shorter than the diffusion length $l_d \sim
t^{1/2}$ only whichever of the species happened to be in the majority
in that region initially will remain. In other words, the two species
asymptotically segregate.  Since the initial difference between the
$A$ and $B$ particle numbers in that region is proportional to
$l_D^{d/2}$, this leads to an asymptotic $t^{-d/4}$ decay
\cite{Toussaint83}. Clearly, for $d < 4$, this dominates the faster
$t^{-1}$ mean-field density decay which assumes well-mixed reactants
throughout the system's temporal evolution. Toussaint and Wilczek
explicitly calculated the amplitude for this decay under the
assumption that the only relevant fluctuations are those in the
initial conditions. These results were since confirmed by exact
methods \cite{Bramson88,Bramson91b, Bramson91a}.

Turning to the field-theoretic RG approach, the action 
\eref{eq:ab_action} for the process $A+B \to 0$ contains diffusive
propagators for both $A$ and $B$ species, possibly with unequal
diffusion constants, together with the interaction vertices $\lambda
\bar a a b$, $\lambda \bar b a b$ and $\lambda\bar a \bar b a b$. 
Power counting reveals $[\lambda] = \kappa^{2-d}$, the same as in the 
$A+A \to 0$ reaction. This implies that the upper critical dimension 
is $d_c = 2$, consistent with the behavior for unequal initial
densities. The renormalization of the $A+B \to 0$ action also follows
similarly to the $A+A \to 0$ case. Surprisingly, however, a full RG
calculation of the asymptotic density in the case of unequal initial
densities has not yet been fully carried through. For the equal 
density case, though, a field theory approach by Lee and Cardy is 
available \cite{Lee95}. The Toussaint-Wilczek analysis reveals a 
qualitative change in the system's behavior in four dimensions, 
whereas the field theory yields $d_c = 2$. The resolution of this 
issue lies in the derivation of an effective theory valid for 
$2 < d \leq 4$, where one must allow for the generation of effective 
initial ($t=0$) `surface' terms, incorporating the fluctuations of the
initial state. Aside from this initial fluctuation term, it was shown 
that the mean-field rate equations suffice \cite{Lee95}. Using the 
field theory approach, Lee and Cardy were also able to demonstrate the 
asymptotic segregation of the $A$ and $B$ species, and thus provided a 
more rigorous justification of the Toussaint-Wilczek result for both 
the $t^{-d/4}$ density decay and amplitude for $2 < d < 4$. For 
$d \leq d_c = 2$, a full RG calculation becomes necessary. Remarkably, 
comparisons with exact results for the decay exponent in one dimension 
\cite{Bramson88,Bramson91b,Bramson91a} show that this qualitative 
change in the system does not lead to any modification in the form of 
the asymptotic density decay exponent at $d = 2$ (and so very unlike 
the case of {\em unequal} initial denities). However, actually 
demonstrating this using field theory methods has not yet been 
accomplished, since this would involve a very difficult 
non-perturbative sum over the initial `surface' terms.

Lastly, we also mention related work by Sasamoto and coworkers
\cite{Sasamoto97} where the $m A + n B \to 0$ reaction was studied 
using field-theoretic techniques, by methods similar to those of 
Ref.~\cite{Lee95}. These authors also found a $t^{-d/4}$ decay rate 
independent of $m$ and $n$ (provided both are nonzero), valid for 
$d < 4 / (m+n-1)$.

\

\noindent
$\bullet$ The segregated $A + B \to 0$ reaction, reaction zones:

\noindent
Two-species reactions can also be studied starting from an initial
condition of a segregated state, where a ($d-1$)-dimensional surface
separates the two species at time $t=0$. Later, as the particles have
an opportunity to diffuse into the interface, a reaction zone forms.
G\'alfi and Racz first studied these reaction zones within the local
mean-field equations, and were able to extract some rich scaling
behavior: the width of the reaction region grows as $w\sim t^{1/6}$,
the width of the depletion region grows, as might be expected, as
$t^{1/2}$, and the particle densities in the reaction zone scale as
$t^{-1/3}$ \cite{Galfi88}.

Redner and Ben-Naim \cite{BenNaim92} proposed a variation of this 
model where equal and opposite currents of $A$ and $B$ particles are
directed towards one another and a steady-state reaction zone is
formed. In this case it is of interest to study how the various
lengths scale with the particle current $J$. Within the local 
mean-field equations they found that the width of the reaction region
grows as $w\sim J^{-1/3}$, whereas the particle densities in the
reaction zone scale as $J^{2/3}$. The above initially segregated
system may be directly related to this steady-state case by observing
that, in the former, the depletion region is asymptotically much 
larger than the reaction zone itself. This means there is a 
significant region where the density evolves only by diffusion, and 
goes from a constant to zero over a range $L \sim t^{1/2}$. Since 
$J \sim -\nabla a$, we find $J \sim t^{-1/2}$, which may be used to 
translate results between these two cases.

Cornell and Droz \cite{Cornell93} extended the analysis of the
steady-state problem beyond the mean field equations and, with RG
motivated arguments, conjectured a reaction zone width $w \sim
J^{-1/(d+1)}$ for the case $d < 2$. Lee and Cardy confirmed this 
result using RG methods \cite{Lee94b}. The essential physics here is 
that the only dimensional parameters entering the problem are the 
reaction rate and the current $J$. However, for $d \leq 2$, RG methods 
demonstrate that the asymptotics are independent of the reaction rate. 
In that case, dimensional analysis fixes the above scaling form (with
logarithmic corrections in $d=2$ \cite{Lee95}). Howard and Cardy
\cite{Howard95} provided explicit calculations for the scaling
functions. However, numerical investigations of the exponent of the
reaction zone width revealed a surprisingly slow convergence to its 
predicted value $w\sim J^{-1/2}\sim t^{1/4}$ in $d = 1$ 
\cite{Cornell95}. The resolution of this issue was provided in
Ref.~\cite{Barkema96}, where the noise-induced wandering of the front
was considered (in contrast to the intrinsic front profile analyzed
previously). There it was shown that this noise-induced wandering
dominates over the intrinsic front width and generates a multiplicative
logarithmic correction to the basic $w\sim J^{-1/2}\sim t^{1/4}$
scaling in $d = 1$.

Remarkably, one can also study the reaction zones in the initially
mixed system with equal initial densities, since it asymptotically 
segregates for $d<4$ and spontaneously forms reaction zones. As shown
by Lee and Cardy \cite{Lee94b}, if one assumes that in the depletion
regions, where only diffusion occurs, the density goes from the bulk
value $t^{-d/4}$ to zero in a distance of order $t^{1/2}$, the current
scales as $J \sim t^{-(d+2)/4}$. From this the scaling of the reaction 
zone width with time immediately follows. As $d \to 4$ from below, the 
reaction zone width approaches $t^{1/2}$, i.e., the reaction zone size 
becomes comparable to the depletion zone, consistent with the 
breakdown of segregation. This analysis also reveals the true critical 
dimension $d_c = 2$, with logarithmic corrections 
$w \sim (t \ln t)^{1/3}$ arising from the marginal coupling in $d = 2$ 
\cite{Lee95}.  

\

\noindent
$\bullet$ Inhomogeneous reactions, shear flow and disorder: 

\noindent 
One important variant of the $A + B \to 0$ reaction, first analyzed by
Howard and Barkema \cite{Howard96b}, concerns its behavior in the
linear shear flow ${\bf v}=v_0 y {\bf \hat x}$, where ${\bf \hat x}$
is a unit vector in the $x$-direction. Since the shear flow tends to
enhance the mixing of the reactants, we expect that the reaction
kinetics will differ from the homogeneous case. A simple
generalization of the qualitative arguments of Toussaint and Wilczek
shows that this is indeed the case. The presence of the shear flow
means that in volumes smaller than $(Dt)^{d/2} [1+(v_0 t)^2/3]^{1/2}$
only the species which was initially in the majority of that region
will remain.  Hence, we immediately identify a crossover time $t_c
\sim v_0^{-1}$.  For $t \ll t_c$ the shear flow is unimportant and the
usual $t^{-d/4}$ density decay is preserved. However for $t \gg t_c$,
we find a $t^{-(d+2)/4}$ decay holding in $d < 2$. Since $d = 2$ is
clearly the lowest possible dimension for such a shear flow, we see
that the shear has essentially eliminated the non-classical
kinetics. These arguments can be put on a more concrete basis by a
field-theoretic RG analysis \cite{Howard96b}, which shows the shear
flow adds terms of the form $\bar a v_0 y \partial_x a$ and $\bar b
v_0 y \partial_x b$ to the action. The effect of these contributions
can then be incorporated into modified propagators, after which the
analysis proceeds similarly to the homogeneous case \cite{Lee95}.

The related, but somewhat more complex example of $A+B\to 0$ in a
quenched random velocity field was considered by Oerding 
\cite{Oerding96}. In this case it was assumed that the velocity at
every point ${\bf r}=(x,{\bf y})$ of a $d$-dimensional system was
either parallel or antiparallel to the $x$ axis and depended only on
the coordinate perpendicular to the flow. The velocity field was
modeled by quenched Gaussian random variables with zero mean, but
with correlator $\langle v({\bf y}) v({\bf y'}) \rangle = f_0 
\delta({\bf y}-{\bf y'})$. In this situation, qualitative arguments 
again determine the density decay exponent. Below three dimensions, a 
random walk in this random velocity field shows superdiffusive 
behavior in the $x$-direction \cite{Honkonen91}. The mean-square 
displacement in the $x$ direction averaged over configurations of 
$v({\bf y})$ is $\langle x^2 \rangle \sim t^{(5-d)/2}$ for $d < 3$. 
Generalizing the Toussaint-Wilczek argument then gives an asymptotic 
density decay of $t^{-(d+3)/8}$. In this case, the system still 
segregates into $A$ and $B$ rich regions, albeit with a modified decay 
exponent for $d<3$. However, to proceed beyond this result, Oerding 
applied RG methods to confirm the decay exponent and also to compute 
the amplitude of the density decay to first order in $\epsilon = 3-d$ 
\cite{Oerding96}. The analysis proceeds along the same lines as the 
homogeneous case \cite{Lee95}, particularly in the derivation of 
effective `initial' interaction terms, although care must also be 
taken to incorporate the effects of the random velocity field, which 
include a renormalization of the diffusion constant. Lastly, we 
mention work by Deem and Park, who analyzed the properties of the 
$A + B \to 0$ reaction using RG methods in the case of long-ranged 
potential disorder \cite{Deem98a}, and in a model of turbulent flow 
\cite{Deem98b}.

\

\noindent
$\bullet$ Reversible reactions, approach to equilibrium:

\noindent
Rey and Cardy \cite{Rey99} studied the reversible reaction-diffusion 
systems $A+A \rightleftharpoons C$ and $A+B \rightleftharpoons C$ 
using RG techniques. Unlike the case of critical dynamics in 
equilibrium systems, the authors found that no new nontrivial 
exponents were involved. By exploiting the existence of conserved 
quantities in the dynamics, they found that, starting from random 
initial conditions, the approach of the $C$ species to its equilibrium
density takes the form $A t^{-d/2}$ in both cases and in all 
dimensions. The exponent follows directly from the conservation laws 
and is universal, whereas the amplitude $A$ turns out to be 
model-dependent. Rey and Cardy also considered the cases of correlated 
initial conditions and unequal diffusion constants, which exhibit more 
complicated behavior, including a nonmonotonic approach to 
equilibrium.

\subsection{Coupled reactions without active phase}
\label{subsec:coupled}

\noindent
The mixed reaction-diffusion system $A + A \to 0$, $A + B \to 0$, 
$B + B \to 0$ was first studied using field-theoretic RG methods by 
Howard \cite{Howard96a}, motivated by the study of persistence 
probabilities (see \sref{subsec:persistence}). The renormalization of 
the theory proceeds again similar to the case of $A + A \to 0$: only 
the reaction rates need to be renormalized, and this can be performed 
to all orders in perturbation theory. For $d \leq d_c = 2$, 
perturbative calculations for the density decay rates were only 
possible in the limit where the density of one species was very much 
greater than that of the other. The density decay exponent of the 
majority species then follows the standard pure annihilation kinetics, 
whereas the minority species decay exponent was computed to 
$O(\epsilon=2-d)$ \cite{Howard96a}. This one-loop exponent turned out 
to be a complicated function of the ratio of the $A$ and $B$ species 
diffusion constants. The calculation of this exponent using RG methods 
has been confirmed and also slightly generalized in 
Ref.~\cite{Rajesh04}. The above mixed reaction-diffusion system also 
provides a good testing ground in which to compare RG methods with the
Smoluchowski approximation, which had earlier been applied to the same 
multi-species reaction-diffusion system \cite{Krapivsky94}. This is a 
revealing comparison as the value of the minority species decay 
exponent is non-trivial for $d \leq 2$, and is no longer fixed purely 
by dimensional analysis (as is the case for the pure annihilation 
exponent for $d < 2$). This difference follows from the existence of 
an additional dimensionless parameter in the multi-species problem, 
namely the ratio of diffusion constants. Nevertheless, in this case, 
it turns out that the Smoluchowski approximation decay exponent is 
identical to the RG-improved tree level result, and provides rather a 
good approximation in $d = 1$ \cite{Krapivsky94,Howard96a}. However, 
this is not always the case for other similar multi-species 
reaction-diffusion models, where the Smoluchowski approximation can 
become quite inaccurate (see Refs.~\cite{Howard96a,Rajesh04} for more 
details).

The same system but with equal diffusion constants was also analyzed 
using RG methods in Ref.~\cite{Konkoli99}, as a model for a steric 
reaction-diffusion system. As pointed out in Refs.~\cite{Howard96a,
Konkoli00}, this model has the interesting property that at large 
times for $d \leq d_c = 2$, the densities of both species always decay
at the same rate, contrary to the predictions of mean-field theory. 
This result follows from the indistinguishability of the two species 
at large times: below the upper critical dimension, the reaction rates 
run to identical fixed points. Since the diffusion constants are also 
equal there is then no way to asymptotically distinguish between the 
two species, whose densities must therefore decay at the same rate. 
The same set of reactions, with equal diffusion constants, was used to 
study the application of Bogolyubov's theory of weakly nonideal Bose 
gases to reaction-diffusion systems \cite{Konkoli03}.

Related models were studied in the context of the mass distribution of
systems of aggregating and diffusing particles \cite{Zaboronski01,
Krishnamurthy02}. In the appropriate limit, the system of 
Ref.~\cite{Krishnamurthy02} reduced to the reactions $A + A \to A$ and
$A + B \to 0$. Progress could then be made in computing to 
$O(\epsilon = 2-d$) the form of the large-time average mass 
distribution, for small masses. Comparisons were also made to 
Smoluchowski-type approximations, which failed to capture an important
feature of the distribution, namely its peculiar form at small masses,
referred to by the authors of Ref.~\cite{Krishnamurthy02} as the
Kang-Redner anomaly. This failure could be traced back to an anomalous
dimension of the initial mass distribution, a feature which, as
discussed in \sref{subsec:relation}, cannot be picked up by
Smoluchowski-type approximations. Howard and T\"auber investigated the
mixed annihilation / `scattering' reactions $A + A \to 0$, 
$A + A \to B + B$, $B + B \to A + A$, and $B + B \to 0$ 
\cite{Howard97}. In this case, for $d < 2$, to all orders in 
perturbation theory, the system reduces to the single-species 
annihilation case. Physically this is again due to the re-entrance 
property of random walks: as soon as two particles of the same species 
approach each other, they will rapidly annihilate regardless of the 
competing `scattering' processes, which only produce particle pairs in 
close proximity and therefore with a large probability of immediate 
subsequent annihilation.

Finally, we mention the multi-species pair annihilation reactions
$A_i + A_j \to 0$ with $1 \leq i < j \leq q$, first studied by 
Ben-Avraham and Redner \cite{Redner86}, and more recently by 
Deloubri\`ere and coworkers \cite{Deloubriere02,Hilhorst04,
Hilhorst04a}. For unequal initial densities or different reaction 
rates between the species, one generically expects the same scaling as 
for $A + B \to 0$ asymptotically (when only the two most numerous, or 
least reactive, species remain). An interesting special case therefore 
emerges when all rates and initial densities coincide. For any $q > 2$ 
and in dimensions $d \geq 2$ it was argued that particle species 
segregation cannot occur, and hence that the asymptotic density decay 
rate for equal initial densities and annihilation rates should be the 
same as for the single-species reaction $A + A \to 0$. In one 
dimension, however, particle segregation does take place for all 
$q < \infty$, and leads to a $q$-dependent power law 
$\sim t^{-(q-1)/2q}$ for the total density \cite{Deloubriere02,
Hilhorst04,Zhong03}. For $q=2$, this recovers the two-species decay 
$\sim t^{-1/4}$, whereas the single-species behavior $\sim t^{-1/2}$ 
ensues in the limit $q \to \infty$ (since the probability that a given 
particle belongs to a given species vanishes in this limit, any 
species distinction indeed becomes meaningless). Other special 
situations arise when the reaction rates are chosen such that certain 
subsets of the $A_i$ are equivalent under a symmetry operation. One 
may construct scenarios where segregation occurs in dimensions $d > 2$ 
despite the absence of any microscopic conservation law 
\cite{Hilhorst04a}.

A variation on this model has a finite number of walkers $N_i$ of each
species $A_i$, initially distributed within a finite range of the
origin. Attention is focused on the asymptotic decay of the 
probability that no reactions have occured up to time $t$. The case of 
$N_i = 1$ for all $i$ reduces to Fisher's vicious walkers 
\cite{Fisher84}, and the case $N_1=1$ and $N_2 = n$ reduces to 
Krapivsky and Redner's lion-lamb model \cite{Krapivsky96}. Applying RG 
methods to the general case, including unequal diffusion constants for 
the different species, Cardy and Katori demonstrated that the 
probability decays as $t^{-\alpha(\{N_i\})}$ for $d<2$, and calculated
the exponent to second order in an $\epsilon = 2-d$ expansion
\cite{Cardy03}.

\subsection{Persistence}
\label{subsec:persistence}

Persistence, in its simplest form, refers to the probability that a
particular event has never occurred in the entire history of an
evolving statistical system \cite{Derrida94}. Persistence
probabilities are often universal and have been found to be nontrivial
even in otherwise well-understood systems. An intensively studied
example concerns the zero-temperature relaxational dynamics of the
Ising model, where one is interested in the persistence probability
that, starting from random initial conditions, a given site has never
been visited by a domain wall. In one dimension, the motion and 
annihilation of Ising domain walls at zero temperature is equivalent 
to an $A+A \to 0$ reaction-diffusion system, where the domain walls 
in the Ising system correspond to the reacting particles. An exact 
solution exists for the persistence probability in this case
\cite{Derrida95}, but, as usual, the solution casts little light on 
the question of universality. A different approach was proposed by 
Cardy who studied, in the framework of the reaction-diffusion model, 
the proportion of sites never visited by any particle \cite{Cardy95}. 
In $d = 1$ (though not in higher dimensions) this is the same quantity 
as the original persistence probability. Furthermore, since Cardy was
able to employ the kind of field-theoretic RG methods discussed in 
this review, the issue of universality could be addressed as well.

Cardy demonstrated that the probability of never finding a particle at
the origin could be calculated within the field-theoretic formalism 
through the inclusion of an operator product $\prod_t 
\delta_{\hat a_0^\dagger\hat a_0,0}$. The subscript denotes that the 
$\hat a_0^\dagger\hat a_0$ operators are associated with the origin, 
and the operator-valued Kronecker $\delta$-function ensures that zero 
weight is assigned to any histories with a particle at the origin. 
This operator has the net effect of adding a term $-h \int_0^t 
\bar\phi(0,t') \phi(0,t') dt'$ to the action, and the persistence 
probability then corresponds to the expectation value $\langle 
\exp( -h \int_0^t \phi(0,t')dt') \rangle$, averaged with respect to 
the modified action. Power counting reveals that $[h] \sim 
\kappa^{2-d}$, so this coupling is relevant for $d < 2$. Cardy showed 
that renormalization of this interaction required both a renormalized 
coupling $h_R$ and a multiplicative renormalization of the field 
$\phi(0,t)$. This results in a controlled $\epsilon = 2-d$ expansion 
for the universal persistence exponent $\theta = 1/2 + O(\epsilon)$ 
\cite{Cardy95}. This compares to the exact result in one dimension by 
Derrida et al., namely $\theta = 3/8$ \cite{Derrida95}. An alternative 
approach to this problem was given by Howard \cite{Howard96a} in the 
mixed two-species reaction $A + A \to 0$, $A + B \to 0$, with immobile 
$B$ particles (see also \sref{subsec:coupled}). In this case 
the persistence probability corresponds to the density decay of 
immobile $B$ particles in $d = 1$, in the limit where their density is 
much smaller than those of the $A$ particles. Howard's expansion 
confirmed the results of Cardy and also extended the computation of 
the persistence probability to $O(\epsilon = 2-d)$. The case of 
persistence in a system of random walkers which either coagulate, with 
probability $(q-2)/(q-1)$, or annihilate, with probability $1/(q-1)$, 
when they meet was also investigated using RG methods by Krishnamurthy 
et al. \cite{Krishnamurthy03}. In one dimension, this system models 
the zero-temperature Glauber dynamics of domain walls in the $q$-state 
Potts model. Krishnamurthy et al. were able to compute the probability 
that a given particle has never encountered another up to order 
$\epsilon = 2-d$.

A further application of field-theoretic methods to persistence
probabilities was introduced by Howard and Godr\`eche \cite{Howard98}
in their treatment of persistence in the voter model. The dynamics of
the voter model consist of choosing a site at random between $t$ and
$t+dt$; the `voter' on that site, which can have any of $q$ possible
`opinions', then takes the opinion of one its $2d$ neighbours, also
chosen at random. This model in $d=1$ is identical to the 
Glauber-Potts model at zero temperature, but can also, in all 
dimensions, be analyzed using a system of coalescing random walkers. 
This again opens up the possiblity for field-theoretic RG 
calculations, as performed in Ref.~\cite{Howard98}. The persistence 
probability that a given `voter' has never changed its opinion up to 
time $t$ was computed for all $d \geq 2$, yielding an unusual 
$\exp[-f(q) (\ln t)^2]$ decay in two dimensions. This result confirmed 
earlier numerical work by Ben-Naim et al. \cite{BenNaim96}.

\section{Active to Absorbing State Transitions}
\label{sec:transitions}

In the previous sections, we have focused on the non-trivial algebraic 
decay towards the absorbing state in diffusion-limited reactions of 
the type $k A \to \ell A$ (with $k \geq 2$ and $\ell < k$), and some 
variants thereof. Universal behavior naturally emerges also near a 
continuous nonequilibrium phase transition that separates an active 
state, with non-vanishing particle density as $t \to \infty$, from an 
inactive, absorbing state. We shall see that {\em generically}, such 
phase transitions are governed by the power laws of the directed 
percolation (DP) universality class \cite{Janssen81,Grassberger82,
Hinrichsen00,Janssen05}.

\subsection{The directed percolation (DP) universality class}
\label{subsec:DP}

A phase transition separating active from inactive states is readily
found when spontaneous particle decay ($A \to 0$, with rate $\mu$) 
competes with the production process ($A \to A+A$, branching rate 
$\sigma$). In this linear reaction system, $a(t) = a(0) \exp [-
(\mu - \sigma) t] \to 0$ exponentially if $\sigma < \mu$. In order
to render the particle density $a$ finite in the active state, i.e.,
for $\sigma > \mu$, we need to either restrict the particle number
per lattice site (say, to $0$ or $1$), or add a binary reaction  
$A+A \to (0,A)$, with rates $\lambda (\lambda')$. The corresponding
mean-field rate equation reads
\begin{equation}
\label{eq:DPrate_eq}
     \partial_t a(t) = (\sigma - \mu) a(t) - (2 \lambda + \lambda') 
     \, a(t)^2 ~ ,
\end{equation}
which for $\sigma > \mu$ implies that asymptotically
\begin{equation}
\label{eq:mfbeta}
     a(t) \to a_\infty = \frac{\sigma - \mu}{2\lambda + \lambda'} ~ ,
\end{equation}
which is approached exponentially $|a(t) - a_\infty| \sim \exp [-
(\sigma - \mu) t]$ as $t \to \infty$. Precisely at the transition 
$\sigma = \mu$, Eq.~\eref{eq:DPrate_eq} yields the binary annihilation 
/ coagulation mean-field power law decay $a(t) \sim t^{-1}$. 
Generalizing Eq.~\eref{eq:DPrate_eq} to a local particle density and 
taking into account diffusive propagation, we obtain with $r = (\mu - 
\sigma) / D$:
\begin{equation}
\label{eq:DPdiff_eq}
     \partial_t a({\bf x},t) = - D (r - \nabla^2) a({\bf x},t) - 
     (2 \lambda + \lambda') \, a({\bf x},t)^2 ~ ,
\end{equation}
wherefrom we infer the characteristic length and diffusive time scales 
$\xi \sim |r|^{-1/2}$ and $t_c \sim \xi^2 / D \sim |r|^{-1}$ which both
diverge upon approaching the critical point at $r = 0$. Upon defining
the critical exponents
\begin{eqnarray}
     &\langle a_\infty \rangle \sim (-r)^\beta \quad (r < 0) ~ , \qquad
     &\langle a(t) \rangle \sim t^{-\alpha} \quad (r = 0) ~ ,\nonumber\\
     &\xi \sim |r|^{-\nu} \quad (r \not= 0) ~ , \qquad
     &t_c \sim \xi^z / D \sim |r|^{- z \nu} \quad (r \not= 0) ~ ,  
\label{eq:crit_exp}
\end{eqnarray}
we identify the mean-field values $\beta = 1$, $\alpha = 1$, 
$\nu = 1/2$, and $z = 2$.

In order to properly account for fluctuations near the transition, we
apply the field theory mapping explained in \sref{sec:mapping}. The
ensuing coherent-state path integral action then reads
\begin{eqnarray}
     \fl \quad S[\tilde\phi,\phi] = \int d^dx \biggl\{ -\phi(t_f) + 
     \int_0^{t_f} dt &&\left[ \tilde\phi \left( \partial_t - D \nabla^2
     \right) \phi - \mu (1 - \tilde\phi) \phi + \sigma (1 - \tilde\phi)
     \tilde\phi \phi \right. \nonumber \\
     &&\left. - \lambda \left( 1 - \tilde\phi^2 \right) \phi^2 - 
     \lambda' \left( 1 - \tilde\phi \right) \tilde\phi \phi^2 \right] 
     - n_0 \tilde\phi(0) \biggr\} ,
\label{eq:DPunshact}
\end{eqnarray} 
which constitutes a {\em microscopic} representation of the stochastic
processes in question. Equivalently, we may consider the shifted action
(with $\tilde\phi = 1 + \bar\phi$)
\begin{equation}
\label{eq:DPshact}
     \fl S[\bar\phi,\phi] = \int \! d^dx \! \int \! dt \left\{ 
     \bar\phi \left[ \partial_t + D (r - \nabla^2) \right] \phi 
     - \sigma \bar\phi^2 \phi + (2\lambda + \lambda') \bar\phi \phi^2 
     + (\lambda + \lambda') \bar\phi^2 \phi^2 \right\} .
\end{equation} 
Since the ongoing particle production and decay processes should 
quickly obliterate any remnants from the initial state, we have dropped
the term $n_0 \bar\phi(0)$, and extended the temporal integral from 
$-\infty$ to $\infty$. The classical field equations $\delta S / 
\delta \phi = 0$ (always solved by $\bar\phi = 0$) and $\delta S / 
\delta \bar\phi = 0$ yield the mean-field equation of motion 
\eref{eq:DPdiff_eq}. 

\begin{figure}[htb]
\begin{center}
\includegraphics[width=2.25in]{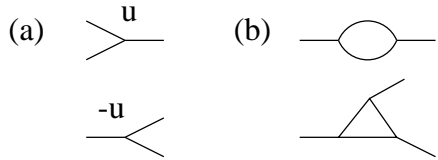}
\caption{DP field theory: (a) vertices, and (b) one-loop Feynman graphs
for the two- and three-point vertex functions.}
\label{fig:dploops}
\end{center}
\end{figure}

Our goal is to construct an appropriate {\em mesoscopic} field theory 
that captures the universal properties at the phase transition. Recall 
that the continuum limit is not unique: We are at liberty to choose the 
scaling dimensions of the fluctuating fields $\bar\phi({\bf x},t)$ and 
$\phi({\bf x},t)$, provided we maintain that their product scales as a 
density, i.e., $[\bar\phi \phi] = \kappa^d$ with arbitrary momentum 
scale $\kappa$. In RG terms, there exists a {\em redundant} parameter 
\cite{Wegner74} that needs to be eliminated through suitable rescaling. 
To this end, we note that the scaling properties are encoded in the 
propagator $G({\bf x},t) = \langle \bar\phi({\bf x},t) \phi(0,0) 
\rangle$. The lowest-order fluctuation correction to the tree-level 
expression
\begin{equation}
\label{eq:dpprop}
     G_0({\bf p},\omega) = {1 \over -i\omega + D (r + p^2)} 
\end{equation}
is given by the Feynman graph depicted in \fref{fig:dploops}(b, top),
which involves the product $\sim - \sigma (2 \lambda + \lambda')$ of 
the two three-point vertices in \eref{eq:DPshact}. Similarly, the
one-loop correction to either of these vertices comes with the very
same factor. It is thus convenient to choose the scaling dimensions of
the fields in such a manner that the three-point vertices attain 
identical scaling dimensions. This is achieved via introducing new
fields $\bar s({\bf x},t) = \bar\phi({\bf x},t) \sqrt{(2 \lambda + 
\lambda') / \sigma}$ and $s({\bf x},t) = \phi({\bf x},t) 
\sqrt{\sigma / (2 \lambda + \lambda')}$, whence
\begin{equation}
\label{eq:DPsact}
     \fl \qquad S[\bar s,s] = \int \! d^dx \! \int \! dt \left\{ \bar s
     \left[ \partial_t + D (r - \nabla^2) \right] s - u (\bar s - s) 
     \bar s s + (\lambda + \lambda') \bar s^2 s^2 \right\} .
\end{equation} 
Here, $u = \sqrt{\sigma (2 \lambda + \lambda')}$ is the new effective 
coupling. Since $[\sigma] = \kappa^2$ and $[\lambda] = \kappa^{2-d} = 
[\lambda']$, its scaling dimension is $[u] = \kappa^{2-d/2}$, and we 
therefore expect $d_c = 4$ to be the upper critical dimension. 
Moreover, $[(\lambda + \lambda') / u] = \kappa^{-d/2}$ scales to zero 
under subsequent RG transformations: compared to $u$, both couplings 
$\lambda$ and $\lambda'$ alone constitute irrelevant parameters which 
will not affect the leading universal scaling properties. 

Upon omitting these irrelevant terms, we finally arrive at the desired 
{\em effective} field theory action
\begin{equation}
\label{eq:DPeffact}
     S_{\rm eff}[\bar s,s] = \int \! d^dx \! \int \! dt \left\{ 
     \bar s \left[ \partial_t + D (r - \nabla^2) \right] s 
     - u (\bar s - s) \bar s s \right\} .
\end{equation}
It displays duality invariance with respect to time (rapidity) 
inversion, $s({\bf x},t) \leftrightarrow - \bar s({\bf x},-t)$.
Remarkably, the action \eref{eq:DPeffact} was first encountered and
analyzed in particle physics under the guise of {\em Reggeon field
theory} \cite{Gribov67,Moshe78}. It was subsequently noticed that it 
actually represents a stochastic (`Gribov') process 
\cite{Grassberger78,Grassberger79}, and its equivalence to the 
geometric problem of directed percolation was established 
\cite{Obukhov80,Cardy80,Janssen81}. In directed bond percolation, 
randomly placed bonds connecting regular lattice sites can only be 
traversed in a given preferred special direction, which is to be 
identified with $t$ in the dynamical problem. Particle decay, 
coagulation, and production respectively correspond to dead ends, 
merging links, or branching of the ensuing percolating structures. 
Near the percolation threshold, the scaling properties of the critical
percolation cluster are characterized by the exponents governing the
divergences of the transverse correlation length $\nu_\perp = \nu$ and
of the longitudinal (in the $t$ direction) correlation length 
$\nu_\parallel = z \nu$. (For more details, we refer the reader to
Refs.~\cite{Hinrichsen00,Janssen05}.)

From our derivation of the effective action \eref{eq:DPeffact} above,
it is already apparent that either pair annihilation or coagulation 
lead to identical critical properties. Instead of these binary 
reactions, we could also have employed site oocupation number 
restrictions to render the particle density finite in the active phase.
Van Wijland has recently shown how such local constraints limiting 
$n_i$ to values of $0, 1$ only can be implemented into the 
second-quantized bosonic formalism \cite{Wijland01}, thus avoiding a 
more cumbersome representation in terms of spin operators. The 
resulting action acquires exponential terms for each field 
$\tilde\phi$. For the competing first-order processes $A \to (0, 2A)$
one eventually obtains
\begin{equation}
\label{eq:DPrestact}
     S_{\rm rest}[\tilde\phi,\phi] = \int d^dx \int dt \left[ - \mu 
     (1 - \tilde\phi) \phi \, e^{-v \tilde\phi \phi} + \sigma (1 - 
     \tilde\phi) \tilde\phi \phi\, e^{-2 v \tilde\phi \phi} \right] ~ ,
\end{equation} 
where we have merely written down the bulk reaction part of the action,
and $v$ is a parameter of scaling dimension $[v] = \kappa^{-d}$ which
originates from taking the continuum limit. Since therefore $v$ will
scale to zero under RG transformations, we may expand the exponentials,
whereupon the leading terms in the corresponding shifted action assume 
the form \eref{eq:DPsact}, with $2 \lambda + \lambda' = (2 \sigma - 
\mu) v \approx \sigma v$ and $\lambda + \lambda' = 4 \sigma v$. Thus
we are again led to the effective DP field theory action 
\eref{eq:DPeffact} (despite the formally negative value for $\lambda$).

Following the procedure outlined in \sref{subsec:spde} \cite{Bausch76},
we find that the field theory action \eref{eq:DPeffact} is equivalent 
to the stochastic differential equation
\begin{equation}
\label{eq:dppde}
     \partial_t s = D \left( \nabla^2 - r \right) s - u s^2 + 
     \sqrt{2 u s} \, \eta ~ ,
\end{equation}
with $\langle \eta \rangle = 0$, $\langle \eta({\bf x},t) \, 
\eta({\bf x}',t') \rangle = \delta({\bf x}-{\bf x}') \, \delta(t-t')$,
or, upon setting $\zeta = \sqrt{2 u s} \, \eta$ in order to eliminate 
the square-root multiplicative noise, $\langle \zeta \rangle = 0$, 
$\langle \zeta({\bf x},t) \, \zeta({\bf x}',t') \rangle = 2 u 
s({\bf x},t) \delta({\bf x}-{\bf x}') \, \delta(t-t')$. We may view 
these resulting terms as representing the leading-order contributions 
in a power-law expansion of the reaction and noise correlation 
functionals $R[s] = r + u s + \ldots$ and $N[s] = u + \ldots$ with 
respect to the density $s$ of activity in 
\begin{equation}
\label{eq:genpde}
     \fl \quad \partial_t s = D \left( \nabla^2 - R[s] \right) s + 
     \zeta ~ , \quad \langle \zeta({\bf x},t) \, \zeta({\bf x}',t') 
     \rangle = 2 s N[s] \delta({\bf x}-{\bf x}') \, 
     \delta(t-t') ~ ,
\end{equation}
which represents the general Langevin description of systems displaying
active and absorbing states \cite{Janssen05}. A factor $s$ has been
factored out of both $R$ and $N$ here, since the stochastic processes
must all cease in the inactive, absorbing phase. These considerations
establish the {\em DP hypothesis}: The critical properties near an
active to absorbing state phase transition should generically be 
governed by the directed percolation scaling exponents, provided the 
stochastic process is Markovian, the order parameter decoupled from any
other slow variable, there is no quenched disorder in the rates, and no
special symmetries require that any of the lowest-order expansion
coefficients $r$ or $u$ vanish \cite{Janssen81,Grassberger82}.  There
is even a suggestion that the glass transition in supercooled liquids
might be governed by a zero-temperature fixed point, with critical exponents
in the DP universality class \cite{Whitelam04}.

\subsection{Renormalization and DP critical exponents}
\label{subsec:DPRG}

The asymptotic scaling behavior of DP can be inferred from the 
renormalized propagator $G({\bf p},\omega) = 
\Gamma^{(1,1)}({\bf p},-\omega)^{-1}$. The tree contribution is given
by \eref{eq:dpprop}; by combining the two three-point vertices in
\fref{fig:dploops}(a) one arrives at the one-loop Feynman graph 
depicted in \fref{fig:dploops}(b, top), whose corresponding analytic
expression reads in Fourier space
\begin{eqnarray}
     2 u^2 \int \frac{d^d p'}{(2 \pi)^d} \int \frac{d\omega'}{2 \pi} \
     &&\frac{1}{-i (\omega'+\omega/2) + D [ r+({\bf p}'+{\bf p}/2)^2 ]}
     \nonumber \\ &&\frac{1}
     {-i (-\omega'+\omega/2) + D [ r+(-{\bf p}'+{\bf p}/2)^2 ]} ~ ,
\label{eq:dp1loop}
\end{eqnarray}
if we split the external momentum and frequency symmetrically inside
the loop. The integration over the internal frequency $\omega'$ is now
readily performed by means of the residue theorem, whereupon we obtain
\begin{equation}
\label{eq:dpgam11}
     \fl \qquad\qquad \Gamma^{(1,1)}({\bf p},\omega) = i \omega + 
     D (r + p^2) + \frac{u^2}{D} \int \frac{d^d p'}{(2 \pi)^d} \ 
     \frac{1}{i \omega / 2 D + r + p^2/4 + {p'}^2} ~ .
\end{equation}
The loop contribution displays IR singularities as $r \to 0$, 
$\omega \to 0$, and ${\bf p} \to 0$. In the ultraviolet, it diverges in
dimensions $d \geq 2$. The leading divergence, however, can be absorbed
into a fluctuation-induced shift of the critical point away from the
mean-field $r = 0$. On physical grounds one must demand 
$G({\bf p}=0,\omega=0)^{-1} = 0$ at criticality. Consequently, the new 
critical point is given self-consistently by
\begin{equation}
\label{eq:dprshift}
     r_c = - \frac{u^2}{D^2} \int \frac{d^d p'}{(2 \pi)^d} \ 
     \frac{1}{r_c + {p}'^2} + O\left( u^4 \right) ~ .
\end{equation}
Fluctuations tend to increase the likelihood of extinction (if the 
density is already low, a chance fluctuation may drive the system into
the absorbing state), and thus reduce the parameter regime of the 
active phase as compared with mean-field theory. In {\em dimensional 
regularization}, one assigns the value
\begin{equation}
\label{eq:dimregint}
     I_s(r) = \int \frac{d^d p}{(2 \pi)^d}\ \frac{1}{(r + p^2)^s} = 
     \frac{\Gamma(s-d/2)}{2^d \pi^{d/2} \Gamma(s)} \ r^{-s+d/2} ~ ,
\end{equation}
also to those momentum integrals that are UV-divergent. The solution 
to \eref{eq:dprshift} then reads explicitly $|r_c| = [2 A_d u^2 / 
(d-2)(4-d) D^2]^{2/(4-d)}$ with $A_d = \Gamma(3-d/2) / 2^{d-1} 
\pi^{d/2}$. The shift of the transition point thus depends 
nonanalytically on $\epsilon = 4 - d$.

Let us introduce the true distance from the critical point $\tau = r - 
r_c$. Upon inserting \eref{eq:dprshift} into \eref{eq:dpgam11}, we find
\begin{equation}
\label{eq:dpsgm11}
     \fl \Gamma^{(1,1)}({\bf p},\omega) = i \omega + D (\tau + p^2) - 
     \frac{u^2}{D} \int \! \frac{d^d p'}{(2 \pi)^d} \, 
     \frac{i \omega / 2 D + \tau + p^2/4}{{p'}^2 \left( i \omega / 2 D 
     + \tau + p^2/4 + {p'}^2 \right)} + O(u^4) .
\end{equation}
The integral here is UV-divergent in dimensions $d \geq 4$. There are
{\em three} such singular terms, proportional to $i \omega$, $D \tau$,
and $D {\bf p}^2$, respectively. Consequently, we require three 
independent multiplicative renormalization factors to render the
two-point function or propagator finite. In addition, the three-point
vertex functions $\Gamma^{(1,2)}$ and $\Gamma^{(2,1)}$ carry 
(identical) UV-singularities for  $d \geq 4$. We thus define 
renormalized parameters $D_R$, $\tau_R$, and $u_R$, as well as 
renormalized fields $s_R$ according to
\begin{equation}
\label{eq:dprencon}
     \fl \qquad\qquad s_R = Z_s^{1/2} s ~ , \quad D_R = Z_D D ~, \quad 
     \tau_R = Z_\tau \tau \, \kappa^{-2} ~ , \quad
     u_R = Z_u u A_d^{1/2} \, \kappa^{(d-4)/2} ~ .
\end{equation} 
As a consequence of rapidity inversion invariance, $\tilde s_R = 
Z_s^{1/2} \tilde s$ as well, whence $\Gamma^{(1,1)}_R = Z_s^{-1} 
\Gamma^{(1,1)}$. In the {\em minimal subtraction} prescription, the $Z$
factors contain merely the $1 / \epsilon$ poles with their residues. 
Choosing the normalization point $\tau_R = 1$, $\omega = 0$, ${\bf p} 
= 0$, we may read off $Z_s$ and the products $Z_s Z_D Z_\tau$, 
$Z_s Z_D$ from the three terms on the right-hand side of 
\eref{eq:dpsgm11}, and therefrom to one-loop order
\begin{equation}
\label{eq:dpzfact}
     \fl \qquad Z_s = 1 - \frac{u^2}{2 D^2} \ 
     \frac{A_d \kappa^{-\epsilon}}{\epsilon} ~ , \quad
     Z_D = 1 + \frac{u^2}{4 D^2} \ 
     \frac{A_d \kappa^{-\epsilon}}{\epsilon} ~ , \quad
     Z_\tau = 1 - \frac{3u^2}{4 D^2} \ 
     \frac{A_d \kappa^{-\epsilon}}{\epsilon} ~ .
\end{equation}
This leaves just $Z_u$ to be determined. It is readily computed from
the three-point function $\Gamma^{(1,2)}$, whose one-loop graph is
depicted in \fref{fig:dploops}(b, bottom), or from $\Gamma^{(2,1)}$.
At the normalization point (NP),
\begin{eqnarray}
     \Gamma^{(1,2)} |_{\rm NP} = - \Gamma^{(2,1)} |_{\rm NP} &&= 
     - 2 u \left( 1 - \frac{2 u^2}{D^2} \int \frac{d^d p}{(2 \pi)^d}
     \ \frac{1}{\left( \tau + p^2 \right)^2} \right) 
     \bigg\vert_{\tau = \kappa^2} \nonumber \\ 
     &&= - 2 u \left( 1 - \frac{2 u^2}{D^2} \, 
     \frac{A_d \kappa^{-\epsilon}}{\epsilon} \right) ~ ,
\label{eq:dpgam12}
\end{eqnarray}
which directly yields the product $Z_s^{3/2} Z_u$, and with 
\eref{eq:dpzfact},
\begin{equation}
\label{eq:dpzfacu}
     Z_u = 1 - \frac{5 u^2}{4 D^2} \ 
     \frac{A_d \kappa^{-\epsilon}}{\epsilon} ~ .
\end{equation}
Since all higher vertex functions are UV-finite, this completes the
renormalization procedure for the DP field theory \eref{eq:DPeffact}.
We identify the effective coupling constant as $v = u^2 / D^2$, with
renormalized counterpart
\begin{equation}
\label{eq:dpzrenv}
     v_R = Z_v v A_d \, \kappa^{d-4} ~ , \qquad Z_v = Z_u^2 / Z_D^2 ~ .
\end{equation} 

We may now write down the DP generalization of the Callan--Symanzik 
equation \eref{eq:CSeq1} for the propagator, recalling that 
$G_R = Z_s G$:
\begin{equation}
\label{eq:DPCSeq1}
     \fl \quad \left[ \kappa \, \frac{\partial}{\partial \kappa} 
     - \zeta_s + \zeta_D D_R \, \frac{\partial}{\partial D_R} 
     + \zeta_\tau \tau_R \, \frac{\partial}{\partial \tau_R} 
     + \beta_v(v_R) \, \frac{\partial}{\partial v_R} \right] 
     G_R({\bf p},\omega,D_R,\tau_R,\kappa,v_R) = 0 ~ ,
\end{equation}
where
\begin{eqnarray}
     &&\zeta_s(v_R) = \kappa \frac{\partial}{\partial \kappa} \ln Z_s
     = \frac{v_R}{2} + O(v_R^2) ~ ,
\label{eq:DPzetas} \\
     &&\zeta_D(v_R) = \kappa \frac{\partial}{\partial \kappa} 
     \ln \frac{D_R}{D} = - \frac{v_R}{4} + O(v_R^2) ~ ,
\label{eq:DPzetad} \\
     &&\zeta_\tau(v_R) = \kappa \frac{\partial}{\partial \kappa} 
     \ln \frac{\tau_R}{\tau} = - 2 + \frac{3 v_R}{4} + O(v_R^2) ~ ,
\label{eq:DPzetat} \\
     &&\beta_v(v_R) = \kappa \frac{\partial}{\partial \kappa} v_R = 
     v_R \left[ - \epsilon + 3 v_R + O(v_R^2) \right]  ~ .
\label{eq:DPbetav}
\end{eqnarray}
In dimensions $d < d_c = 4$ ($\epsilon > 0$), the $\beta$ function 
\eref{eq:DPbetav} yields a nontrivial stable fixed point
\begin{equation}
\label{eq:DPfp}
     v_R^* = \frac{\epsilon}{3} + O(\epsilon^2) ~ .
\end{equation}
Solving \eref{eq:DPCSeq1} with the method of characteristics,
$\kappa \to \kappa \ell$, and using the form \eref{eq:dpprop},
we find in its vicinity the scaling law
\begin{equation}
\label{eq:DPscal1}
     \fl G_R({\bf p},\omega,D_R,\tau_R,\kappa,v_R)^{-1}\! \sim p^2 D_R 
     \ell^{\zeta_s(v_R^*) + \zeta_D(v_R^*)} \ \hat\Gamma\!\left( 
     \frac{\bf p}{\kappa \ell},\frac{\omega}{D_R \ell^{\zeta_D(v_R^*)}
     (\kappa \ell)^2},\tau_R \ell^{\zeta_\tau(v_R^*)},v_R^* \right)\! ,
\end{equation}
with $\hat\Gamma$ representing a dimensionless scaling function. Upon
employing the matching condition $\ell^2 = p^2 / \kappa^2$, this
yields
\begin{equation}
\label{eq:DPscal2}
     G_R({\bf p},\omega,D_R,\tau_R,\kappa,v_R^*)^{-1}\! \sim D_R 
     \kappa^2 |{\bf p}|^{2 - \eta} \ \hat\Gamma\!\left( 1,
     \frac{\omega}{D_R |{\bf p}|^z}, \tau_R |{\bf p}|^{-1/\nu},v_R^* 
     \right) \! , 
\end{equation}
with the {\em three} independent scaling exponents
\begin{eqnarray}
     &&\eta = - \zeta_s(v_R^*) - \zeta_D(v_R^*) = - \frac{\epsilon}{12}
     + O(\epsilon^2) ~ ,
\label{eq:DPeta} \\
     &&z = 2 + \zeta_D(v_R^*) = 2 - \frac{\epsilon}{12} + O(\epsilon^2)
     ~ , 
\label{eq:DPzet} \\
     &&\nu^{-1} = - \zeta_\tau(v_R^*) = 2 - \frac{\epsilon}{4} 
     + O(\epsilon^2) ~ .
\label{eq:DPnup}
\end{eqnarray}
Alternatively, with $\ell = |\tau_R|^\nu$ we arrive at 
$G_R \sim |\tau|^{- \gamma}$, where
\begin{equation}
\label{eq:DPgam}
     \gamma = \nu (2 - \eta) = 1 + \frac{\epsilon}{6} + O(\epsilon^2) ~ .
\end{equation}

In a similar manner, we obtain in the active phase for the average
$\langle s_R \rangle$:
\begin{equation}
\label{eq:DPCSeq2}
     \fl \ \left[ \kappa \frac{\partial}{\partial \kappa} 
     - \frac{\zeta_s}{2} + \zeta_D D_R \, 
     \frac{\partial}{\partial D_R} + \zeta_\tau \tau_R \, 
     \frac{\partial}{\partial \tau_R} + \beta_v(v_R) \, 
     \frac{\partial}{\partial v_R} \right] 
     \langle s_R(t,D_R,\tau_R,\kappa,v_R) \rangle = 0 ~ ,
\end{equation}
whose solution near the stable fixed point $v_R^*$ reads
\begin{equation}
\label{eq:DPscal3}
     \fl \qquad \langle s_R(t,D_R,\tau_R,\kappa,v_R) \rangle \sim 
     \kappa^{d/2} \ell^{[d - \zeta_s(v_R^*)]/2} \ \hat s\!\left( 
     D_R \kappa^2 \ell^{2 + \zeta_D(v_R^*)} t, 
     \tau_R \ell^{\zeta_\tau(v_R^*)},v_R^* \right) , 
\end{equation}
where $\langle s_R \rangle = \kappa^{d/2} \hat s$. Consequently,
matching $\ell = |\tau_R|^\nu$ and $\ell = (t / D_R \kappa^2)^{-1/z}$, 
respectively, gives $\langle s_R \rangle \sim |\tau_R|^\beta$ and
$\langle s_R \rangle \sim t^{- \alpha}$, with
\begin{eqnarray}
     &&\beta = \frac{\nu [d - \zeta_s(v_R^*)]}{2} = 
     \frac{\nu (d + \eta + z - 2)}{2} = 1 - \frac{\epsilon}{6} 
     + O(\epsilon^2) ~ ,
\label{eq:DPbet} \\
     &&\alpha = \frac{\beta}{z \nu} = 1 - \frac{\epsilon}{4} 
     + O(\epsilon^2) ~ .
\label{eq:DPalp}
\end{eqnarray}

Field-theoretic tools in conjunction with the RG therefore allow us to
define the generic universality class for active to absorbing state
phase transitions, derive the asymptotic scaling laws in the vicinity 
of the critical point, and compute the critical exponents 
perturbationally by means of a systematic expansion about the upper 
critical dimension $d_c = 4$. In higher dimensions $d > 4$, the only 
fixed point is $v_R = 0$, and we recover the mean-field scaling 
behavior. Precisely at the upper critical dimension, there appear
logarithmic corrections. These, as well as the two-loop results for the
critical exponents and the scaling behavior of various other 
observables, are presented in Ref.~\cite{Janssen05}. Monte Carlo
simulations have determined the numerical values for the DP critical 
exponents in dimensions $d < 4$ to high precision \cite{Hinrichsen00,
Odor04}, and confirmed the logarithmic corrections predicted by the RG
\cite{Janssen04a,Luebeck04}.

\subsection{Variants of directed percolation processes}
\label{subsec:DPvar}

\noindent
$\bullet$ Multi-species DP processes:

\noindent
We argued in \sref{subsec:DP} that absorbing to active phase transitions
should generically be described by the critical exponents of DP. This
far-reaching assertion is based on the structure of Eq.~\eref{eq:genpde},
identifying the field $s$ as some coarse-grained `activity' density 
\cite{Janssen81,Grassberger82}. It is indeed very remarkable that the DP
universality class extends to multiple species of reacting agents. 
Consider, for example, the reactions $A \rightleftharpoons A + A$, 
$A \to 0$, coupled to a similar system $B \rightleftharpoons B + B$, 
$B \to 0$ via the processes $A \to B + B$, $A + A \to B$, and its obvious
extension to additional reactants. Inclusion of higher-order reactions 
turns out not to change the critical properties, since the corresponding 
couplings are all irrelevant under the RG. One then arrives at the 
following effective Langevin description for coupled coarse-grained 
density fields $s_i$ \cite{Janssen97,Janssen01}:
\begin{eqnarray}
\label{eq:muldpe}
     &&\partial_t s_i = D_i \left( \nabla^2 - R_i[s_i] \right) 
     s_i + \zeta_i ~ , \quad 
     R_i[s_i] = r_i + \sum_j g_{ij} \, s_j + \ldots 
\label{eq:mulpde} \\
     &&\langle \zeta_i({\bf x},t) \, \zeta_j({\bf x}',t') \rangle = 2 
     s_i N_i[s_i] \delta_{ij} \delta({\bf x}-{\bf x}') \delta(t-t') ~ , ~
     N_i[s_i] = u_i + \ldots , 
\label{eq:muldpn}
\end{eqnarray}
generalizing Eq.~\eref{eq:genpde}. As Janssen has demonstrated, the
ensuing renormalization factors are all given precisely by those of the
single-species process, whence the critical point is generically 
described by the ordinary DP scaling exponents.

Yet if first-order particle transmutations $A \to B$, etc. are added
(notice these are also effectively generated by the above reactions),
leading to additional terms $\sim \sum_{j \not= i} g_j s_j$ in 
Eq.~\eref{eq:muldpe}, one finds that the ensuing  RG flow typically 
produces asymptotically unidirectional processes. One then encounters 
{\em multicritical} behavior, if several control parameters $r_i$ vanish 
simultaneously \cite{Taeuber98a,Goldschmidt98,Taeuber98b,Goldschmidt99}. 
While the critical exponents $\eta$, $\nu$, $z$, and $\gamma$ remain 
unchanged, there emerges in this situation a {\em hierarchy} of order 
parameter exponents $\beta_k = 1/2^k - O(\epsilon)$ on the $k$th level, 
e.g., $\beta_1 = \beta = 1 - \epsilon/6 + O(\epsilon^2)$, $\beta_2 = 
1/2 - 13 \epsilon/96 + O(\epsilon^2)$, and similarly for the decay 
exponents $\alpha_k = \beta_k / z \nu$. The crossover exponent 
associated with the multicritical point can be shown to be $\phi = 1$ 
to all orders in the perturbation expansion \cite{Janssen01}.
 
\

\noindent
$\bullet$ Dynamic isotropic percolation (dIP):

\noindent
An alternative mechanism to generate novel critical behavior in a 
two-species system operates via a passive, spatially fixed and initially
homogeneously distributed species $X$ that couples to the diffusing and
reproducing agents $A \to A + A$ through the decay processes $A \to X$
and $X + A \to X$. Upon integrating out the fluctuations of the
inert species $X$, and expanding about the mean-field solution, the 
resulting effective action eventually becomes
\begin{equation}
\label{eq:dIPact}
     \fl \qquad S_{\rm eff}[\bar s,s] = \int \! d^dx \! \int \! dt 
     \left\{ \bar s \left[ \partial_t + D (r - \nabla^2) \right] s 
     - u {\bar s}^2 \, s + D u \bar s s \int_{-\infty}^t s(t') dt'  
     \right\} ,
\end{equation} 
which corresponds to a stochastic differential equation
\begin{equation}
\label{eq:dipde}
     \partial_t s = D \left( \nabla^2 - r \right) s - D u s 
     \int_{-\infty}^t s(t') dt' + \sqrt{2 u s} \, \eta ~ ,
\end{equation}
with $\langle \eta \rangle = 0$, $\langle \eta({\bf x},t) \, 
\eta({\bf x}',t') \rangle = \delta({\bf x}-{\bf x}') \, \delta(t-t')$ as
in DP. Thus, the induced decay rate is proportional to the product of 
the densities of active agents $s$ and the `debris' 
$D \int_{-\infty}^t s(t') dt'$ produced by decayed agents $A$. More
generally, the action \eref{eq:dIPact} describes the {\em general
epidemic process} \cite{Grassberger83,Janssen85,Cardy85,Janssen86}, in 
contrast with the {\em simple epidemic process} represented by DP. 

We may now consider the quasistatic limit of the field theory 
\eref{eq:dIPact} by introducing the fields $\bar \varphi = \bar s(t \to 
\infty)$, and $\varphi = D \int_{-\infty}^\infty s(t') dt'$, whereupon we
arrive at the action
\begin{equation}
\label{eq:sIPact}
     \fl \qquad S_{\rm qst}[\bar \varphi,\varphi] = \int \! d^dx \, 
     \bar \varphi \left[ r - \nabla^2 - u (\bar \varphi - \varphi) 
     \right] \varphi ~ .
\end{equation} 
Note, however, that as a manifestation of its dynamic origin, this 
{\em quasi-static} field theory must be supplemented by causality rules. 
The action \eref{eq:sIPact} then describes the scaling properties of 
critical {\em isotropic} percolation clusters \cite{Benzoni84}. Thus, as
first remarked by Grassberger \cite{Grassberger83}, the general epidemic 
process is governed by the static critical exponents of isotropic 
percolation. These are readily obtained by means of a RG analysis in an 
$\epsilon$ expansion about the upper critical dimension $d_c = 6$. The 
one-loop diagrams are precisely those of \fref{fig:dploops}, with the 
static propagator $G_0({\bf p}) = (r + p^2)^{-1}$. The explicit 
computation proceeds as outlined in \sref{subsec:DPRG} (for more details,
see, e.g., Ref.~\cite{Janssen05}), and yields $\eta = - \epsilon / 21 + 
O(\epsilon^2)$, $\nu^{-1} = 2 - 5 \epsilon / 21 + O(\epsilon^2)$, and 
$\beta = 1 - \epsilon / 7 + O(\epsilon^2)$, with $\epsilon = 6 - d$. 

In order to characterize the dynamic critical properties, however, we
must return to the full action \eref{eq:dIPact}. Yet its structure once
again leads to the Feynman graphs depicted in \fref{fig:dploops}, but 
with the second vertex in (a) carrying a temporal integration. One then
finds $z = 2 - \epsilon / 6 + O(\epsilon^2)$, which completes the
characterization of this {\em dynamic isotropic percolation} (dIP) 
universality class \cite{Janssen85,Cardy85,Janssen86}. Precisely as for 
DP processes, multi-species generalizations generically yield the same
critical behavior, except at special multicritical points, characterized
again by a crossover exponent $\phi = 1$ \cite{Janssen01}.

\

\noindent
$\bullet$ L\'evy flight DP:

\noindent
Long-range interactions, as can be modeled by L\'evy flight contributions
$D_L p^\sigma$ to the propagators, may modify the critical behavior of
both DP and dIP \cite{Hinrichsen99,Janssen99}. Two situations must be 
distinguished \cite{Janssen98}: for $2 - \sigma = O(\epsilon)$, a double 
expansion with respect to both $\epsilon$ and $2 - \sigma$ is required; 
on the other hand, if $2 - \sigma = O(1)$, the ordinary diffusive 
contribution $D p^2$ to the propagator becomes irrelevant, whereas the 
non-analytic term $D_L p^\sigma$ acquires no fluctuation corrections, 
whence $Z_s Z_{D_L} = 1$ to all orders in perturbation theory. 
Subsequent scaling analysis yields the critical dimensions 
$d_c = 2 \sigma$ for DP and $d_c = 3 \sigma$ for dIP, respectively. 
To one-loop order, one then finds at the new long-range fixed points, 
with $\epsilon = d - d_c$: $\eta = 2 - \sigma$, and for DP: $z = \sigma 
- \epsilon/7 + O(\epsilon^2)$, $\nu^{-1} = \sigma - 2 \epsilon/7 + 
O(\epsilon^2)$, $\beta = 1 - 2 \epsilon/7 \sigma + O(\epsilon^2)$; for
dIP: $z = \sigma - 3\epsilon/16 + O(\epsilon^2)$, $\nu^{-1} = \sigma -  
\epsilon/4 + O(\epsilon^2)$, $\beta = 1 - \epsilon/4 \sigma 
+ O(\epsilon^2)$. 

\

\noindent
$\bullet$ DP coupled to a non-critical conserved density (DP-C):

\noindent
A variant on the DP reaction scheme $A \leftrightarrow A+A$ with 
decay $A \to 0$ is to require the $A$ particle decay to be catalyzed 
by an additional species $C$, via $A + C \to C$. The $C$ particles 
move diffusively and are conserved by the reaction. In the population 
dynamics language, the $C$ particles can be said to poison the $A$ 
population \cite{Kree89}. Related is a model of infection dynamics, 
$A + B \to 2 B$, $B\to A$, where $A$ and $B$ respectively represent 
healthy and sick individuals \cite{vanWijland98,Oerding00}. The 
latter model reduces to the former in the case of equal diffusion
constants $D_A = D_B$ (see Ref.~\cite{Janssen05} for details).

These systems exhibit, like DP, an upper critical dimension $d_c=4$.  
Below this dimension there exist three different regimes, depending on
whether the ratio of diffusion constants greater than, equal to, or 
less than unity. For the case $D_A > D_B$, the resulting RG flows 
run away, indicating a fluctuation-induced first-order transition
\cite{Oerding00}. For the case $D_A = D_B$ the critical exponents are 
given by $z=2$, $\nu = 2/d$ (exact), and $\beta = 1 - \epsilon/32
+ O(\epsilon^2)$; while for $D_A < D_B$ a distinct fixed point is 
obtained \cite{vanWijland98} with exact values $z=2$, $\nu=2/d$, and 
$\beta=1$.

\subsection{Branching and annihilating random walks (BARW)}
\label{subsec:BARW}

Branching and annihilating random walks are defined through diffusing
particles $A$ subject to the competing branching reactions 
$A \to (m+1) A$ (with rate $\sigma$) and annihilation processes 
$k A \to 0$ (rate $\lambda$) \cite{Grassberger84}. The corresponding 
mean-field rate equation for the particle density reads
\begin{equation}
\label{eq:bawmfr}
     \partial_t a(t) = \sigma \, a(t) - k \lambda \, a(t)^k ~ ,
\end{equation}
with the solution
\begin{equation}
\label{eq:bawmfs}
     a(t) = {a_\infty \over \left( 1 + \left[ (a_\infty/a_0)^{k-1} - 1 
     \right] e^{-(k-1) \sigma t} \right)^{1/(k-1)}} ~ ,
\end{equation} 
which for $t \gg 1 / \sigma$ approaches the finite density $a_\infty = 
(\sigma / k \lambda)^{1/(k-1)}$, independent of the initial density 
$a_0$. Mean-field theory therefore predicts only an active phase, 
provided $\sigma > 0$. At $\sigma_c = 0$, there exists a `degenerate' 
critical point, whose critical exponents are given by the pure 
diffusion-limited annihilation model, $\eta = 0$, $\nu = 1/2$, $z = 2$, 
$\gamma = 1$, $\alpha = \beta = 1 / (k-1)$. 

This mean-field scenario should hold in dimensions $d > d_c = 2 / (k-1)$,
as determined from the scaling dimension of the annihilation process. The
branching rate with scaling dimension $[\sigma] = \kappa^2$ constitutes 
a relevant parameter. However, as initially established in Monte Carlo 
simulations \cite{Grassberger84,Hinrichsen00,Odor04}, for $k=2$ 
fluctuations invalidate this simple picture, rendering BARW considerably 
more interesting \cite{Cardy96,Cardy98}. Indeed, one has to distinguish 
the cases of odd and even offspring numbers $m$: Whereas the active to 
absorbing transition in BARW with odd $m$ is described by the DP critical
exponents, provided $\sigma_c > 0$, BARW with even $m$ define a genuinely
different {\em parity-conserving} (PC) universality class, named after 
its unique feature of conserving the particle number parity under the 
involved reactions.

\begin{figure}[htb]
\begin{center}
\includegraphics[width=2.25in]{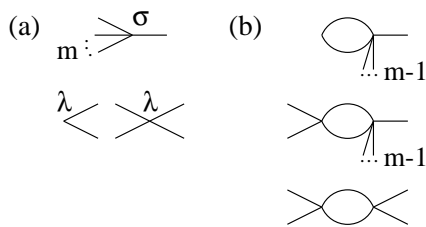}
\caption{BARW field theory: (a) branching (top) and annihilation (bottom)
vertices; (b) one-loop Feynman diagrams generating the $A \to (m-1)A$ 
process, and renormalizing the branching and annihilation rates, 
respectively.}
\label{fig:bawloops}
\end{center}
\end{figure}

The field-theoretic representation for BARW with $k=2$ and $m$ offspring
particles, omitting the temporal boundary terms, reads
\begin{equation}
\label{eq:barwmicac}
     \fl \qquad S[\tilde\phi,\phi] = \int d^dx \int dt \left[ 
     \tilde\phi \left( \partial_t - D \nabla^2 \right) \phi 
     + \sigma (1 - \tilde\phi^m) \tilde\phi \phi 
     - \lambda \left( 1 - \tilde\phi^2 \right) \phi^2 \right] ,
\end{equation}
and the corresponding vertices are shown in \fref{fig:bawloops}(a). Upon
combining the branching with the pair annihilation processes as in the
top one-loop diagram in \fref{fig:bawloops}(b), we see that in addition
to the original $A \to (m+1) A$ reaction all lower branching processes
$A \to (m-1) A, (m-3) A, \ldots$ become generated. In a first 
coarse-graining step, all these reactions then must be added to the 
`microscopic' field theory \eref{eq:barwmicac}. Furthermore, upon
inspecting the renormalization of the branching and annihilation rates by
the one-loop Feynman graphs depicted in \fref{fig:bawloops}(b), we see
that identical loop integrals govern the corresponding UV singularities, 
but the renormalization of the branching process with $m$ offspring 
carries a relative combinatorial factor of $m(m+1)/2$. Since the loop
contributions carry a negative sign, the resulting downward shift of the
branching rate's scaling dimension is lowest for $m = 1$ and $m = 2$,
respectively. For {\em odd} offspring number $m$, the most relevant
emerging branching process thus is $A \to A+A$, but in addition the
spontaneous decay $A \to 0$ is generated \cite{Cardy96,Cardy98}. 
Consequently, the {\em effective} field theory that should describe BARW 
with odd $m$ becomes \eref{eq:DPeffact}, and the phase transition is 
predicted to be in the DP universality class. This is true {\em provided}
the induced particle decay rate may overcome the generated or 
renormalized branching rate with single offspring, and thus shift the 
critical point to $\sigma_c > 0$. Within a perturbational analysis with
respect to the annihilation rate $\lambda$, this happens only in low 
dimensions $d \leq 2$ \cite{Cardy96,Cardy98}. In a recent nonperturbative
numerical RG study, however, the emergence of an inactive phase and DP
critical behavior was found in higher dimensions as well \cite{Canet04,
Canet04a}.

It now becomes apparent why BARW with {\em even} offspring number should
behave qualitatively differently: in this case, spontaneous particle 
decay processes $A \to 0$ cannot be generated, even on a coarse-grained 
level. This is related to the fact that in the branching processes 
$A \to (m+1) A$ as well as the pair annihilation $A + A \to 0$ the 
particle number {\em parity} remains conserved; correspondingly, there 
are two distinct absorbing states for even-offspring BARW, namely the 
strictly empty lattice if the initial particle number $N_0$ is even, and 
a single remaining particle, if $N_0$ is odd. In the field theory 
\eref{eq:barwmicac}, this conservation law and symmetry are reflected in
the invariance with respect to simultaneously taking $\phi \to - \phi$ 
and $\tilde\phi \to - \tilde\phi$. This invariance must be carefully 
preserved in any subsequent analysis. Performing the field shift 
$\tilde\phi = 1 + \bar\phi$ masks the discrete inversion symmetry; worse,
it becomes lost entirely if afterwards, based on mere power counting 
arguments, only the leading powers in $\tilde\phi$ are retained, whence
one would be erroneously led to the DP effective action. It is therefore
safest to work with the unshifted action \eref{eq:barwmicac}, but adding
the generated branching processes with $m-2, m-4, \ldots$ offspring
particles. As explained before, the most relevant branching reaction will
be the one with two offspring. Setting $m = 2$ in the action 
\eref{eq:barwmicac} indeed yields a renormalizable theory, namely the
effective action for the PC universality class, with the particle 
production processes with higher offspring numbers constituting 
irrelevant perturbations.

The bare propagator of this theory is similar to Eq.~\eref{eq:dpprop},
\begin{equation}
\label{eq:bawprop}
     G_0({\bf p},\omega) = {1 \over -i\omega + \sigma + D p^2} 
\end{equation}
but contains the branching rate $\sigma$ as a mass term. The branching 
rate also appears in the three-point vertex, \fref{fig:bawloops}(a,top).
Since we need to follow the RG flow of the renormalized reaction rates
\begin{equation}
\label{eq:barwren}
     \sigma_R = Z_\sigma \sigma / D \, \kappa^2 ~ , \quad
     \lambda_R = Z_\lambda \lambda C_d / D \kappa^{2-d} ~ ,
\end{equation} 
with $C_d = \Gamma(2-d/2) / 2^{d-1} \pi^{d/2}$, we must set the 
normalization point either at finite external momentum $p = 2 \kappa$ or 
frequency / Laplace transform variable $i \omega = s = 2 D \kappa^2$.
From the one-loop Feynman graphs in \fref{fig:bawloops}(b) that 
respectively describe the propagator, branching vertex, and annihilation
vertex renormalizations, one then finds that the UV singularities, for
{\em any} value of $\sigma$, can be absorbed into the $Z$ factors
\begin{equation}
\label{eq:bawzfac}
     \fl \qquad Z_\sigma = 1 - \frac{3 C_d}{2-d} \, 
     \frac{\lambda / D}{(\kappa^2 + \sigma/D)^{1-d/2}} ~ , \quad
     Z_\lambda = 1 - \frac{C_d}{2-d} \, 
     \frac{\lambda / D}{(\kappa^2 + \sigma/D)^{1-d/2}} ~ ,
\end{equation}
which are functions of both $\sigma/D$ and $\lambda/D$, as in other 
crossover theories with relevant parameters. With $\zeta_\sigma = 
\kappa \partial_\kappa \ln (\sigma_R / \sigma)$ and $\zeta_\lambda = 
\kappa \partial_\kappa \ln (\lambda_R / \lambda)$, we are thus led to the
coupled RG flow equations \cite{Cardy96,Cardy98}
\begin{eqnarray}
     &&\ell \frac{d \sigma_R(\ell)}{d\ell} = \sigma_R(\ell) 
     \zeta_\sigma(\ell) = \sigma_R(\ell) \left( - 2 + 
     \frac{3 \lambda_R(\ell)}{[1 + \sigma_R(\ell)]^{2-d/2}} \right) ~ ,\\
\label{eq:bawflos}
     &&\ell \frac{d \lambda_R(\ell)}{d\ell} = \lambda_R(\ell) 
     \zeta_\lambda(\ell) = \lambda_R(\ell) \left( d - 2 + 
     \frac{\lambda_R(\ell)}{[1 + \sigma_R(\ell)]^{2-d/2}} \right) ~ .
\label{eq:bawflol}
\end{eqnarray}

\begin{figure}[htb]
\begin{center}
\includegraphics[width=2.0in]{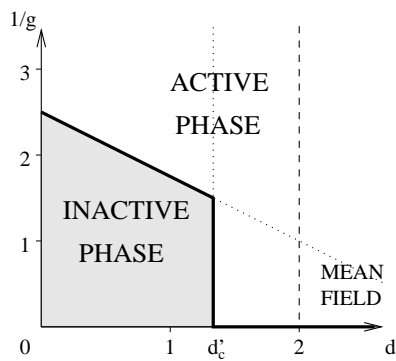}
\caption{Phase diagram and unstable RG fixed point $1/g_R^*$ for 
even-offspring BARW (PC universality class) as function of dimension $d$
(from Ref.~\cite{Cardy98}).}
\label{fig:bawcpd}
\end{center}
\end{figure}

The effective coupling controlling the RG flows, to one-loop order at 
least, is $g_R = \lambda_R / (1 + \sigma_R)^{2-d/2}$. For 
$\sigma_R = 0$, that is for the pure pair annihilation model, according 
to Eq.~\eref{eq:bawflol} $g_R \to g_R^* = 2 - d$, which after trivial
rescaling corresponds to the annihilation fixed point \eref{eq:fpoint}.
For $\sigma > 0$, however, we expect $\sigma_R(\ell) \to \infty$, 
whereupon the RG $\beta$ function for the coupling $g_R$ becomes
\begin{equation}
\label{eq:pcbetag}
     \beta_g(g_R) \to g_R \left[ \zeta_\lambda - \left( 2 - \frac{d}{2}
     \right) \zeta_\sigma \right] = g_R \left[ 2 - \frac{10-3d}{2} g_R
     \right] ~ , 
\end{equation}
which yields the Gaussian fixed point at $g_R = 0$ and a critical fixed 
point $g_c^* = 4 / (10-3d)$. Yet since the bare reaction rate 
corresponding to the pure annihilation fixed point is already infinite,
see \sref{subsec:renor} following Eq.~\eref{eq:gR}, we must demand on 
physical grounds that $g_c^* \leq  2 - d$, whence we infer that the 
critical fixed point comes into existence only for $d < d_c' = 4/3$. If 
initially $g_R < g_c^*$, $g_R(\ell) \to 0$, consistent with 
$\sigma_R(\ell) \sim \ell^{-2} \to \infty$ as $\ell \to 0$. This 
Gaussian fixed point, characterized by naive scaling dimensions, 
describes the active phase with exponential correlations. On the other 
hand, for $g_R > g_c^*$, and provided that $d < d_c'$, $\sigma_R(\ell) 
\to 0$ and $g_R(\ell) \to 2-d$, which describes an inactive phase that 
in its entirety is governed by the pure annihilation model power laws. 
The phase transition in the PC universality class, which apparently has 
no counterpart in mean-field theory, is thus triggered through 
fluctuations that drive the branching rate irrelevant. In contrast to 
equilibrium systems, fluctuations here open up a novel phase rather 
than destroying it, and we may view the new borderline dimension $d_c'$ 
as an `inverted lower' critical dimension, since the phase transition 
only exists for $d < d_c'$. The phase diagram as function of spatial 
dimension, within the one-loop approximation, is summarized in 
\fref{fig:bawcpd}.

In order to obtain the asymptotic scaling behavior for the particle 
density, we write down the solution of its RG equation in the vicinity of
an RG fixed point $g_R^*$, which reads
\begin{equation}
\label{eq:bawrgeq}
     a(t,D_R,\sigma_R,\lambda_R,\kappa) = \kappa^d \ell^d ~ a\!\left(
     D_R \kappa^2 \ell^2 t, \sigma_R \ell^{\zeta_\sigma(g_R^*)}, 
     \lambda_R \ell^{\zeta_\lambda(g_R^*)} \right) ,
\end{equation}
since there is no renormalization of the fields or diffusion constant to
one-loop order, which immediately implies $\eta = 0$ and $z = 2$. For
$d_c' < d \leq 2$ the branching rate $\sigma_R$ plays the role of 
the critical control parameter $\tau_R$ in DP, and $g_R^* = 2 - d = 
\epsilon$ is the annihilation fixed point. In an $\epsilon$ expansion 
about the upper critical dimension $d_c = 2$, we thus obtain the scaling 
exponents
\begin{equation}
\label{eq:bawsce}
     \nu^{-1} = - \zeta_\sigma(g_R^*) = 2 - 3 \epsilon  ~ , 
     \quad z = 2 ~ , \quad\alpha = d/2 ~ , \quad 
     \beta = d \nu = z \nu \alpha ~ ,
\end{equation}
the latter via matching $\sigma_R \ell^{\zeta_\sigma(g_R^*)} = 1$. 
Notice that $\nu$ diverges as $\epsilon \to 2/3$ or $d \to d_c'$. 

Yet the PC phase transition at $\sigma_c > 0$ can obviously not be 
captured by such an $\epsilon$ expansion. One is instead forced to 
perform the analysis at fixed dimension, without the benefit of a small 
expansion parameter. Exploiting the mean-field result for the density in 
the active phase, for $d < d_c'$ we may write in the vicinity of $g_c^*$:
\begin{equation}
\label{eq:pcrgeq}
     \fl \qquad \qquad a(t,D_R,\sigma_R,\lambda_R,\kappa) = \kappa^d 
     \frac{\sigma_R}{\lambda_R} ~ \ell^{d + \zeta_\sigma(g_c^*) - 
     \zeta_\lambda(g_c^*)} ~ \tilde a\!\left( \sigma_R t \kappa^2 
     \ell^{2 + \zeta_\sigma(g_c^*)}, 
     \varepsilon_R \ell^{\zeta_\varepsilon(g_c^*)} \right) ,
\end{equation}
where $\varepsilon \propto g_c^* - g$ constitutes the control parameter 
for the transition, and $\zeta_\varepsilon = d \beta_g / d g_R$. Now
setting $\varepsilon_R \ell^{\zeta_\varepsilon(g_c^*)} = 1$, we obtain
with $\zeta_\sigma(g_c^*) = -2(4-3d)/(10-3d)$, $\zeta_\lambda(g_c^*) = -
(4-d)(4-3d)/(10-3d)$, and $\zeta_\varepsilon(g_c^*) = -2$, the critical
exponents \cite{Cardy96,Cardy98}
\begin{equation}
\label{eq:epcsce}
    \fl \quad \nu = \frac{2 + \zeta_\sigma(g_c^*)}
    {- \zeta_\varepsilon(g_c^*)} = \frac{3}{10-3d} ~ , \quad z = 2 ~, 
    \quad \beta = \frac{d + \zeta_\sigma(g_c^*) - \zeta_\lambda(g_c^*)}
    {- \zeta_\varepsilon(g_c^*)} = \frac{4}{10-3d} ~ .
\end{equation}
Note that the presence of the dangerously irrelevant parameter 
$1 / \sigma_R$ precludes a direct calculation of the power laws precisely
at the critical point (rather than approaching it from the active phase),
and the derivation of `hyperscaling' relations such as 
$\beta = z \nu \alpha$. Numerically, the PC critical exponents in one
dimension have been determined to be $\nu \approx 1.6$, $z \approx 1.75$,
$\alpha \approx 0.27$, and $\beta \approx 0.92$ \cite{Hinrichsen00,
Odor04}. Perhaps not too surprisingly, the predictions \eref{eq:epcsce} 
from the uncontrolled fixed-dimension expansion yield rather poor values
at $d = 1$. Unfortunately, an extension to, say, higher loop order, is
not straightforward, and an improved analytic treatment has hitherto not 
been achieved.

\subsection{BARW variants and higher-order processes}

\noindent
$\bullet$ L\'evy flight BARW:

\noindent
Simulations clearly cannot access the PC borderline critical dimension 
$d_c'$. This difficulty can be overcome by changing from ordinary 
diffusion to L\'evy flight propagation $\sim D_L p^\sigma$. The existence
of the power-law inactive phase is then controlled by the L\'evy exponent
$\sigma$, and in one dimension emerges for $\sigma > \sigma_c = 3/2$
\cite{Vernon01}.

\

\noindent
$\bullet$ Multi-species generalizations of BARW:

\noindent
There is a straightforward generalization of the two-offspring BARW to a
variant with $q$ interacting species $A_i$, according to $A_i \to 3 A_i$
(rate $\sigma$), $A_i \to A_i + 2 A_j$ ($j \not= i$, rate $\sigma'$), and 
$A_i + A_i \to 0$ only for particles of the same species. Through simple 
combinatorics $\sigma_R / \sigma_R' \to 0$ under renormalization, and the 
process with rate $\sigma'$ dominates asymptotically. The coarse-grained
effective theory then merely contains the rate $\sigma_R'$, corresponding
formally to the limit $q \to \infty$, and can be analyzed exactly. It 
displays merely a degenerate phase transition at $\sigma_c' = 0$, similar
to the single-species even-offspring BARW for $d > d_c'$, but with 
critical exponents $\nu = 1/d$, $z = 2$, $\alpha = d/2$, and $\beta = 1 
= z \nu \alpha$ \cite{Cardy96,Cardy98}. The situation for $q = 1$ is thus 
qualitatively different from any multi-species generalization, and cannot
be accessed, say, by means of a $1/q$ expansion.

\

\noindent
$\bullet$ Triplet and higher-order generalizations of BARW:

\noindent

Invoking similar arguments as above for $k=3$, i.e., the triplet 
annihilation $3 A \to 0$ coupled to branching processes, one would expect
DP critical behavior at a phase transition with $\sigma_c > 0$ for any
$m {\rm mod} 3 = 1,2$. For $m = 3, 6, \ldots$, however, special critical
scenarios might emerge, but limited to mere logarithmic corrections,
since $d_c = 1$ in this case \cite{Cardy98}. Simulations, however,
indicate that such higher-order BARW processes may display even richer
phase diagrams \cite{Odor04}.

\

\noindent
$\bullet$ Fission / annihilation or the pair contact process with 
diffusion (PCPD)

\noindent

One may expect novel critical behavior for active to bsorbing state
transitions if there is no first-order process present at all. This 
occurs if the branching reaction competing with $A + A \to (0,A)$ is 
replaced with $A + A \to (n+2) A$, termed fission/annihilation reactions 
in Ref.~\cite{Howard97}, but now generally known as pair contact process 
with diffusion (PCPD) \cite{Henkel04}. Without any restrictions on the 
local particle density, or, in the lattice version, on the site 
occupation numbers, the density obviously diverges in the active phase, 
whereas the inactive, absorbing state is governed by the power laws of 
the pair annihilation/coagulation process \cite{Howard97}. By introducing
site occupation restrictions, or alternatively, by adding triplet 
annihilation processes, the active state density becomes finite, and the 
phase transition continuous. In a field-theoretic representation, one 
must also take into account the infinitely many additional fission 
processes that are generated by fluctuations. Following 
Ref.~\cite{Wijland01}, one may construct the field theory action for the 
restricted model version, whence upon expanding the ensuing exponentials,
see \eref{eq:DPrestact}, one arrives at a renormalizable action. Its RG 
analysis however leads to runaway RG trajectories, indicating that this 
action cannot represent the proper effective field theory for the PCPD 
critical point \cite{Janssen04}. Since Monte Carlo simulation data for 
this process are governed by long crossover regimes, the identification 
and characterization of the PCPD universality class remains to date an 
intriguing open issue \cite{Henkel04}.

\subsection{Boundaries}
\label{subsec:boundaries}

\begin{figure}[htb]
\begin{center}
\includegraphics[width=2.25in,angle=0]{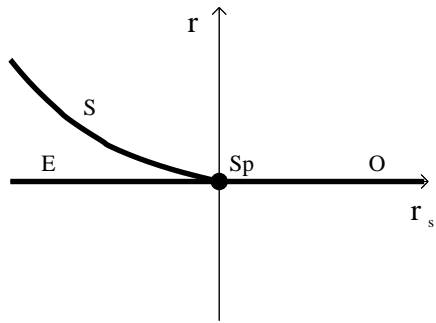}
\caption{Schematic mean field phase diagram for boundary DP. The
transitions are labeled by O=ordinary, E=extraordinary, S=surface,
and Sp=special.}
\label{fig:dpps}
\end{center}
\end{figure}

In equilibrium critical phenomena it is well-known that, close to
boundaries, the critical behavior can be different from that in the bulk 
(see Refs.~\cite{Diehl86,Diehl97} for comprehensive reviews).  As we will
see, a similar situation holds in the case of nonequilibrium 
reaction-diffusion systems (see also the review in 
Ref.~\cite{Froedjh01}). Depending on the values of the boundary and bulk
reaction terms, various types of boundary critical behavior are possible.
For example, if the boundary reaction terms ensure that the boundary, 
independent of the bulk, is active, while the bulk is critical, then we 
have the so-called extraordinary transition. Clearly, by varying the 
boundary/bulk reaction rates three other boundary transitions are 
possible: the ordinary transition (bulk critical, boundary inactive), 
the special transition (both bulk and boundary critical, a multicritical 
point), and the surface transition (boundary critical, bulk inactive).  
Defining $r$ and $r_s$ as the deviations of the bulk and boundary from 
criticality, respectively, a schematic boundary phase diagram is shown 
in \fref{fig:dpps}. In this review, for reasons of brevity, we will 
concentrate on the case of DP with a planar boundary \cite{Janssen88,
Froedjh98,Howard00}. Other cases ($A + A \to \emptyset$ with a boundary 
and boundary BARW) will be dealt with more briefly.

\

\noindent 
$\bullet$ Boundary directed percolation:

\noindent
In this section, we will focus on the ordinary transition in boundary DP.
The mean-field theory for this case was worked out in 
Ref.~\cite{Howard00}, while the field theory was analyzed to one-loop 
order in Ref.~\cite{Janssen88}.

As we have discussed earlier, the field theory for bulk DP is described 
by the action \eref{eq:DPeffact}. Consider now the effect of a 
semi-infinite geometry $\{x=(x_{\parallel},z),~ 0\leq z<\infty\}$,
bounded by a plane at $z=0$. The complete action for bulk and boundary is
then given by $S = S_{\rm eff}+S_{\rm bd}$, where
\begin{equation}
\label{eq:actionsurface}
     S_{\rm bd}= \int d^{d-1}x \int dt ~ D r_s \, \bar s_s \, s_s ~ ,
\end{equation}
with the definitions $s_s = s (x_{\parallel},z=0,t)$ and $\bar s_s = \bar
s(x_{\parallel},z=0,t)$. This boundary term is the most relevant 
interaction consistent with the symmetries of the problem, and which
respects the absorbing state criterion. Power counting indicates that
the boundary coupling has scaling dimension $[r_s] \sim \kappa$, and is 
therefore relevant. The presence of the wall at $z=0$ enforces
the boundary condition
\begin{equation}
     \left. \partial_z s\right|_{z=0}= r_s s_s.
\end{equation}
This condition guarantees that a boundary term of the form $\bar s \,
\partial_z s$ is not required, even though it is marginal according to 
power counting arguments.

Since $[r_s] \sim \kappa$ the only possible fixed points of the 
renormalized coupling are $\Delta_{s_R} \to 0$ or $\pm \infty$. Here we
focus on the case $r_{s R} \to \infty$, corresponding to the
ordinary transition. At this fixed point, the propagator in the presence 
of a boundary $G_{0 s}$ can be written entirely in terms of the bulk 
propagator $G_0$:
\begin{equation}
      G_{0 s}(x_{\parallel},z,z',t) = G_0(x_{\parallel},z,z',t) -
      G_0(x_{\parallel},z,-z',t) ~ .
\end{equation}
Due to the above boundary condition, which implies that 
$G_{0 s}(x_{\parallel},z,z',t)|_{z=0} = 0$, we see that the appropriate
boundary fields for the ordinary transition are not $s_s$, $\bar s_s$,
but rather $s_{\perp} = \partial_z s|_{z=0}$, and $\bar s_{\perp} =
\partial_z \bar s|_{z=0}$. For example, in order to compute the order
parameter exponent $\beta_1$ at the boundary, defined by 
$s_R(z=0,r) \sim |\tau_R|^{\beta_1}$ ($\tau_R < 0$), we must investigate
how $s_{\perp} = \partial_z s|_{z=0}$ scales. In mean-field theory,
straightforward dimensional analysis yields $\beta_1 = 3/2$ 
\cite{Howard00}. Of course, to go beyond this simple mean-field picture 
and to incorporate fluctuations, we must now employ the machinery of the 
field-theoretic RG.

Because of the presence of the surface, we expect to find new divergences
which are entirely localized at the surface. These divergences must be 
absorbed into new renormalization constants, in addition to those 
necessary for renormalization of the bulk terms. At the ordinary 
transition, the new divergences can be absorbed by means of an additional
surface field renormalization, yielding the renormalized fields:
\begin{equation}
     s_{{\perp}_R} = Z_0 Z_s^{1/2} s_{\perp} ~ , \qquad 
     \bar s_{{\perp}_R} = Z_0 Z_s^{1/2} \bar s_{\perp} ~ .
\end{equation}
Note that the same factor $Z_0$ enters both renormalized surface fields, 
similar to the bulk field renormalization. The fact that one independent 
boundary renormalization is required translates into the existence of one
independent boundary exponent, which we can take to be $\beta_1$, defined
above.

Consider now the connected renormalized correlation function 
$G_R^{(N,M)}$, composed of $N$ $\{s$, $\bar s\}$ fields and $M$ 
$\{s_{\perp}$, $\bar s_{\perp}\}$ fields. The renormalization group
equation then reads (excepting the case $N=0$, $M=2$ for which there
is an additional renormalization):
\begin{equation}
     \fl \left(\kappa \frac{\partial}{\partial \kappa} - \frac{N+M}{2} \,
     \zeta_s - M \zeta_0 + \zeta_D D_R \frac{\partial}{\partial D_R} +
     \zeta_{\tau} \tau_R \frac{\partial}{\partial \tau_R} + \beta_v(v_R)
     \frac{\partial}{\partial v_R} \right) G_R^{(N,M)} = 0 ~ ,
\end{equation}
with the definitions \eref{eq:DPzetas}--\eref{eq:DPbetav} and 
$\zeta_0 = \kappa \, \partial_\kappa \ln Z_0$. Solving the above equation
at the bulk fixed point using the method of characteristics, combined 
with dimensional analysis, yields
\begin{equation}
     \fl \quad G_R^{(N,M)}(\{x,t\},D_R,\tau_R,\kappa,v_R^*) \sim 
     |\tau_R|^{(N+M) \beta + M \nu (1-\eta_0)} \, \hat G^{(N,M)}\!\left(
     \left\{ \frac{\kappa x}{|\tau_R|^{-\nu}}, 
     \frac{\kappa^2 D_R t}{|\tau_R|^{- z \nu}} \right\}\right) .
\end{equation}
With $\epsilon=4-d$, and defining $\eta_0 = \zeta_0(v_R^*) = \epsilon/12 +
O(\epsilon^2)$ (the value of the $\zeta_0$ function at the bulk fixed 
point), we see that at the ordinary transition
\begin{equation}
      \beta_1 = \beta + \nu (1-\eta_0) = \frac{3}{2} - 
      \frac{7\epsilon}{48} + O(\epsilon^2) ~ ,
\end{equation}
where we have used some results previously derived for bulk DP. The 
general trend of the fluctuation correction is consistent with the 
results of Monte Carlo simulations in two dimensions \cite{Froedjh98} and
series expansions in $d=1$ \cite{Jensen99}, which give $\beta_1=1.07(5)$ 
and $\beta_1=0.73371(2)$, respectively. For dIP, an analogous analysis 
yields the boundary density exponent $\beta_1 = 3/2 - 11 \epsilon / 84 + 
O(\epsilon^2)$ \cite{Janssen05}. 

One unsolved mystery in boundary DP concerns the exponent 
$\tau_1 = z \nu - \beta_1$, governing the mean cluster lifetime in the 
presence of a boundary \cite{Froedjh98}. This exponent has been 
conjectured to be equal to unity \cite{Essam96}; series expansions
certainly yield a value very close to this ($1.00014(2)$) 
\cite{Jensen99}, but there is as yet no explanation (field-theoretic
or otherwise) as to why this exponent should assume this value.

\

\noindent
$\bullet$ Boundaries in other reaction-diffusion systems:

\noindent
Aside from DP, boundaries have been studied in several other 
reaction-diffusion systems. BARW (with an even number of offspring) with 
a boundary was analyzed using field-theoretic and numerical methods in 
Refs.~\cite{Froedjh01,Howard00,Lauritsen98}. As in the bulk case, the 
study of boundary BARW is complicated by the presence of a second 
critical dimension $d_c'$ which prevents the application of controlled 
perturbative $\epsilon$ expansions down to $d = 1$. Nevertheless some 
progress could still be made in determining the boundary BARW phase 
diagram \cite{Howard00}. The situation is somewhat more complicated than 
in the case of DP, not only because the location of the bulk critical 
point is shifted away from zero branching rate (for $d<d_c'$), but also 
because the parity symmetry of the bulk can be broken but {\em only} at 
the boundary. The authors of Ref.~\cite{Howard00} proposed that the 
one-dimensional phase diagram for BARW is rather different from that of 
mean-field theory: If a symmetry breaking $A \to 0$ process is present on
the boundary, then only an ordinary transition is accessible in $d=1$; 
whereas if such a reaction is absent then only a special transition is 
possible. Furthermore, an exact calculation in $d=1$ at a particular 
plane in parameter space allowed the authors of Ref.~\cite{Howard00} to 
derive a relation between the $\beta_1$ exponents at the ordinary and 
special transitions. It would be very interesting to understand this 
result from a field-theoretic perspective, but until controlled 
perturbative expansions down to $d=1$ become possible, such an 
understanding will probably remain elusive. More details of these results
can be found in Refs.~\cite{Froedjh01,Howard00}.

Richardson and Kafri \cite{Richardson99b,Kafri99} analyzed the presence 
of a boundary in the simpler $A+A \to 0$ reaction. For $d\leq 2$, they 
found a fluctuation-induced density excess develops at the boundary, and
this excess extends into the system diffusively from the boundary. The 
(universal) ratio between the boundary and bulk densities was computed to
first order in $\epsilon=2-d$. Since the only reaction occurring both on 
the boundary and in the bulk is the critical $A+A \to 0$ process, this 
situation corresponds to the special transition.

\section{Open Problems and Future Directions}
\label{sec:future}

As we have seen, enormous progress has been made over the last decade or 
so in understanding fluctuations in reaction-diffusion processes. Many 
systems are now rather well understood, thanks to a variety of 
complementary techniques, including mean-field models, Smoluchowski 
approximations, exact solutions, Monte Carlo simulations, as well as
the field-theoretic RG methods we have predominantly reviewed in this
article. However, we again emphasize the particular importance of RG
methods in providing the only proper understanding of universality. 
Despite these undoubted successes, we believe that there are still many 
intriguing open problems:

\

\noindent
$\bullet$ 
Already for the simple two-species pair annihilation process $A+B\to 0$, 
field-theoretic RG methods have not as yet been able to properly analyze 
the asymptotic properties in dimensions $d<2$ in the case of equal 
initial densities \cite{Lee95}. Moreover, the standard bosonic field 
theory representation appears not to capture the particle species 
segregation in multi-species generalizations adequately 
\cite{Hilhorst04}. A viable description of topological constraints in one
dimension, such as induced by hard-core interactions that prevent 
particles passing by each other, within field theory remains a challenge.

\

\noindent
$\bullet$ Branching-annihilating random walks (BARW) with an even number 
of offspring particles is still poorly understood in $d=1$, due to the
existence of the second critical dimension $d_c'$ \cite{Cardy96,Cardy98}.
A systematic extension of the one-loop analysis at fixed dimension to 
higher orders has not been successfully carried out yet. Ideally one 
would like to find a way of circumventing this difficulty, in particular 
to understand why certain one-loop results (for the exponent $\beta$ and 
the value of $d_c'$ \cite{Vernon01}) appear to be exact, even when the 
two-loop corrections are known to be nonzero. Nonperturbative numerical
RG methods might be of considerable value here \cite{Canet04,Canet04a}.
There is also an interesting suggestion for a combined Langevin 
description of both DP and PC universality classes \cite{Hammal04}, but
the ensuing field theory has yet to be studied by means of the RG.

\

\noindent
$\bullet$ Despite intensive work over recent years the status of the
the pair contact process with diffusion (PCPD) \cite{Howard97} is still 
extremely unclear. In particular, even such basic questions as the 
universality class of the transition, remain highly controversial. Since 
simulations in this model have proved to be very difficult, due to 
extremely long crossover times, it appears that only a significant 
theoretical advance will settle the issue. However, the derivation of an 
appropriate effective field theory remains an unsolved and highly 
nontrivial task \cite{Janssen04}. Other higher-order processes also 
appear to display richer behavior than perhaps naively expected 
\cite{Odor04}.

\

\noindent
$\bullet$ Generally, the full classification of scale-invariant behavior
in diffusion-limited reactions remains a formidable program, especially 
in multi-species systems; see Refs.~\cite{Hinrichsen00,Odor04} for an
overview of the current data from computer simulations. To date, really 
only the many-species generalizations of the pair annihilation reaction
as well as the DP and dIP processes are satisfactorily understood.

\

\noindent
$\bullet$ An important, yet hardly studied and less resolved issue is the
effect of disorder in the reaction rates, especially for active to
absorbing state transitions. A straightforward analysis of DP with random
threshold yields runaway RG flows \cite{Janssen97a}, which seem to 
indicate that the presence of disorder does not merely change the value 
of the critical exponents, but may lead to entirely different physics 
(see, e.g., Ref.~\cite{Vojta04}). This may in turn require the further 
development of novel tools, e.g., real-space RG treatments directly aimed 
at the strong disorder regime \cite{Hooyberghs03,Hooyberghs04}.

\

\noindent
$\bullet$ In contrast with the many theoretical and computational 
successes, the subject of fluctuations in reaction-diffusion systems is 
badly in need of experimental contact. Up to this point, the impact of 
the field on actual laboratory (as opposed to computer) experiments has 
been very limited. In this context, the example of Directed Percolation 
(DP) seems especially relevant. DP has been found to be ubiquitous in 
theory and simulation, but is still mostly unobserved in experiments, 
despite some effort. Ideally, one would like to understand why this is 
the case: could it be due to disorder or to the absence of a true 
absorbing state? 

\

\noindent
$\bullet$ There are a number of additional extensions of the 
field-theoretic approach presented here that could further improve our
understanding of reaction-diffusion systems. For example, Dickman and 
Vidigal have shown how to use this formalism to obtain the full 
generating function for the probability distribution of simple 
processes \cite{Dickman02}; Elgart and Kamenev have used the field 
theory mapping to investigate rare event statistics \cite{Elgart04}; 
and Kamenev has pointed out its relation to the Keldysh formalism 
for quantum nonequilibrium systems \cite{Kamenev01}. Path-integral
representations of stochastic reaction-diffusion processes are now
making their way into the mathematical biology literature 
\cite{Pastor01,Jarvis04}.

\

\noindent 
We believe that these questions and others will remain the object of
active and fruitful research in the years ahead, and that the continued 
development of field-theoretic RG methods will have an important role to 
play.

\ack 
We would like to express our thanks to all our colleagues and
collaborators who have shaped our thinking on reaction-diffusion
systems, particularly Gerard Barkema, John Cardy, Stephen Cornell,
Olivier Deloubri\`ere, Michel Droz, Per Fr\"ojdh, Ivan Georgiev,
Melinda Gildner, Claude Godr\`eche, Yadin Goldschmidt, Malte Henkel,
Henk Hilhorst, Haye Hinrichsen, Hannes Janssen, Kent Lauritsen, Mauro
Mobilia, Tim Newman, Geza \'Odor, Klaus Oerding, Beth Reid, Magnus
Richardson, Andrew Rutenberg, Beate Schmittmann, Gunter Sch\"utz,
Steffen Trimper, Daniel Vernon, Fred van Wijland, and Mark
Washenberger. MH acknowledges funding from The Royal Society (U.K.).
UCT acknowledges funding through the U.S. National Science Foundation,
Division of Materials Research under grants No. DMR 0075725 and
0308548, and through the Bank of America Jeffress Memorial Trust,
research grant No. J-594. BVL acknowledges funding from the U.S.
National Science Foundation under grant No. PHY99-07949. 

\end{document}